\documentclass[prodmode,acmtap]{acmlarge}

\acmVolume{2}
\acmNumber{3}
\acmArticle{1}
\articleSeq{1}
\acmYear{2010}
\acmMonth{5}

\usepackage[ruled]{algorithm2e}
\usepackage{graphicx}
\usepackage{caption}
\usepackage{algorithmic}
\usepackage{amsmath}

\usepackage{graphicx}
\usepackage{caption}
\usepackage{algorithmic}
\usepackage{amsmath}
\usepackage{import}
\usepackage{csquotes}

\usepackage{url}
\usepackage{float}

\newcommand{\ignore}[1]{}
\newcommand{\ainote}[1]{\textcolor{blue}{\small [AI: #1]}}

\usepackage{url}
\usepackage{float}
\SetAlFnt{\algofont}
\SetAlCapFnt{\algofont}
\SetAlCapNameFnt{\algofont}
\SetAlCapHSkip{0pt}
\IncMargin{-\parindent}

\title{A Survey on Privacy and Security in Online Social Networks}
\author{Imrul Kayes \affil{University of South Florida}
Adriana Iamnitchi\affil{University of South Florida}
}
\begin{abstract}
Online Social Networks (OSN) are a permanent presence in today's personal and professional lives of a huge segment of the population, with direct consequences to offline activities. 
Built on a foundation of trust--users connect to other users with common interests or overlapping personal trajectories--online social networks and the associated applications extract an unprecedented volume of personal information. 
Unsurprisingly, serious privacy and security risks emerged, positioning themselves along two main types of attacks: attacks that exploit the implicit trust embedded in declared social relationships; and attacks that harvest user's personal information for ill-intended use. 
This article provides an overview of the privacy and security issues that emerged so far in OSNs.
We introduce a taxonomy of privacy and security attacks in OSNs, we overview existing solutions to mitigate those attacks, and outline challenges still to overcome. 
\end{abstract}

\category{H.5.2}{Computer-Communication Networks}{Distributed Systems, Distributed applications}

\terms{Privacy and Security}
\keywords{Privacy, Security, Online social networks}

\begin{document}

%\begin{bottomstuff}
%This work is supported by the Widget Corporation Grant \#312-001.\\
%Author's address: D. Pineo, Kingsbury Hall, 33 Academic Way, Durham,
%N.H. 03824; email: dspineo@comcast.net; Colin Ware, Jere A. Chase
%Ocean Engineering Lab, 24 Colovos Road, Durham, NH 03824; email: cware@ccom.unh.edu;
%Sean Fogarty, (Current address) NASA Ames Research Center, Moffett Field, California 94035.
%\end{bottomstuff}

\maketitle

% Head 1

\section{Introduction}

Online Social Networks (OSNs) have become a mainstream cultural phenomenon for millions of Internet users.
Combining user-constructed profiles with communication mechanisms that enable users to be pseudo-permanently ``in touch'', OSNs leverage users' real-world social relationships and blend even more our online and offline lives.
As of 2014, Facebook had 1.32 billion monthly active users %\footnote{http://newsroom.fb.com/company-info/} 
and it was the second most visited site on the Internet~\cite{Alexa2014Users}. 
Twitter, a social micro-blogging platform, claims over 500 million users~\cite{Holt2013Twitter}.
According to Nielsen's 2012 survey, social networking was the fourth most popular online activity~\cite{Nielsen2009OSN}.

Perhaps more than previous types of online applications, OSNs are blending in real life:  companies are mining trends on Facebook and Twitter to create viral content for shares and likes; employers are checking Facebook, LinkedIn and Twitter profiles of job candidates\footnote{http://goo.gl/kHJFI5}; law enforcement organizations are gleaning evidence from OSNs to solve crimes\footnote{http://www.cnn.com/2012/08/30/tech/social-media/fighting-crime-social-media/}; activities on online social platforms change political regimes~\cite{lotan2011arab} and swing election results\footnote{http://goo.gl/9A6FR}.
%Eventually, OSNs are bringing serious real-life consequences. 
%For example, while at one moment a photo of taking alcoholic shots at a party with friends may seem harmless, the photo may  give a negative impression about the person to an employer doing a background check. 

Because users in OSNs are typically connected to friends, family, and acquaintances, a common perception is that OSNs provide a more secure, private and trusted internet-mediated environment for online interaction~\cite{Cutillo2009Safebook}. 
%However, OSN privacy breaches and security issues make the news frequently. 
In reality, however, OSNs have raised the stakes for privacy protection because of the availability of an astonishing amount of personal user data which would not have been exposed otherwise. 
More importantly, OSNs expose now information from multiple social spheres -- for example, personal information on Facebook and professional activity on LinkedIn -- that, aggregated, leads to uncomfortably detailed profiles~\cite{Nissenbaum2004ContextOnline}.  % for \ainote{the high degree of the connectivity (e.g., personal data on Facebook or profession data in LinkedIn of a person) associated with the user data. -- to edit}

Unwanted disclosure of user information combined with the OSNs-induced blur between the professional and personal aspects of user lives allow for incidents of dire consequences. 
The news media covered some of these, such as the case of a teacher suspended for posting gun photos~\cite{Dam2009Facebook} or employee fired for commenting on her salary compared with that of her boss~\cite{Daily2001Banker}), both on Facebook. 
On top of this, social networks themselves intentionally (e.g., Facebook Beacon controversy~\cite{Dwyer2011Privacy}) or unintentionally (e.g., published anonymized social data used for de-anonymization and inference attacks~\cite{Narayanan2011Link}) are contributing to breaches in user privacy. 
Moreover, the high volume of personal data, either disclosed by the technologically-challenged average user or due to OSNs' failure to provide sophisticated privacy tools, %to protect users information 
have attracted a variety of organizations (e.g., GNIP\footnote{http://gnip.com/}, 80legs\footnote{http://80legs.com/}) that aggregate and sell user's social network data. 
%These organizations seek to aggregate users' social network data using publicly available social network APIs, and pose a serious threat on users' privacy due to user profiling from multiple social services.
In addition, the trusted nature of OSN relationships has become an effective mechanism for spreading spam, malware and phishing attacks. 
Malicious entities are launching a wide range of attacks by creating fake profiles, using stolen OSN account credentials sold in the underground market~\cite{Staff2010Sale} or deploying automated social robots~\cite{Wagner2012socialbot}.

This paper provides a  comprehensive review of solutions to privacy and security issues in OSNs. 
While previous literature reviews on OSN privacy and security are focused on specific topics, such as privacy preserving social data publishing techniques~\cite{zheleva2011privacy}, social graph-based techniques for mitigating Sybil attacks~\cite{Yu2011Sybil}, or OSN design issues for security and privacy requirements~\cite{zhang2010privacy}, we address a larger spectrum of security and privacy problems and solutions.
First, we introduce a taxonomy of attacks based on OSNs' stakeholders. % and with whom those stakeholders interact. %AI: I don't know/remember what you mean here.
We broadly categorize attacks as attacks on users and attacks on the OSN and then refine our taxonomy based on entities that perform the attacks.
These entities might be human (e.g., other users), computer programs (e.g., social applications) or organizations (e.g., crawling companies).
Second, we present how various attacks are performed, what counter-measures are available, and what are the challenges still to overcome.

\section{A Taxonomy of Privacy and Security Attacks in Online Social Networks}

We propose a taxonomy of privacy and security attacks in online social networks based on the stakeholders of the OSN and the forms of attack targeted at the stakeholders. %, when they are named. %with whom the stakeholders interact in the social network.
%A stakeholder is a person with an interest or concern in something.
We identify two stakeholders in online social networks: the OSN users and the OSN itself.
On one hand, OSN users share an astonishing amount of information ranging from personal  to professional; the misuse of this information can have significant consequences. %Their concern is that by sharing this information they are not putting themselves at risk. \marginpar{revise}
On the other hand, OSN services handle users' information and manage all users' activities in the network, being responsible for the correct functioning of its services and maintaining a profitable business model.  
Indirectly, this translates into ensuring that their users continue to happily use their services without becoming victims of malicious actions.

The distinction we make between OSN users and the OSN itself as stakeholders is defined by scale: isolated attacks on users may not affect the wellbeing of the OSN. 
However, a large attack on user population may translate into reputation damage, service disruption, or other consequences with direct effect on the OSN.  

We thus classify online social network privacy and security issues into the following attack categories (summarized in Table~\ref{tab:attacks}). 

\begin{enumerate}
\item Attacks on Users: these attacks are isolated, targeting a small population of random or specific users. We identify various such attacks based on the attacker:
	\begin{enumerate}
	\item Attacks from other users:
Users might put themselves at risk by interacting with other users, specially when some of them are strangers or mere acquaintances.
Moreover, some of these users may not even be human (e.g.,  social robots~\cite{Hwang2012Bot}), or may be crowdsourcing workers strolling and interacting with  users for mischievous purposes~\cite{Stringhini2013Follower}. 
Therefore, the challenge is to protect users and their information from other users.

	\item Attacks from social applications:
For enhanced functionality, users may interact with various third-party-provided social applications linked to their profiles.
%Social applications are written by third party developers and they are linked to the profiles of users.
To facilitate the interaction between OSN users and these external applications, the OSN provides application developers an interface through which to access user information. 
Unfortunately, OSNs put users at risk by disclosing more information than necessary to these applications.
Malicious applications can collect and use users' private data for undesirable purposes~\cite{Felt2008Privacy}.

	\item Attacks from the OSN:
Users' interactions with other users and social applications are facilitated by the OSN services, in exchange for, typically, full control over user's information published on the OSN. 
%Users trust the service and contribute their personal data.
%OSN states in  user agreement that all the user provided data becomes the property of the OSN.
While this exchange is explicitly stated in Terms of Service documents that the user must agree with (and supposedly read first), in reality few users understand the extent of this exchange ~\cite{fiesler2014copyright} and most users do not have a real choice if they don't agree with the exchange. 
Consequently, the exploitation by the OSN of user's personal information is seen as a breech of trust, and many solutions have been proposed to hide personal information from the very service that stores it. 

%Also, an OSN itself does not guarantee any security of user information (e.g.,  insider attackers from the OSNs (e.g., employee) could see or even modify a user's personal information~\cite{Megan2007Facebook} or hackers could steal user information).
%So, one challenge is to keep user information protected from the OSN.

 %\emph{AI: I think this is a gray area: the OSN states in user agreement that all the user provided data becomes the property of the OSN, I suppose. Then, whatever the OSN is doing with it cannot be really called an attack, can it? It may be a trust breach, but is it an attack? To revisit this after the corresponding section where the attack is described. }%IK: A body of research is recognizing this as an attack, they are saying solutions are needed. See how much/types of solutions are available in the relevant section. } 
% \ainote{can you give me a logical response to the point I raised about user agreements? There are some arguments that can help in Nissembaum's paper on privacy online, I seem to remember.}

	\item De-anonymization and inference attacks:
OSN services  publish social data for others (e.g., researchers, advertisers) to analyze and use for other purposes.
Typically, this data is anonymized to protect user information. %users and/or replacing their personally-identifiable attributes with random attributes.
%But social network data publishing involves privacy disclosure risks: 
However, an attacker can de-anonymize social data and infer attributes that the user did not even mention in the OSN (such as sexual or political orientation inferred from the association with other users).

	\end{enumerate}

\item Attacks on the OSN: these attacks are aimed at the service provider itself, by threatening its core business. 
\begin{enumerate}
	\item Sybil Attacks:
Sybil attacks are characterized by users assuming multiple identities to manipulate the outcome of a service~\cite{Douceur2002Sybil}. 
Not specific to OSNs, Sybil attacks were used, for example, to determine the outcome of electronic voting~\cite{RILEYYoutube}, to artificially boost the popularity of some media~\cite{ratkiewicz2011detecting}, or to manipulate social search results~\cite{Matt2011GoogleSearch}. 
However, OSNs have also become vulnerable to Sybil attacks: by controlling many accounts, Sybil users are illegitimately increasing their influence and power in the OSNs~\cite{YuSybilGuard2006}.
%\ainote{you need references in this section. For example, cite the paper that introduced the concept of Sybil attacks in this case.}

	\item Crawling attacks:
%OSNs provide APIs to crawl publicly viewable data.
%Unfortunately, 
Large-scale distributed data crawlers from professional data aggregators exploit the OSN-provided APIs or scrape publicly viewable profile pages to build databases from user profiles and social links.
Professional data aggregators sale such databases to insurance companies, background-check agencies, credit-ratings agencies, or others~\cite{Bonneau2009Prey}. 
Crawling users' data from multiple sites and multiple domains increases profiling accuracy. % and endangers social network users.
This profiling might lead to ``public surveillance'',  where an overly curious agency (e.g., government) could monitor individuals in public through a variety of media~\cite{Nissenbaum2004Context}.

	\item Social Spam:
Social spam are contents or profiles that an OSN's ``legitimate'' users don't wish to receive~\cite{Heymann2007Spam}.
%Social spam is unwanted content designed to mislead users. \ainote{is this correct? Marketing is still spam, even if not necessarily misleading.} \ainote{you perhaps need to define what \textbf{social} spam is, not only spam}
Spam undermines resource sharing and hampers interactivity among users by contributing phishing attacks, unwanted commercial messages, and promoting websites.
Social spam spreads rapidly via OSNs due to the embedded trust relationships among online friends, which motivates a user to read messages or even click on links shared by her friends. 
%These messages and links are perfect vehicles for social spam. \ainote{they are vehicles? aren't messages the spam itself?}

	\item Distributed Denial-of-service attacks (DDoS).
DDoSes are common forms of attacks, where a service is sent a large amount of seemingly inoffensive service requests that overload the service and deny access to it~\cite{Mirkovic2004DOS}. 
As many popular services, OSNs are also subjected to such coordinated, distributed attacks.

	\item Malware Attacks:
Malware is the collective name for programs that gain access, disrupt computer operation, gather sensitive information, or damage a computer without the knowledge of the owner.
ONSs are being exploited for propagating malware~\cite{Facebook2012Koobface}.
Like social spam, malware propagation is rapid due to the  trust relationships in social networks.
	\end{enumerate}

\end{enumerate}

%We discuss the other entities with whom users and OSNs interact and risks associated with the interactions.

%%\begin{itemize}

%%\item \emph{Attacks from other users}: 

%%\item \emph{Attacks from social applications}: 

%%\item \emph{Attacks from the OSN}: 

%%\item \emph{De-anonymization and inference attacks}: 

%%\item \emph{Crawling attacks}: 

%%\item \emph{Sybil, spam, DDoS and malware attacks}: 

%%\end{itemize}

\begin{table}
\centering
\begin{tabular}{|l|l|}
\hline
Attacks on Users & Attacks on the OSN\\
\hline
\hline

Attacks from other users&Sybil attacks\\
\hline
Attacks from social applications&Crawling attacks\\
\hline
Attacks from the OSN & Social spam\\
\hline
De-anonymization and inference attacks  & Distributed Denial-of-service attacks (DDoS)\\
\hline
& MalwareAttacks\\
\hline
\end{tabular}
\caption{Categories of attacks.}
\label{tab:attacks}
\end{table}

%\begin{figure}[htbp]
%\centering
%\includegraphics[height=7.5cm]{figureDraw/InforShare}
%\caption{Interactions in an online social network. AI: this picture can be significantly improved: it has some inconsistencies and also it can carry much more information (although I am not sure what information it needs). First, the users who attack are part of the OSN, so they should be represented the same as the other users -- but perhaps in a different color? -- and within the square of the OSN. Second, perhaps ``researchers'' are not really the attackers, are they? Perhaps we can call them ``analytics'' or something else? Third, what is the meaning of the arrows? It should be explicitly stated in the figure caption -- especially since it's not intuitive -- some are bi-directed, some are uni-directed.}
%\label{fig:dataShare}
%\end{figure}

The rest of the paper is organized as follows.
Mitigating attacks on users (Sections~\ref{otherusers} to~\ref{inference}) include discussions of attacks from other users (Section~\ref{otherusers}), from social applications (Section~\ref{apps}), from the OSN itself (Section~\ref{osn}), and de-anonymization and inference attacks (Section~\ref{inference}). 
Mitigating attacks on the OSN (Sections~\ref{sybil} to~\ref{malware}) includes a discussion of Sybil attacks (Section~\ref{sybil}), crawling attacks (Section~\ref{crawl}), social spam (Section~\ref{spams}), distributed denial-of-service attacks (Section~\ref{ddos}) and malware (Section~\ref{malware}). 
Finally, we conclude the paper in Section~\ref{summary}.

%Section~\ref{otherusers} discusses mitigating attacks from other users.
%Section~\ref{apps} discusses mitigating attacks from social applications. 
%Section~\ref{osn} discusses mitigating attacks from the OSN. 
%We discuss mitigating de-anonymization and inference attacks in Section~\ref{inference}.

\section{MITIGATING ATTACKS FROM OTHER USERS}
\label{otherusers}

%Ouline
%What type of data OSNs reveals 
Users reveal an astonishing amount of personally identifiable information on OSNs, including physical, psychological, cultural and preferential attributes. 
For example, Gross and Acquisti's study~\cite{Gross2005InformationRevelation} shows that 90.8\% of Facebook profiles have an image, 87.8\% of profiles have posted their birth date, 39.9\% have revealed phone number, and 50.8\% profiles show their current residence.
The study also shows that the majority of users reveal their political views, dating preferences, current relationship status, and various interests (including music, books, and movies). 

%How this information revelation is related to privacy
Due to the diversity and specificity of the personal information shared on OSNs, users put themselves at risk for a variety of cyber and physical attacks.
Stalking, for example, is a common risk associated with unprotected location  information\footnote{http://www.theguardian.com/technology/2012/feb/01/social-media-smartphones-stalking}. 
%A potential adversary could exploit user reported location information to stalk.
Demographic re-identification was shown to be doable: 87\% of the US population can be uniquely identified by gender, ZIP code and full date of birth~\cite{Sweeney2000Uniqueness}. 
Moreover, the birth date, hometown, and current residence posted on a user's profile are enough to estimate the user's social security number and thus expose the user to identity theft~\cite{Gross2005InformationRevelation}.
Unintended revealing of personal information brings other online risks, including scraping and harvesting~\cite{Lindamood2009IPI,Strufe2010PPB}, social pushing~\cite{Jagatic2007Phishing}, and automated social engineering~\cite{Bilge2009Contacts}.

%Given the amount of personal data users share in OSNs, an important question is how users data will be protected from 
Given the amount of sensitive information users expose on OSNs and the different types of relationships in their online social circles, the challenge OSNs face is to provide the correct tools for users to protect their own information from others while taking full advantage of the benefits of information sharing. 
This challenge translates into a need for fine-grained settings, that allow flexibility within a type of relationships (as not all friends are equal~\cite{Banks2009AllFriends,Cummings2002QOS}) and flexibility with the diversity of personal data. 
However, this fine granularity in classifying bits of personal information and social relationships leads to an overwhelmingly complex cognitive task for the user. 
Such cognitive challenges worsen an already detrimental user tendency of ignoring settings all together, and blindly trusting the default privacy configurations that serve the OSN's interests rather than the user's.

Solutions to these three challenges are reviewed in the remainder of this section.
Section~\ref{sec:fine-grained} surveys solutions that allow fine tunings in setting protection of personal data. 
The complexity challenge is addressed in the literature on two planes: by providing a visual interface in support of the complex decision that the user has to make (Section~\ref{sec:visual}) and by automating the privacy settings (Section~\ref{sec:automated}). 
To address the problem of users not changing the platform's default settings, researchers proposed various solutions presented in Section~\ref{sec:default}.

%Privacy is understood as an individual's right to determine to what extent her data shall be communicated to others~\cite{Nissenbaum2010ContextBook}. 
%Charles Fried defines privacy as not simply an absence of information about us in the minds of others, rather the control we have over information about ourselves~\cite{FriedPrivacy}.
%Regan considers privacy as the right to control information about and access to oneself~\cite{ReganPrivacy}.

\ignore{
\ainote{what about the following paragraph instead of the one above: if it makes sense, maybe you want to finish it up and polish it?}
Practically all OSNs provide some tools for user privacy protection from other OSN users and from the outside world, yet the number and impact of privacy incidents is overwhelming~\cite{trrue?}.
We identify two main, interrelated reasons addressed extensively in the literature: inadequate tools and user behavior.
To address the lack of inadequate tools, researchers proposed ... (and cite all papers on this topic).

However, appropriate tools require to be used in order to be effective. 
Yet studies~\cite{} have shown that tools are not used because they are too complicated.  
Even worse, users are often unaware of or unconcerned by the privacy risks they are exposed to. 

\ainote{There might be a tradeoff here between tools and behavior -- we need perhaps to articulate it. }
}

%---------------------------------------------
\subsection{Fine-grained Privacy Settings} 
\label{sec:fine-grained}
%---------------------------------------------

Fine-grained privacy advocates~\cite{Krishnamurthy2008Characterizing,Simpson2008FineGrained} argue that fine-grained privacy controls are crucial features for privacy management.
Krishnamurthy et al.~\cite{Krishnamurthy2008Characterizing} introduce privacy ``bits''---pieces of user information grouped together for setting privacy controls in OSNs.
In particular, they categorize a user's data into multiple pre-defined bits, namely thumbnail (e.g., user name and photo); greater profile (e.g., interests, relationships and others); list of friends; user-generated content (such as photos, videos, comments and links) and comments (e.g., status updates, comments, testimonials and tags about the user or user content).
Users can share these bits with a wide range of pre-defined users, including friends, friends of friends, groups, and all.
Current OSN services (e.g., Facebook and Google+) have implemented this idea by allowing users to create their own social circles and to define which pieces of information can be accessed by which circle. 

%%Moreover, given that all friends are not equal, a user might want to further sub-divide friends into smaller groups and share sensitive information with more trusted friends or groups.

To help users navigate the amount of social information necessary for setting correct fine-grained privacy policies, researchers suggest various ways to model the social graph.
One model is based on ontologies that exploits the inherent level of trust associated with relationship definition to specify privacy settings.  
Kruk~\cite{Kruk2004FOAF} proposes Friend-of-a-Friend (FOAF)-Realm, an ontology-based access control mechanism that uses RDF to describe relations among users.
The system uses a generic definition of relationships (``knows'') as a trust metric and generate rules that control a friend's access to resources based on the degree of separation in the social network.
Choi et al. ~\cite{Choi2006Trust} propose a more fine-grained approach, which considers named relationships (e.g., ``worksWith'', ``isFriendOf'', ``knowsOf'') in modeling the social network and the access control. 
A more nuanced trust-related access control model is proposed  by Carminati et al. ~\cite{Carminati2006Trust} based on relationship type, degree of separation, and a quantification of trust between users in the network.

%An inherent challenge in trust-based privacy models is that the trust threshold values should be smoothed as much as possible. 
%In practice, it might be difficult to comprehend and specify appropriate trust thresholds without  prior threshold value tuning experiments.

%Along with the degree of separation, relationship type could be also used  to quantify the level of relationship between two users.
%Trust-based privacy solutions compute a trust level (often a quantitative value) among two individuals and use it in defining privacy settings.

For more fine-grained ontology-based privacy settings, semantic rules have been used. 
Rule-based policies represent the social knowledge base in an ontology and define policies as Semantic Web Rule Language (SWRL) rules.
SWRL\footnote{http://www.w3.org/Submission/SWRL/} is a  language for the Semantic Web, which can represent rules as well as logic.
Researchers used SWRL to express access control rules that are set by the users. 
Finally, access request related authorization is provided by reasoning on the social knowledge base. 
Systems that leverage OWL and SWRL to provide rule-based access control framework are ~\cite{Elahi2008Rule,Carminati2009Rule,Masoumzadeh2011Rule}.
Although conceptually similar, ~\cite{Carminati2009Rule} provides richer OWL ontology and different types of policies; access control policy, admin policy and filtering policy. 
A more detailed semantic rule-based model is developed by Masoumzadeh and Joshi~\cite{Masoumzadeh2011Rule}.
Rule-based privacy models have two challenges to overcome. 
First, authorization is provided by forward reasoning on the whole knowledge base, challenging scalability with the size of the knowledge base. 
Second, rule management is complex and requires a team of expert administrators~\cite{EngelmoreAI}.

Role and  Relationship-Based Access Control (ReBAC) are other types of fine-grained privacy models that employ roles and relationships in modeling the social graph. 
The working principle of these models is two-fold: 1) track roles or relationships between resource (e.g., photos) owner and the resource accessor; 2) enforce access control policies in terms of the roles or relationships. 
Fong~\cite{Fong2011Relation} proposes a ReBAC model based on the context-dependent nature of relationships in social networks. 
This model targets social networks  that are poly-relational (e.g., teacher-student relationships are distinct from child-parent relationships), directed (e.g., teacher-student relationships are distinct from student-teacher relationships) and tracks multiple access contexts that are organized into a tree-shaped hierarchy.
When access is requested in a context, the relationships from all the ancestor contexts are combined with the relationships in the target access context to construct a network on which authorization decisions are made.
Giunchiglia et al.~\cite{Giunchiglia2008RelBAC}  propose RelBac, another relation-based access control model to support sharing of data among large groups of users.
The model defines permissions as relations between users and data, thus separating them from roles.
The entity-relationship model of RelBac enables description logics and as well as the reasoning for access control policies.
%The formalization of the RelBac model as an entity-relationship model allows for its direct translation into description logics, which also allows reasoning for access control policies. 

In practice, many online social networks (such as Facebook) have already implemented fine-grained controls.
A study of Bonneau et al.~\cite{Bonneau2010PrivacyJungle} on 29 general purpose online social network sites shows that 13 of them offer a line-item setting where individual data items could be set with different visibility. 
These line-item settings are granular (one data item is one `bit') and flexible (users can change social circles).

%---------------------------------------------
\subsection{View-centric Privacy Settings}
\label{sec:visual}
%---------------------------------------------

Lack of appropriate visual feedback has been identified as one of the reasons for confusing and time consuming privacy settings~\cite{Strater2008StrategisandStruggles}.
View-centric privacy solutions %s~\cite{Lipford2008Understanding,Paul2012C4PS} 
are built on the intuition that a better interface for setting privacy controls can impact users' understanding of privacy settings and thus their success in correctly exercising privacy controls.
These solutions visually inform the user of the setting choices and consequences of his choices. 

%allow users to have a better mental model of privacy.
%It is known that  the 
%Current approaches to privacy settings (\ainote{such as what?}) are considered difficult to understand for users and, to some extent, fundamentally flawed~\cite{Madejski2011Failure}. \ainote{why are they considered fundamentally flawed?}
%Hence, view-centric privacy settings could provide more usability.

In~\cite{Lipford2008Understanding}, the authors propose an alternative interface for Facebook privacy settings.
This interface is a collection of tabbed pages, where each page shows a different view of the profile as seen by a particular audience (e.g., friends, friends of friends, etc), along with controls for restricting the information shared with that group. 
While this solution provides visual feedback on how other users will see her profile, it's  management is tedious for users with many groups. 
%A user can browse those tabbed pages and see how her profile will be presented to other users.
%%This approach poses a challenge in sites like Facebook where group-based access control is already in place (e.g., one can create a group by adding a list of frinds). %\ainote{This assumes that the audience mentioned above is defined by Facebook groups -- is this what the paper states? If so, then we should use the same term for audience and groups.}
%%If a user has a moderate number of groups, he will be presented a lot of tabs in the privacy setting.
%%Going through all the tabs and defining privacy settings for each data item would be tedious.

A simpler interface is proposed by \emph{C4PS} (Colors for Privacy Settings)~\cite{Paul2012C4PS}, which applies color coding for different privacy visibilities to minimize the cognitive overhead of the authorization task.
This approach applies four color schemes for different groups of users; red--visible to nobody; blue--visible to selected friends; yellow--visible to all friends; and green--visible to everyone.
A user can change the privacy setting for a specific data item by clicking the buttons on the edge of the attribute. 
The color of the buttons shows the visibility of the data.
If users click ``selected friend'' (blue) button, a window will open in which friends or groups (a pre-defined set of friends) are granted access to the data item. 

A similar approach is implemented in today's most popular OSNs in different ways. 
For example, Facebook provides a dropdown of viewers (e.g., only me, friends, and public) with icons as visual feedback. 
In the custom setting, users can set more granular scales, e.g., share the data item with friends of friends, friends of those tagged and restrict sharing with specific people or lists of people. 

%---------------------------------------------
\subsection{Automated Privacy Settings}
\label{sec:automated}
%---------------------------------------------

Automated privacy settings methods employ machine learning %in an attempt to decrease users' explicit acts of access authorization % by applying methods from machine learning.
%The ultimate goal of this type of solutions~\cite{Fang2010Wizard,Adu2008Social,Danezis2009Privacy} is 
to automatically configure a user's privacy setting with minimal user effort.

Fang and  Lefevre's \emph{privacy wizard}~\cite{Fang2010Wizard} iteratively asks a user about his privacy preferences (\emph{allow} or \emph{deny}) for specific (\emph{data item, friend}) pairs. % $(i,j) \in I\times F$, where $I$ is the set of data items in the user's profile and $F$ is the user's set of friends.
The wizard constructs a classifier from these preferences, which automatically assigns privileges to the remaining of the user's friends.
The classifier considers two types of features: community structure (e.g., to which community a friend  of the user belongs) and profile information (such as age, gender, relationship status,  education, political and religion views, work history).
The classifiers employed (NaiveBayes, NearestNeighbors and Decision Tree) use uncertainty sampling~\cite{Lewis1994SAT}, an active learning paradigm, acknowledging the fact that users may quit labeling friends at any time.

%\ainote{If it's only two authors, then it's Fang and Lefevre. If it's more than two authors, then it's Fang et al. I've noticed this before in your survey, please fix all instances. Also, there is absolutely no oder in References, which makes it impossible to find anything. Is this the format? Can you order it alphabetically, even if the format doesn't specify it? I find it maddening and totally useless to have 100+ references unsorted} privacy wizard~\cite{Fang2010Wizard} iteratively asks a user's preferences (\emph{allow or deny}) for specific (\emph{data item, friend}) pairs $(i,j) \in I\times F$, where $I$ is the set of data items in the user's profile and $F$ is the set of friends of the user.

\emph{Social Circles}~\cite{Adu2008Social} is an automated grouping technique that analyzes the users' social graph to identify ``social circles'', clusters of densely and closely connected friends.
The authors posit social circles as uniform groups from the perspective of privacy settings. 
The assumption is that users will share the same information with all friends in a social circle.
Hence, friends are automatically categorized into social circles for different circle-specific privacy policy settings.
To find the social circles, they used a  $(\alpha, \beta)$ clustering algorithm proposed in~\cite{Mishra2007Clustering}.
While convenient, this approach limits users' flexibility in changing the automate settings.
% \ainote{``might'' or ``will''? Also, I am not sure this is an intuition or an assumption -- depending on how much freedom to change this setting the user has. Which is an important question for this subsection, anyway: are users allowed to change the automatic settings? Or are they stuck with them? This is perhaps something to clarify in the first paragraph} 

Danezis~\cite{Danezis2009Privacypolicy} aims to infer the \emph{context} within which user interactions happen, and enforces policies to prevent users that are outside that context from seeing the interaction.
Conceptually similar to Social Circles, contexts are defined as cohesive groups of users, e.g., groups that have many links within the group and fewer links with non-members of the group. 
The author used a greedy algorithm to extract the set of groups from a social graph.

An inherent tradeoff for this class of solutions is ease of use vs. flexibility: while the average user might be satisfied with an automatically-generated privacy policy, the more savy user will want more transparency and possibly more control. 
To this end, the privacy wizard~\cite{Fang2010Wizard} provides for advanced users the visualization of a decision tree model and tools to change it.
Another challenge for some of these solutions is bootstrapping: a newcomer in the online social network has no history of interactions to inform such approaches.
%%There are several challenges this class of solutions still has to address.
%%First, reduced user input may be a good choice for average users, but advanced users might be frustrated with the lack of transparency on how the automated privacy has been generated.
%%Fang et al.'s privacy wizard~\cite{Fang2010Wizard} provides a visualization of a decision tree model for advanced users.
%%Users can change the decision tree and generate custom privacy settings.
%%However, even for advanced users, this decision tree is overly complex to handle. 
%%Second, bootstrapping privacy settings is a concern: a newcomer in the online social network has no history of interactions to inform such approaches.
%This user is not well informed of the system to generate privacy policies.
%And finally, the dynamic nature of the social networks in terms of friendship creation and content generation will pose challenges on when and how frequent the systems have to generate privacy policies.  
%\ainote{develop this. It reads more like you trying hard to find issues with this approach. Can you cite behaviors in OSN analysis that would really pose a challenge in this direction?}

%Tools are not enough, the consequence of default privacy and solutions to the problems

%---------------------------------------------
\subsection{Default Privacy Settings}
\label{sec:default}
%---------------------------------------------

%\ainote{shouldn't section titles be with capital letters? check the sample pdf, please}
%%%Providing the best privacy solution will not work, if users do not take advantages of the solutions. 
%\ainote{couldn't the automated privacy settings be used as default and personalized privacy settings? What is really the difference between them?}
%%%Unfortunately, this is often the case in OSNs. 
%%%Although increasing privacy awareness~\cite{Hargittai2010Facebook,acquisti2006imagined}, there is a disconnect between users desire to protect privacy and their behaviors. \ainote{any support for this claim? Is it directly relevant for the discussion? What is the point you're trying to make, anyway?}

Studies have shown that users on OSNs often do not take advantage of the privacy controls available. 
For example, more than 99\% Twitter users retained the default privacy setting where their name, list of followers, location, website, and biographical information are visible~\cite{Krishnamurthy2008Twitter}. 
Similarly, the majority of Facebook users has default settings~\cite{Gross2005Revelation,acquisti2006imagined,Krishnamurthy2008Characterizing}. %\ainote{are both of these correct citations?} % show that t or permissive privacy settings. 
Under-utilization of privacy options are mostly due to poor privacy setting interface~\cite{Lipford2008Understanding}, intricate privacy settings~\cite{Madejski2011Failure}, %permissive default settings~\cite{Gross2005Revelation} \ainote{logic?}
and inherent trust in OSNs~\cite{acquisti2006imagined,Boyd2004Friendster}.
The problem with not changing the default settings is that they almost always tend to be more open that the users would prefer~\cite{Liu2011Facebook}.
To overcome this situation, approaches to automatically generate more appropriate default privacy settings have been proposed. %, either from some example settings~\cite{Squicciarini2010Prima,Shehab2010PolicyMgr}, or using real world social theories~\cite{Kayes13Aegis,Kayes13Out_of_the_wild}.
 
\emph{PriMa}~\cite{Squicciarini2010Prima} automatically generates privacy policies, acknowledging the fact that the average user will find the task of personalizing his access control policies overwhelming, due to growing complexity of OSNs and the diversity  of user content.
The policies in PriMa are generated based on the average privacy preference of similar and related users, the accessibility of similar items from similar and related users, closeness of owner and accessor (measured by the number of common friends), the popularity of the owner (i.e., popular users have sensitive profile items), etc. 
%Access control policies for profile items are finally generated aggregating these factors.
%However, this approach is vulnerable to highly volatile policies due to changes in these factors. \ainote{are you saying this or is it acknowledged in the paper? I doubt it is a significant problem because 1) SNs are rather stable; 2) I doubt people will fuss with their settings too much as to cause frequent changes in other user's settings. I would cut these observations if you cannot back them up with some numbers or intuition that addresses these arguments (unless the authors acknowledge this limitation, in which case you should say it explicitly.}
However, a large number of factors and their parametrized tuning contribute to longer policy generation and enforcement time.
%Unfortunately, this limitation is not addressed, so it is difficult to judge their impact on practice. 
A related approach, \emph{PolicyMgr}~\cite{Shehab2010PolicyMgr}, uses supervised learning of user-provided example policy settings and builds classifiers that are then used for automatically generating access control policies. % to regulate access to user profile objects.

\emph{Aegis}~\cite{Kayes13Aegis,Kayes13Out_of_the_wild} is a privacy framework and implementation that leverages the \emph{`Privacy as Contextual Integrity'} theory proposed by Nissenbaum~\cite{Nissenbaum2004Context} for generating default privacy policies.
Unlike the approaches just presented above, this solution does not need user input or access history. 
Instead, it aggregates social data from different OSNs in an ontology-based data store and then applies the two norms of Nissembaum's theory to regulate the flow of information between social spheres and access to information within a social sphere.

\section{Mitigating Attacks From Social Applications}
\label{apps}

Social applications, written by third-party developers and running on OSN platforms, provide enhanced functionality linked to a user profile. 
For example, Candy Crush Saga\footnote{https://www.facebook.com/candycrushsaga} (a social game) and Horoscopes\footnote{https://www.facebook.com/dailyhoroscopes} (users can check horoscope) are two popular social applications on Facebook.

%\ainote{Open Social is not an OSN, is it? That's what you suggested below (text now commented) In fact, do you have examples of 3-party apps running on different OSNs? If so, make sure the discussion below -- about how they work -- is general enough. If not, feel free to be specific to each platform (2 platforms at most, I'd think).}.
%Two major social networking platforms Facebook and Open Social allow third-party applications on their platforms.

\begin{figure}[htbp]
\centering
\includegraphics[height=4.3cm]{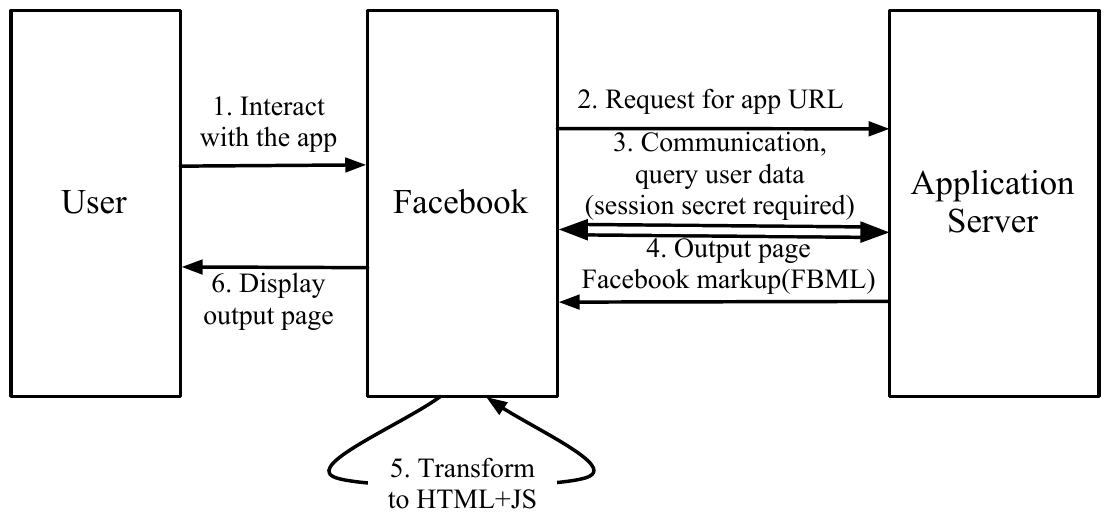}
\caption{Data flow in a Facebook application.}
\label{fig:FacebookDF}
\end{figure}

%\ainote{to rewrite for flow: follow some logical sequence of actions} 
The social networking platform works as a proxy between users and applications and mediates the communication between them.
To better understand this proxy, we show data flow between a third-party social application and the Facebook platform in Figure~\ref{fig:FacebookDF}.
An application is hosted on a third-party server and runs on user's data that are taken from the Facebook platform.
When a user installs the application on Facebook, it takes permission from the user to use some of her profile information.
Application developers write the application pages of an application using  Facebook mark-up language (FBML)---a subset of HTML and CSS extended with proprietary Facebook tags.

When a user interacts with an application, such as clicks an application icon on Facebook to generate horoscopes (step 1 on Figure~\ref{fig:FacebookDF}), Facebook requests the page from the third-party server where the application is actually hosted (step 2).
The application requests the user's profile information using secret communication with Facebook (step 3).
The application uses the information (e.g., birth date may be used to create horoscopes) and returns a FBML page to Facebook (step 4).
Facebook finally transforms the application page from the server by replacing the FBML page with standard HTML, JavaScript (step 5), and transmits the output page to the end user (step 6).

OSN users are facing multiple risks while using social applications.
First, an application might be malicious; it could collect a high volume of user data for unwanted usage.
For example, to show this vulnerability, BBC News developed a malicious application that could collect large amounts of user data in only three hours~\cite{Kelly2008Risk}.

%Social networking platforms put users at risk by disclosing a large amount of personal information to these applications. \ainote{You may want to be more specific here: the problem may be that Facebook gives more information than what the user agreed to give. If so, say it explicitly and cite sources. Otherwise, if Facebook is giving away what the user agreed to, I don't see the problem.}
%A malicious application, thus, could collect a high volume of user data for unwanted usage.
%For example, to show this vulnerability, BBC News developed a malicious application that could harvest large amounts of user profile data in just three hours~\cite{Kelly2008Risk}.

Second, application developers can violate developer policies to control user data.
Application developers are supposed to abide by a set of rules set by the OSNs,  called \textit{``developer policies''}.
Developer polices are intended to prohibit application developers from misusing personal information or forwarding it to other parties.
However, reported incidents~\cite{Mills2008Risk,Steel2010RIsk} show that applications violate these developer policies.
For example, a Facebook application, ``Top Friends'' enabled everyone to view the birthday, gender and relationship status of all Top Friends users, even though those users kept their privacy for those information to private~\cite{Mills2008Risk}, violating the developer policies that private information of friends are not accessible. %\ainote{what's wrong with that, assuming that the application clearly stated its intent at installation and the user agreed with this? Does it violate a developer's policy? If so, which one? Be specific.}
The Wall Street Journal finds evidence  that  Facebook applications  transmit identifying information to advertising and tracking companies~\cite{Steel2010RIsk}.

%In another risk direction \ainote{if this is the 3rd type of risk, then structure the text with first, second, third and state above that there are three main types of risks associated with 3-rd party applications running on OSNs}, 
Finally, third-party social applications can query more data about a user from an OSN, regardless whether needed or not for proper operation. 
A study by Felt and Evans~\cite{Felt2008Privacy} of 150 of the top applications on Facebook shows that most of the applications only needed user name, friends, and their networks. 
However, 91\% of social networking applications have accessed data that they do not need for operation.
This violates the principle of least privilege~\cite{Saltzer1975Least}, which states that every user should only get the minimal set of access rights that enables him to complete his task.

%Applications-aware privacy research solutions attempt to
We identified three classes of solutions that attempt to minimize the privacy risks stated above: (i)~by anonymizing social data made available to applications (Section~\ref{appn-anno}); (ii)~by defining and enforcing more granular privacy policies that the third-party applications have to respect (Section~\ref{privacy_policy}); and (iii)~by providing third-party platforms for executing these applications and limiting the transfer of the social data from applications to other parties (Section~\ref{platform}).

\subsection{Anonymizing Social Data For Third-party Applications}
\label{appn-anno}

Privacy-by-proxy~\cite{Felt2008Privacy} uses special markup tags that abstract user data and handle user input.
Third-party applications do not have access to users' personal data, rather they use users' IDs and tags to display data to users.
For example, to display a user's hometown, an application would use a tag $<$hometown id=``3125''/$>$.
The social network server would then replace the tag with real data value (e.g.,  ``New York'') while rendering the corresponding page to the user.
However, applications might rely on private data for operations, for example a horoscope application might require users' gender information.
A conditional tag handles this dependency (e.g., $<$if-male$>$ tag can choose the gender of an avatar). 
%Social graph information embodied in the user's friend list are provided to the applications using graph anonymization.  
Privacy-by-proxy ensures privacy by limiting what applications can access, which might also limit the social value and usability of  the applications. 
Data availability through proxy also means that application developers have to expose the business logic to social network sites (in a form of Javascript to end users).
This might discourage third-party developers in the first place.
Moreover, applications could still develop learning mechanisms to infer attributes of a user.
For example, developers might include scripting code in the personal data dependent conditional execution blocks (if-else) that could send information to an external server when the block executes.

\begin{figure}[htbp]
\centering
\includegraphics[height=5cm]{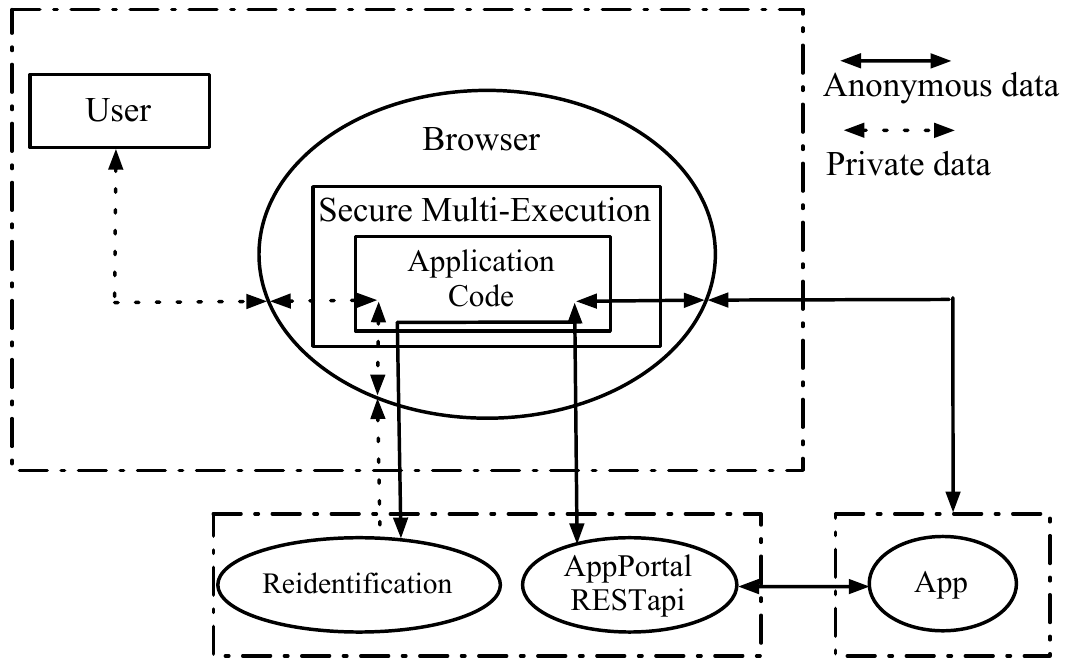}
\caption{Data flow in a  PESAP aware browser~\cite{Reynaert2012PESAP}.}
\label{fig:PESAP}
\end{figure}

Similar to Privacy-by-proxy, PESAP~\cite{Reynaert2012PESAP} provides anonymized social data  to applications.
However, PESAP secures the information flow inside the browser, so that  applications cannot do information leakage though outgoing communications with other third-parties.
The anonymization is provided by encrypting the IDs of the entities of the social graph with an application-specific symmetric key.
Applications use a REST API to get access to the anonymized social graph.
PESAP provides a re-identification end-point in order to enable users to see the personal information of their friends in the context of social applications.
Secure information flow techniques protect the private information in the browser of a user. 
This is done by a dynamic, secure multi-execution flow technique~\cite{Devriese2010Multiexecution}, which analyzes information flow inside a browser and ensures that the flow complies with certain policies.
The multi-execution flow technique labels the inputs and the outputs of the system with security labels and runs a separate sub-execution of the program for each security label.
The inputs have designated security labels and can be accessed by a sub-execution having the same or a higher security label.
Figure~\ref{fig:PESAP} shows the data flow in a PESAP-aware browser.

%The goal is to track the personal information inside the application's code in the browser, and to prevent information leakage though outgoing communication.
% 

\subsection{Enforcing Additional Privacy Policies To Social Applications}
\label{privacy_policy}

Besmer et at.~\cite{Besmer2009Framework} propose an access control framework for applications, which adds a new user-application policy layer on the top of the user-user policy to restrict the information applications can access.
Upon installing an application, a user can specify which profile information the application could access.
However, the framework still uses user-to-user policy to additionally govern an application's access to friends' information on behalf of 
the user (Alice's installed applications will not get her friend Bob's private data if user-user policy of Bob denies Alice to do so).
An additional \emph{friendship-based protection} restricts the information the application can request of a user's friends.
For example, Alice installs an application which requests her friend Bob's information and Bob did not install the application.
Consider that Bob's default privacy policy is very permissive.
But  Alice is a privacy conscious and she allows applications to access only the Birth Date attribute.
According to friendship-based protection, when the application will request Bob's information via Alice, it will only be able to get Bob's birth date. 
So, friendship-based protection enables Alice's privacy policies to extend to Bob.
%\ainote{I don't understand this example: does Alice's privacy policy now extends to Bob's? I understand Alice's information other than her birthdate to be hidden to the application. What happens if Bob has the application installed as well?}
The model works well for privacy-savvy concerned users who make informed decisions about an application's data usage while installing an application.
An additional functionality could be a set of restrictive default policies for average users.
%However, the framework lacks default policies for applications. 
%Average users will still not be able to protect themselves from third-rd party applications. %vulnerabilities.

\subsection{Third-party Platforms For Running Social Applications}
\label{platform}
\begin{figure}[htbp]
\centering
\includegraphics[height=6cm]{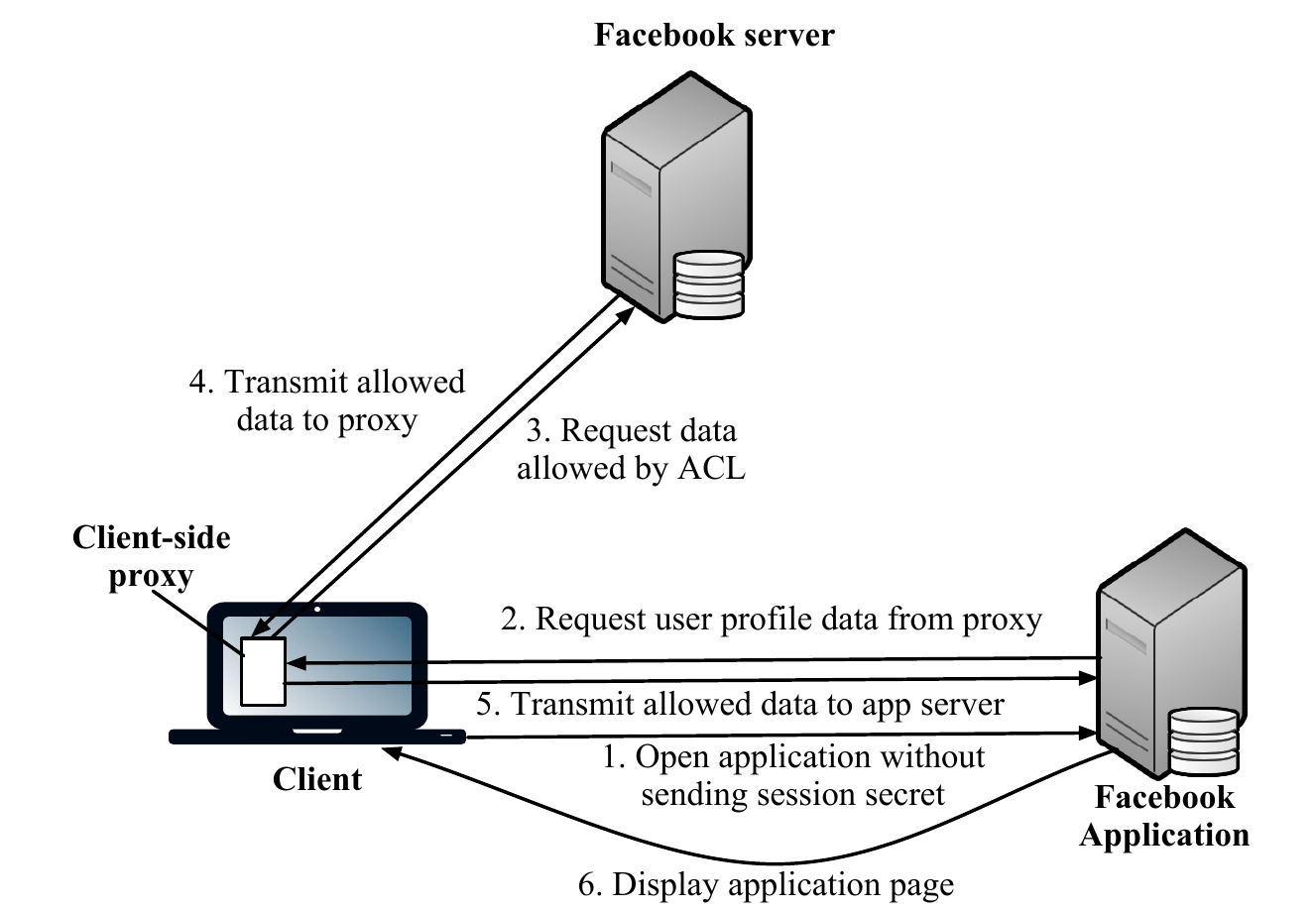}
\caption{A data-flow between applications and server with PoX~\cite{Egele2012Pox}.}
\label{fig:PoX}
\end{figure}

%PoX~\cite{Egele2012Pox} is a proxy that runs on client side and provides fine-grained access control capabilities over which parts of a social network user's private information can be accessed by third-party applications.
Egele et al.~\cite{Egele2012Pox} note that, since popular OSN services such as Facebook did not implement user-defined access control mechanisms to date, pragmatic solutions should not rely on the help of OSNs.
They introduce PoX, %a proxy that overcomes the design challenges; it's 
a browser extension for Facebook applications that runs on a client machine and  works as a  proxy to provide fine-grained access controls.
PoX works as a reference monitor which sits between applications and the Facebook server and controls an application's access to users' data stored on the server. 
In so doing, an application requests the proxy for users' profile data. 
Upon receiving the request, the proxy performs access control checks based on user-provided privacy settings.
If the request is allowed, the proxy signs the access request with its  key, sends the request to the OSN server, and finally replays the result from the server to the application. 
This application to server data flow is shown in Figure~\ref{fig:PoX}. %, where PoX is working as a reference monitor. 
An application developer needs to use the PoX server-side library instead of the Facebook server-side library.
One potential challenge is to motivate application developers to write PoX-aware applications  when existing mechanisms (e.g., Facebook application environment) are perfectly in place.
%\ainote{are the pictures cut -and-paste from papers or redrawn? I don't think we're allowed -- due to copy right -- to copy-paste them, but we may deal with this for camera ready}

%One issue limit the applicability of PoX.
%It is difficult to motivate application developers to write PoX-aware applications  when existing mechanisms (e.g., Facebook application environment) are perfectly in place.
%First, PoX is browser and machine dependent, which will put extra burden on users in terms of installations and maintenance, and note that this is just for the use of the social applications. \ainote{I do not understand this argument. Everybody has browsers and plugins, no? So?}
%Second, an application developer needs to use the PoX server-side library instead of the Facebook server-side library.
%It is difficult to motivate application developers to write PoX-aware applications  when existing mechanisms (e.g., Facebook application environment) are perfectly in place.

xBook~\cite{Singh2009xBook} is a restricted ADSafe-based JavaScript framework that provides a server-side container in which applications are hosted and a client-side environment to render the applications to users.
xBook is different than PoX in that it not only controls third-party applications' access to user data (which PoX also does), but also it  limits what applications do with the data. 
Applications are developed as a set of components; a component is a smallest  granular building block of codes monitored by xBook.
A component also reveals the information that the component can access and the external entity with which it communicates. 
During the deployment of an application in xBook, an application developer requires to specify these information.
 %information required for each component and the external entities a particular component needs to communicate with. \ainote{oops. Isn't this what you just said?}
From the specification, xBook generates a manifest for the application.
A manifest is a set of statements that specifies what user data the application will use and with which external services it will share the data.
At the time of installing the application, the manifest will be presented to the user.
In this way, a user will be able to make a more informed decision before installing  an application.
Although xBook controls third-party applications' access to user data and limits application's data usage, it has to deal with two challenges.
First, the platform itself has to be trusted by users and by applications, as it is expected to protect users' personal data and enable third-party applications to execute.
Second, hosting and executing applications in xBook requires resources (storage, computation and maintenance) that may be difficult to provide in the absence of a business model. 
%Although, It's difficult to envision incentives for all these activities. \ainote{this is related to the point you just made, about need of resources.}

\section{Protecting User Data from the OSN}
%Mitigating Attacks From the OSN} 
%\ainote{not the best title, maybe}
\label{osn}

The ``notice-and-consent'' approach to online privacy is the status-quo for practically all online services, OSNs included. 
This approach informs the user of the privacy practices of the service and provides the user a choice whether to engage in the service or not. 

The limitations of this approach have been acknowledged for long. 
First, the long and abstruse privacy policies offered for reading are virtually impossible to understand, even if the user is willing to invest the time for reading them. 
For example, on August 2014, we found 4389 words on Facebook's privacy policies and 3473 words on Twitter's privacy policies.
Second, such policies always leave room for future modifications; therefore, the user is expected to read them repeatedly in order to practice informed consent.
And third, long as they are, these privacy policies tend to be incomplete\footnote{http://goo.gl/yXqI1s}, as they often cannot include all the parties to which user's private information will be allowed to flow (such as advertisers). 
Consequently, generally people do not read the Terms of Service and when they do, they do not understand them~\cite{fiesler2014copyright}. 

A second serious deterrent for users protecting their online privacy is the ``take-it-or-leave-it'' ``choice'' the users are offered. 
While it may seem as a free choice, in reality the cost of not using the online service (whether email, browsing, shopping, etc) is unacceptably high. 

Cornered in this space of falsely informed and lack of choice, users may look for solutions that allow them to use the online service without paying the information cost associated with it.
Researchers built  on this intuition in two directions. 
The first direction tends to hide the user information from the very service that stores it (Section~\ref{osn_leverage}). 
The second taps into different business models than the ones that make a living from user's private information and replaces the centralized service provider with a fully decentralized solution that is privacy-aware by design (Section~\ref{decentralization}). 

\subsection{Protection by Information Hiding}% Leveraging the OSNs}
\label{osn_leverage}
%The surge of popular OSNs (e.g., Facebook) is such that it would be unrealistic to think that people will leave those platforms just for privacy reasons.
%Solutions that protect users by leveraging the OSNs enable users to use OSNs, while keeping personal information protected from the OSN. 
%They allow a user to be a part of the network, but at the same time protect her from the network provider.

%It appears that users are also concerned about the OSN's access to their personal information.
This line of work is empirically supported by the Acquisti and Gross's study~\cite{acquisti2006imagined} that shows that while 60\% of users trust their friends  completely with their private and personal information, only 18\% of users trust Facebook to the same degree.

The general approach for hiding information from the OSN is based on the observation that OSNs can run on \emph{fake} data. 
If the operations that OSNs perform on the fake data are mapped back to original data, users can still use the OSNs without providing them real information.
Fake data could be ciphertext (encrypted) or obtained by substituting the original data with pre-mapped data from a dictionary.
Encrypted data can be stored on a user's trusted device (including third-party servers or a friend's computer).
Access controls are provided by allowing authorized users (e.g., friends) to get the original data from the fake data.
%limiting the ability to recover the real data from the fake data to authorized users only (e.g., friends). 
Different implementations of this idea are presented next. 
%The benefit of this design is that users can still use popular OSNs, while keeping personal information protected from the OSN.
%Moreover, 

\emph{flyByNight}~\cite{Lucas2008FlybyNight} is a Facebook application that enables users to communicate on Facebook  without storing a recorded trace of their communication in Facebook.
%When a user uses the application for the first time, 
The flyByNight Facebook application  generates a public/private key pair and a password during configuration.
The password is used as a key to encrypt the private key and the key is stored on flyByNight server.
When a user installs the application, it downloads a client-side JavaScript from the FlyByNight server.
This JavaScript does key generation and cryptographic operations.
The application knows a user's friends and their public keys who have also installed the flyByNight application.
To send messages to friends, a user enters the message into the application and selects the recipient friends. 
The client-side JavaScript encrypts the content of the message with other users' public keys, tags the encrypted message with the Facebook ID numbers of their recipients, and sends them to a  flyByNight  message database server. 
The encrypted messages reside on the flyByNight server.
When a user reads a message, she provides the password to get the private key (stored in the flyByNight key database).
The private key is used to decrypt the message.
flyByNight operates under the regulation of Facebook, as it is a Facebook application.
%However, flyByNight totally depends on Facebook for all operations.
%It can operate perfectly as long as Facebook didn't pay attention to it.
It is possible that the computation load on the Facebook servers due to encryption, as well as the suspicious lack of communication among users might attract Facebook's attention and lead to deactivating the application.  
In the worst case, users lose their ability of hiding their communication, but previous messages remain hidden from the OSN.

\emph{Persona}~\cite{Baden2009Persona} hides user data from the OSN by combining attribute-based encryption (ABE) and public key cryptography.
The core functionalities of current OSNs such as profiles, walls, notes, etc., are implemented in Persona as applications.
Persona uses an application ``Storage'' to enable users to store personal information, and share them with others through an API.
%Considering the popularity of OSNs, the authors acknowledge that it will be unrealistic to assume that any privacy-enabled networks (including Persona) will replace existing OSNs.
%So, they enable Persona to be inter-operable with Facebook as applications.
Persona application in Facebook is similar to any third-party Facebook application, where users log-in by authenticating to the browser extension.
The browser extension translates Persona's special markup language.
User information is stored in Persona storage services rather than on Facebook and other Persona users can access the data given that they have the necessary keys and access rights. 
Similar to the flyByNight, Persona's operation depends on the OSN, as core functionalities are implemented as applications.
%In the worst case, Facebook could remove the application, considering under utilization of their features.

\emph{NOYB}~\cite{Guha2008Noyb} distorts user information in an attempt to hide real identities from the OSN, allowing only trusted users (e.g., friends) to get access to the restored, correct information. %is based on the observation that if a user's information are scattered and replaced with other users' information, then even though the information are known, OSNs will not be able to combine these pieces to conclude about the person.
%Only trusted parties such as a user's friend will be able to combine the user's information.
%In so doing, 
To implement this idea, NOYB splits a user's information into atoms.
For example, Alice's name, gender and age (Alice, F, 26) are split into two atoms: (Alice, F) and (26).
Instead of encrypting the information, NOYB replaces a user's atom with  pseudorandomly picked  another user's atom.
So, Alice's first atom is substituted with, for example, the atom (Athena, F) from Athena's profile, and the second atom with Bob's atom from the same class (38). %, from Athena and Bob respectively.
All atoms from the same class for all users are stored in a dictionary.
%NOYB encrypts the index of the user's atom in this dictionary and uses the ciphered index to pick the replacement atom from the dictionary.
NOYB uses ciphered index of a user's atom to substitute an atom from this dictionary. 
Only an authorized friend knows the encryption key and can reverse the encryption.  
A proof-of-concept implementation of NOYB as a Firefox browser plugin adds a button to ego's Facebook profile that encrypts his information and another button on alter's page that decrypts alter's profile.
The cleverness of NOYB is that it stores legitimate atoms of information in plain text, thus not raising the suspicions of the OSN. %One of the benefits of  NYOB is that it does not store cipher-text in an OSN.
%It stores a cipher atom, which is also a legitimate atom.
%If the cipher-text was stored, OSNs could easily find that users are encrypting their data by looking for the traits of the cipher-text (a problem of flyByNight).
The challenge, however, is the scalability of the dictionaries: the dictionaries are public, contain atoms from both NOYB users and non-users, and are maintained by a third party with unspecified business/incentive model. 
%Considering the size of the social network (Facebook has more that one billion users), a trusted third-party has to be in place to properly maintain the huge dictionaries, even though the trusted party might not get any incentives by maintaining it. 
%Moreover, NOYB remains silent regarding a user's day-to-day produced data (e.g., comments, notes, likes etc.).
%Using NOYB to protect all information of a user from an OSN would mean that a parallel OSN operation has to run.
%Another problem of NOYB is that key management protocol is out-of-band, which is a big problem for usability. 
%Also, the anonymity of the atoms depends on the number of its users.

\begin{figure}[htbp]
\centering
\includegraphics[height=6cm]{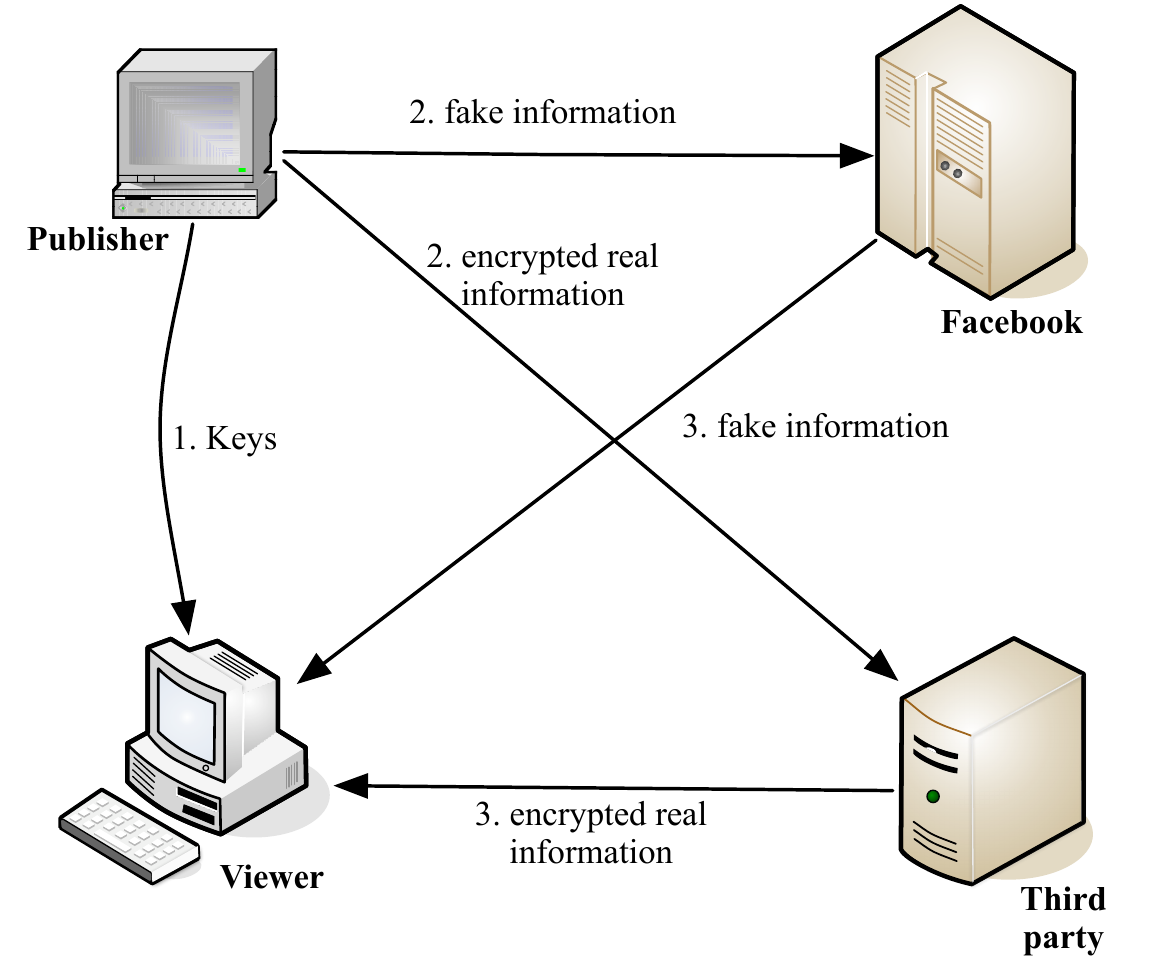}
\caption{FaceCloak architecture~\cite{Luo2009FaceCloak}.}
\label{fig:FaceCloak}
\end{figure}

\emph{FaceCloak}~\cite{Luo2009FaceCloak}, implemented as a Firefox browser extension, protects user information by storing fake data in the OSN. 
Unlike NOYB, it does not replace a user's information with another user's information, rather it uses dictionaries and random Wikipedia articles as replacements.
% in three phases: the setup phase, the encryption phase and the decryption phase.
%When a user sets up FaceCloak in her browser, FaceCloak generates three keys: a master key, a personal index key, and an access key.
%It provides mechanisms to distribute master key and personal index key to her friends.
%But the access key is stored on the user's computer and never distributed. 
%A master key is used to derive a symmetric encryption key for encrypting user information and a MAC key protects the integrity of this information. 
%Users' information is stored in a third-party server using index-value pairs---a value consists of the encrypted personal information and its MAC.
%An index of encrypted information is the cryptographic hash of the personal index key and an identifier that depends on the type of  information.
%The access key is used to update the existing data. (See the architecture of FaceCloak in Figure~\ref{fig:FaceCloak})
A user, say Alice, can protect  information from the OSN by using a special marker pre-defined by FaceCloak (Ò@@Ó in their implementation).
When Alice submits the form to the OSN, FaceCloak intercepts the submitted information, replaces the fields that start with the special marker by appropriate fake text and stores the fake data in the OSN.
It uses a  dictionary (for profile information) and random Wikipedia articles (for walls and notes) to provide fake data.
Now, using Alice's master key and personal index key, FaceCloak does the encryption of the real data, computes  MAC keys, computes the index, and sends them to a third-party server.
Now consider one of Alice's friends Bob, who has installed FaceCloak in his browser, and Bob wants to see Alice's information.
After downloading Alice's page (which also includes fake data from the OSN), FaceCloak computes indexes of relevant fields using master and personal index key of Alice.
Then it downloads the corresponding values from the third-party server.
Upon receiving the value, FaceCloak checks the integrity of the received cipher-text, decrypts it, and substitutes the real data for the fake data. 
If the value is not found, then the data is left unchanged. 
See the architecture of FaceCloak in Figure~\ref{fig:FaceCloak}.
FaceCloak depends on a ``parallel'' centralized infrastructure to store the encrypted data, which means that a third-party has to maintain all users' data, probably without getting any benefits from it.
And, users have to trust the reliability of the third-party server, which also represents a single point of failure.

%FaceCloak suffers from multiple problems.
%The worst one is that friends of a user will not be able to see her data if they do not install the browser extension.
%Maintaining dictionaries and their cryptographic operations might slow down the overall operation performed by a user.
%Unfortunately, the paper does not provide any performance study to assess the concern.

\emph{Virtual Private Social Networks) (VPSN)}~\cite{Conti2011VPS}, unlike  flyByNight, FaceCloak, and NOYB, does not require third-party services to protect users' information from an OSN.
Instead, they leverage the computational and storage resources of the OSN users to store real profile data of other users, while storing fake profile data on Facebook. 
%Moreover, it does not require collaboration from the OSN and unnecessary cryptographic overhead to interact with the OSN.
%VPSN creates virtual networks leveraging the architecture and the infrastructure of an already existing social network and requires system resources to be distributed among VPSN members.
\emph{FaceVPSN} is a Firefox browser extension that implements VPSN for Facebook.
In FaceVPSN, user Alice changes her profile information  to some fake information and stores the fake information in Facebook and sends by email her correct and fake profiles in a prespecified XML format to her friends.
%Later, she invites her friend Bob to FaceVPSN and also sends him an email.
%The email contains an XML file consists of details of fake and real information, and Facebook identification of Alice.
In order to access Alice's real profile, her friends have to have FaceVPSN installed (as a regular Firefox extension) and use its GUI to add Alice's XML file.  
%Bob requires to install FaceVPSN just like any other Firefox extension and he has to add the XML file of Alice to the list of preferences using Graphical User Interface (GUI) of FaceVPSN.
When Alice's friend  Bob requests Alice's Facebook page, Facebook sends an HTML response that has Alice's fake data from Facebook. 
FaceVPSN's JavaScript code is triggered when ``page load'' event is fired. 
The JavaScript code of FaceVPSN searches the profile information of Alice in Bob's stored XML file and replaces the fake information with real information.

Unlike other solutions presented above, FaceVPSN does not risk being suspended by the OSN (since it is not an application running with the OSN's support). 
Like FaceCloak, however, FaceVPSN requires a user's friends to install the FaceVPSN extension in order to see the user's profile.
Moreover, FaceVPSN demands a high degree of user interaction that might affect usability.
In particular, upon the addition of a new contact to the friend list, the user has to explicitly exchange profile information with the new friend and upload it into the FaceVPSN application. 
On top of it, every change of profile information has to be emailed as an XML file to all friends,  and the friends are required to go through the XML update process in order to see the changes.
This entire process affects usability, given the high number of friends a user might have in OSNs (e.g., half the Facebook users have more than 200 friends, and 15\% have more than 500 friends\footnote{http://www.pewresearch.org/fact-tank/2014/02/03/6-new-facts-about-facebook/})

%FaceVPSN claims that the architecture could be used as a generic secure vault for users' information.
%However, one concern is that whether it could be used for user generated contents also.
%For example, it will be difficult  for a user to imagine that she has to send a file to all of her friends after posting a status update or making a comment on Facebook. 

While the various implementations of the idea of hiding the personal information from the OSN have different tradeoffs, as discussed above, there are also risks associated with the approach itself. 
%However, there are tradeoffs if users want to use the approaches discussed above.
First, because the OSN operates on fake data (whether encrypted or randomized), it will not be able to provide some personalized services such as social search and recommendation.
Second, users end up transferring their trust from the OSN to either a third-party server or friends' computers for unclear benefits. 
The third-party server provides yet another service whose terms of use are probably presented in yet another incomprehensible Terms of Service document, with an opt-out ``choice''. 
Friends' computers require extra care for fault tolerance and malicious attacks. 

%====================================
\subsection{Protection Via Decentralization}
\label{decentralization}
%====================================

An alternative to obfuscate information from the OSN is to migrate to another service that is especially designed for user privacy protection. 
Research in this area explored the design space of decentralized (peer-to-peer) architectures for managing user information, thus avoiding the centralized service with a global view of the entire user population.
The typical overlay used in most of these solutions is based on distributed hash tables, preferred over unstructured overlays for their performance guarantees. 
In addition, data is encrypted and only authorized users get access to the plain text. 
In this section, we discuss decentralized solutions for OSNs.
There are three dimensions that differentiate the solutions: (1) how the distributed hash table has been implemented (e.g., OpenDHT, FreePastry, Likir DHT)? (2) where to store users' content (e.g., nodes run by the user, by the friends or cloud infrastructures)? (3) how to manage encryption keys for access controls (e.g., public-key infrastructure, out-of-band)? 

%What differentiates such solutions is the management of encryption keys and resource management in a potentially unreliable environment. \ainote{Imrul, help! Is the management of encryption keys? Or what is really different?}

%\ignore{
%Privacy breaches are possible in existing OSNs due to the centralized nature of the system.
%Centralized architectures require all user information in OSNs to flow through their central servers. 
%An alternative, decentralized social network architecture might be a solution to the problem.
%Decentralized design-based solutions envision no centralized authority.%, thus no data exploitation from the authority.
%\ainote{what are we trying to say: attacks from outside are potentially more damaging because all information is there? or the OSN itself has all info, thus more power? perhaps the later, no? The problem with decentralization is that it has multiple places to defend, therefore intrinsically is not more secure. So we need to better explain the motivation. }
%\ainote{it would be good to list here the approaches that are covered below to give intuition.}
%}

\emph{PeerSooN}'s~\cite{Buchegger2009Peerson} architecture has two-tiers.
One tier, implemented using OpenDHT, serves as a look-up service to find a user.
It stores users' meta-data for example, the IP address, information about files, and notifications for users. 
A peer can connect to another peer asking the look-up service directly to get all required information.
The second tier is formed by peers and it contains users' data, such as user profiles. 
Users can exchange information either through the DHT (e.g., a message is stored within the DHT if receiver of a message is offline) or directly between their devices.
The system assumes a public-key infrastructure (PKI) for privacy protection.
A user encrypts data with the public keys of the intended audience, i.e., the friends of the user. 
%PeerSooN suffers from the OpenDHT in that  a maximum time period of seven days of storage.
%Moreover, PeerSooN does not consider replication of user information, thus, a user's information becomes unavailable to her friends when she is offline.

\emph{Safebook}~\cite{Cutillo2009Safebook,Cutillo2009Decentralization}  is a decentralized OSN, which uses a peer-to-peer architecture to get rid of a central, omniscient authority.
Safebook has three main components: a trusted identification service for certification of public keys and the assignment of pseudonyms; 
matryoshkas, a set of concentric shells around each user, which serve to replicate the profile data and anonymizes traffic; and a peer to peer substrate (e.g., DHT) for the location of matryoshkas that  enables access to profile data and exchange messages. 
%The system implements different OSN functionalities similar to the existing centralized OSNs such as  account creation, data publication, data retrieval, contact request and acceptance, message management, but offers stronger  protection from the existing OSNs. 

\emph{LifeSocial.KOM}~\cite{Graffi2011LifeSocial} is another P2P-based OSN.
It implements common functionalities in OSNs using OSGi-based\footnote{http://www.osgi.org/Specifications/HomePage} software components called ``plugins''.
%such as user profiles, friend lists, photo albums, user groups, live chat, and status updates using OSGi-based software components, called ``plugins''.
As a P2P overlay, it uses FreePastry for interconnecting the participating nodes and PAST for reliable, replicated data storage. 
The system uses cryptographic public keys as user ID.
To protect privacy, a user encrypts a private data object (e.g., profile information) with a symmetric cryptographic key. 
She then encrypts the symmetric cryptographic key individually with the public keys of authorized users (e.g., her friends) and appends to the data object. 
The object and the list of  encrypted symmetric keys are also signed by the user and they are stored in the P2P overlay.
Other users in the system can authenticate the data object by using the public key of the author.
But only authorized users (e.g., friends) can decrypt the symmetric key and thus, the content of the object.

\emph{LotusNet}~\cite{Aiello2012Lotusnet} is a framework for the implementation of a P2P based OSN  on a Likir DHT~\cite{Aiello2008Likir}.
It binds a user identity to both overlay nodes and published resources for robustness of the overlay network and secures identity based resource retrieval. 
Users' information is encrypted and stored  in the Likir DHT.
Access control responsibility is assigned to overlay index-nodes. 
Users issue signed grants to other users for accessing their data.
DHT returns the stored data to the requestor only if the requestor can provide a proper grant, signed by the data owner.

\emph{Vis-a-Vis}~\cite{Shakimov2011visavis} targets high content availability. 
Users store their personal data in Virtual Individual Servers (VISes), which are kept on the user's computer.
The server data are also replicated on a cloud infrastructure so that the data is available from the cloud when a user's computer is offline.
Users can share information with other users using peer-to-peer overlay networks that connect VISes of the users.
The cloud service needs to be rented (considering the high volume of the data users store in OSNs), which makes the scheme monetary dependent.

\emph{Prometheus}~\cite{Kourtellis2010Middleware} is a peer-to-peer social data management system for socially-aware applications.
It does not implement traditional OSN functionalities (e.g., profile creation, management, contacts, messaging, etc.), rather it manages users' social information from various  sources and exposes APIs for social applications.
Users' social data are encrypted and stored in a group of trusted peers selected by users for high service availability.
Prometheus architecture is based on Pastry, a DHT-based overlay, and it uses Past  to replicate social data.
An inference on social data is subject to user defined access control policy enforced by the trusted peers.
Prometheus relies on  a public-key infrastructure (PKI) for user authentication and message confidentiality.

The toughest challenge for decentralized OSNs is to convince traditional OSN users to migrate to their systems.
%User's migration to a new architecture is a big question.
Centralized social networks  have large, established user bases and they are accessible from anywhere.
Moreover, they  already have a mature infrastructure, making good revenues from users' data and  maintaining excellent usability. 
However, decentralized OSNs are also becoming popular, specially among privacy-aware users.
For example, Diaspora\footnote{https://joindiaspora.com/} is a fully operating open source, stable and  decentralized OSN, which relies on user contributed local servers to provide all the major centralized OSN functionalities.

%It will be a real challenge to convince existing centralized OSN users to migrate to some new architectures, even though better privacy is ensured.

\section{Mitigating De-anonymization and Inference Attacks}
\label{inference}

%The analysis of social network data sheds light on the hidden social patterns.
Analysis of social data has become immensely popular in a variety of domains, such as  biological systems, organizational studies, information science, communication studies, economics, political science, social psychology, development studies, anthropology and sociolinguistics. 
%It has become easier to collect digital signatures of social networks, thanks to the technological progress.
Researchers and agencies collect or purchase social data to do the analysis.
For example, Kwak et al.~\cite{Kwak2010NewsMedia} collected the entire Twitter network as of  2010: 41.7 million Twitter profiles, 1.47 billion follower-following relations, and 106 million tweets.
In addition, some organizations and OSN service providers publish social data for others to analyze.
For example, the Federal Energy Regulatory Commission published a repository of approximately $500,000$ email messages of  Enron Corporation, which has been frequently analyzed for research~\cite{klimt2004enron,Shetty2005ENron}.

However, publishing and allowing the collection of social network data involves privacy disclosure risks.
%An attacker may violate the privacy of targeted individuals using published or collected network data as a background knowledge.
%A telling example of such privacy violations .
For example, in 2006 AOL released an anonymized dataset  of twenty million search keywords for over 650,000 users~\cite{Arrington2006AOL}.
The dataset was published for research purpose and novel findings emerged (e.g.,~\cite{Nunes2008AOLresearch,Adar2007AOLresearch,Jansen2010AOLresearch}). 
However, despite the fact that the data released was anonymized, users' privacy was compromised.
To make the point, the New York Times identified an individual from this dataset by cross referencing users with phonebook listings.

%~\cite{Barbaro2006AOLScandal}. 

Privacy attacks in published or collected social network data can be categorized into two categories:  de-anonymization attacks and inference attacks. 
In the following, we discuss the types of de-anonymization attacks, how these attacks take place, and what solutions were proposed to combat such attacks.
More in-depth discussion on de-anonymization can be found in~\cite{zheleva2011privacy}. 
In this work, we include the latest work on de-anonymization attacks.

\subsection{De-anonymization Attacks}
In de-anonymization attacks, an attacker uses external background knowledge and published social data to de-anonymize/identify users in the social graph, and thus learn sensitive user information.

%\subsubsection{Privacy Breaches Due to De-anonymization Attacks}

We categorize privacy breaches due to de-anonymization attacks into four classes:  
\begin{enumerate}
\item \textit{Identity disclosure} reveals the identity of a user and makes him vulnerable in the real world.
For example, although a published dataset on the disease-infection network could advance research on how the disease transmits in communities, an adversary (e.g., an insurance company) that can identify an individual and his disease could exploit this information in unintended ways (for example, for denying insurance). %\ainote{this is not a cyber-attack as the danger claimed in the previous sentence}

%\ainote{what to do with these cites? where do they belong? they were lumped together in the previous version. Say what they do: do they address this type of attack? Or define this type of attack? Or?}~\cite{Narayanan2009De-anonymizing,Wondracek2010De-anonymize,Hay2007Anonymizing,Liu2008anonymization}, 

\item \textit{Social attributes disclosure} refers to the disclosure of sensitive data associated with a user.
For example, disclosure of a user's date of birth, gender and home address could allow the inference of the user's social security number (SSN) and hence could lead to identity theft~\cite{Gross2005InformationRevelation}.

%\ainote{what to do with these cites? where do they belong? they were lumped together in the previous version. Say what they do}~\cite{Campan2009Data,Zhou2011k,Truta2010Avoiding}, 

\item \textit{Relationship disclosure} refers to the situation when the relationships of a user are exposed and this information exploited.
For example, two nodes (e.g., companies) in a transaction network are connected by an edge and a weight (e.g., transaction expense) if they are involved in a financial transaction.
An adversary, for example a competitor company, can detect whether two target companies have done a financial transaction if it can infer whether an edge exists between the two companies in the network.
The adversary can learn the transaction expense from the edge weight and can exploit that information to get advantages.
%\ainote{it is unclear what the damage is. it does not apply only to financial transactions, no? can you give a better example?}

%\ainote{what to do with these cites? where do they belong? they were lumped together in the previous version. Say what they do}~\cite{Backstrom2007WAT,Zheleva2008Preserving,Liu2008Privacy,Korolova2008LPS}

\item \textit{Social graph property disclosure}  refers to the disclosure of various graph metrics, such as degree, betweenness centrality, closeness centrality, or clustering co-efficient.
An attacker can find out the most central users in the network and can make the network structurally vulnerable.
For example, an attacker can identify  and remove the highest betweenness centrality nodes  to disrupt communications between other nodes in the network~\cite{Newman2010Book}.
%\ainote{what to do with these cites? where do they belong? they were lumped together in the previous version. Say what they do}~\cite{Backstrom2007WAT,Hay2008RSR,Ving2008Randomizing}.

\end{enumerate}

\subsubsection{De-anonymization Attack Techniques}
\begin{figure}[htbp]
\centering
\includegraphics[height=8.5cm]{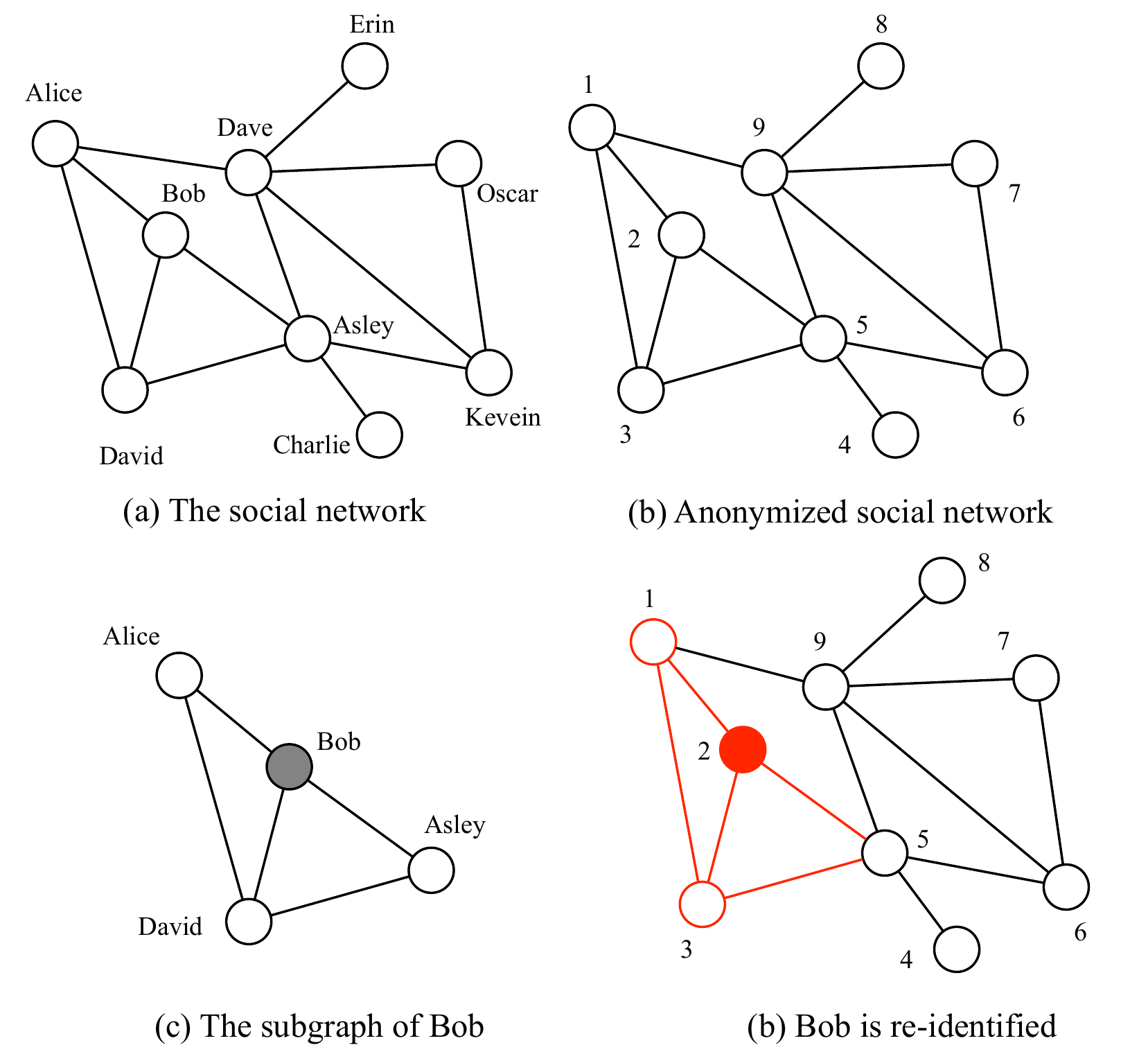}
\caption{Anonymization and de-anonmymization attacks.} %\ainote{Kevein in here should be Kevin, I suspect?} \ainote{Is this a good example? First, it's more complicated than needed -- what are you trying to convey? Second, I think it's wrong: if only Bob's neighborhood is known, Bob can also be node 6. Find a simple and more clear example: maybe know that somebody is very popular and friends with X and infer from there. Assuming too much knowledge weakens the message}}
\label{fig:anonymizationExample}
\end{figure}

Anonymization is usually done by substituting personally identifying information associated with each user with a random ID~\cite{Wu2010Anonymization}.
However, this substitution is not sufficient to preserve users' privacy.
For example, consider a social network in Figure~\ref{fig:anonymizationExample}(a) that has been anonymized in Figure~\ref{fig:anonymizationExample}(b) by replacing user names with random IDs.
Now, if an attacker knows that Alice, David and Asley are friends of Bob and Alice-David and Asley-David are also friends, a subgraph shown in Figure~\ref{fig:anonymizationExample}(c), then the attacker can uniquely identify the subgraph in the anonymized network (shown in Figure~\ref{fig:anonymizationExample}(d)).
So, the attacker will be able to re-identify Bob in the anonymized and published social network.

Researchers have shown different techniques to perform de-anonymization attacks.
Backstorm et al.~\cite{Backstrom2007WAT} present two types of attacks---\textit{active} and \textit{passive}. %to show that both types of attacks can reveal true identities of targeted users on anonymized social networks.
In active attacks, the attacker is assumed to be able to modify the network prior to social network data release.
The attacker chooses an arbitrary set of target individuals (whose privacy she wants to compromise), creates  a small number of new user accounts, makes connections with target individuals (thus forms edges), and establishes a highly distinguishable pattern comprising nodes and edges among the new accounts.
The attacker can then efficiently find the subgraph in the released anonymized network, thus can expose the identities of the target individuals.

In passive attacks, an attacker does not have to create new accounts or connections.
The intuition is that most nodes in social networks form small uniquely identifiable subgraphs.
So, the attacker simply has to form a coalition with other users.
The attacker recruits $k-1$ number of his neighbors and forms a coalition of size $k$.
The users in the coalition know names of their neighbors outside of the coalition.
Finally, the attacker tries to identify the subgraph (formed by the coalition) in the published social network, and compromises the privacy of neighboring nodes.

Narayanan and Shmatikov~\cite{Narayanan2009De-anonymizing} demonstrate the feasibility of a large-scale de-anonymization attack under the assumption that  the attacker has background knowledge of a different network whose membership partially overlaps with the target network.
Using a de-anonymization algorithm, the authors show that a third of the common users of both Twitter and Flickr can be identified in the anonymous Twitter  social graph with a low (12\%) error rate.

Both attacks~\cite{Backstrom2007WAT,Narayanan2009De-anonymizing} presented above use a subgraph as background knowledge. % of an attacker.
However, the way an attacker achieves this knowledge is different.
In~\cite{Backstrom2007WAT}, an attacker creates the background knowledge by adding nodes and edges in the social graphs or forming a coalition among nodes.
In~\cite{Narayanan2009De-anonymizing}, the authors propose to collect network data by crawling OSNs, or deploying third-party malicious applications.
%However, the growing concerns over third-party social applications~\cite{Wang2013AppPrivacy}, online public data crawling~\cite{mondal2012genie} and users' privacy awareness~\cite{Dey2012FacebookPrivacy} might make it difficult for an attacker to obtain the background knowledge. 

\subsubsection{Privacy Preserving Anonymization Methods}

In order to combat de-anonymization attacks, a social network should be anonymized properly before publishing.
We categorize privacy preserving social network data anonymization methods into two categories: (1) edge modification-based approaches and (2) clustering-based generalization.
\\

\textit{Edge modification-based approaches:}
Edge modification-based approaches preserve privacy by modifying the social graph structure via addition, deletion or randomization of the edges.
Zhour and Pei~\cite{Zhou2011k} consider a de-anonymization attack, where an attacker, equipped with the background knowledge about the target's 1-hop neighbors, attempts to re-identify the target in the anonymized dataset using neighborhood matching.
Their anonymization method is inspired by $k$-anonymity model.
Although not targeted to social network data publishing, $k$-anonymity model ensures that each user's information in a released dataset cannot be identified from at least $k-1$ other individuals in the dataset~\cite{Sweeney2002k} .

Zhour and Pei extend  $k$-anonymization to social networks.
The goal of the anonymization is to ensure that even knowing the neighborhood of a node, an attacker will not be able to re-identify the node in the anonymized dataset with confidence higher than $\frac {1}{k}$. 
Let we have a social network $G=(V,E)$ and the anonymized network $\hat{G}=(\hat{V}, \hat{E})$, where there exists a bijection function $f: V \rightarrow \hat{V}$ and for each $(u,v) \in E$, $(f(u),f(v)) \in \hat{E}$.
The authors assume that an attacker has the knowledge of the neighborhood subgraph of a node $u \in V(G)$, denoted by $Neighbor_{G}(u)$.
The goal of the $k$-anonymization is to ensure that there exist at least $k-1$ other nodes $v_1,v_2,v_3, \ldots ,v_{k-1} \in V(G)$ such that $Neighbor_{\hat{G}}(f(v_1))$$,\ldots ,$$Neighbor_{\hat{G}}(f(v_{k-1}))$ are isomorphic.
The anonymization method first extracts the neighborhoods of all nodes in the network.
Then it greedily combines nodes into groups and anonymizes the neighborhoods of the nodes in the group, until $k$-anonymity conditions are met.
To anonymize neighborhoods of two nodes such as $Neighbor_{G}(u)$ and $Neighbor_{G}(v)$, the method first finds all perfect matches of neighborhood components in $Neighbor_{G}(u)$ and $Neighbor_{G}(v)$.
For those unmatched components, it tries to pair similar components based on anonymization cost and anonymizes them (this might involve  an addition of an edge between two nodes).

Liu and Terzi~\cite{Liu2008anonymization} propose $k$-degree anonymity to combat de-anonymization attacks.
They assume that an attacker has the background knowledge of the degree of a target node.
The attacker could search the degrees of the nodes in the published network and could re-identify a target node.
$k$-degree anonymity ensures that for every node $u$ in the graph, there exist at least $(k-1)$ other nodes having the same degree as $u$.
Similar to the work of Zhour and Pei, even having the degree background knowledge, an attacker will not be able to re-identify the node in the anonymized dataset with confidence higher than $\frac {1}{k}$.
Their anonymization algorithm has two steps.
In the first step, it starts from a degree sequence ${\bf d}$ of the original network $G(V,E)$ and constructs a new degree sequence ${\bf \hat {d}}$ that is $k$-degree anonymous, so that degree anonymization cost is minimized.
%$D_{A}({\bf \hat {d}}-${\bf d}$ )=L_1({\bf \hat {d}}-${\bf d}$)$ 
In the second step, the algorithm constructs a graph $\hat{G}=(\hat{V}, \hat{E})$ such that ${\bf d_{\hat {G}}} = {\bf \hat {d}}$, $\hat {V}=V$ and $\hat {E} =E$.
To solve the first step, the authors used a dynamic programming method, while the second step is based on a set of graph construction algorithms given a degree sequence with constraints.
To construct the new degree sequence the algorithm uses a randomized edge swap transformation strategy.
\\

\textit{Clustering-based generalizations:}
Clustering-based generalizations cluster nodes and edges into groups and anonymize a subgraph into a super-node ~\cite{Zhou2008Anonymization}.
So, details about users are hidden.

Hay et al.~\cite{Hay2008RSR} propose a vertex clustering-based generalization approach to combat de-anonymization attacks.
They model an attacker's background knowledge as the access to an entity that answers a restricted knowledge query about a target node in the network.
They assume three types of queries: \textit{vertex refinement queries} return the local structure of a node in an iterative refined way (e.g., degree of a node, the set of neighbors degrees of a node); \textit{subgraph queries} confirm a subgraph around a target node; \textit{hub fingerprint queries} for a target node returns the vectors of distances between the node and a set of hubs (note that in social networks a hub is defined as a node with high betweenness and degree  centrality~\cite{Newman2010Book}).
The anonymity method is based on structural similarity.
The intuition is that  structurally similar nodes may be indistinguishable to an attacker.
The anonymity method generalizes a social graph by grouping nodes into partitions and publishes the number of nodes in each partition including the densities of edges across and within the partitions.
The size of the partition is at least $k$ (a positive integer), which is similar to $k$-anonymity in relational data.
The method used a simulated annealing algorithm for partitioning~\cite{Russell1995Artificial}.

Zheleva and Getoor~\cite{Zheleva2008Preserving} consider a \textit{link re-identification} attack, where nodes have multiple types of edges and an attacker attempts to re-identify sensitive edges.
As a background knowledge, they assume that an attacker can predict a sensitive edge based on other non-sensitive edges.
% has  an accurate probabilistic model which can predict the existence of a sensitive link (edge) based on other non-sensitive edges.
They describe five anonymization techniques: i) remove all sensitive edges; (ii) remove some non-sensitive edges which significantly contribute to the prediction of a sensitive edge; (iii) collapse the anonymized nodes into a single node for each equivalence class (they assume that nodes are clustered into equivalence classes) and publish the count of same types of edges between two equivalence class nodes; (iv) similar technique as (iii), but it needs the equivalence class nodes to have the same constraints as any two nodes in the  social network; (v) remove all edges.

Campan and Truta ~\cite{Campan2009Data} propose an \emph{edge generalization} technique, leveraging the $k$-anonymity model.
In their model, each node is similar to  at least other $(k-1)$ nodes considering attributes and associated structured information (e.g., neighborhood structure of nodes).
Nodes are partitioned into clusters and nodes from a cluster are combined into one single node.
Edges between two clusters are collapsed into a single edge.
An edge between two clusters are labeled with the number of edges between them.
While this approach is similar to~\cite{Zheleva2008Preserving}, two major differences are: (i)~\cite{Campan2009Data} considers all relationships are the same type, but  in~\cite{Zheleva2008Preserving} there are different types of relations; (ii) ~\cite{Campan2009Data} considers both  generalization and structural information loss while clustering.

The main challenge for anonymization methods is providing sufficient anonymity while preserving (all) the relevant structural properties of the network. 
Without preserving enough of the structural properties of the original network, publishing anonymized social network datasets loses its value. 
%We identify following challenges researches  still to overcome to combat de-anonymization attacks.
%%First, it is difficult to anonymize a social network keeping structural properties intact.
%%For example, existing anonymization techniques, whether based on graph modifications or generalizations, change the topology of the social graph.
%%Hence, structural properties of the social graph e.g., centralities and clustering coefficients are lost, at least depleted significantly.
%Second, modeling background knowledge of an attacker is difficult. \ainote{don't buy this in the lack of more information/support}
%and anonymizing a network to combat a wide variety of attacks is challenging.
%Existing anonymiztion approaches consider limited background knowledge of an attacker (e.g., degree sequence or 1-hop neighborhood).
%However, an attacker, equipped with a combination of different backgrounds, might challenge those anonymiztion methods. 

\subsection{Mitigating Inference Attacks}
The goal of an inference attack is to infer undisclosed private information about a user using other published details of that user.
For example, a person might not want to state her political affiliation in Facebook because of privacy concerns.
But if he is a member of  ``ban the same sex marriage'' group, then from this group membership an inference may be possible regarding his political affiliation.

Zheleva and Getoor~\cite{Zheleva2009ToJoin} study four social networks (Facebook, Flickr, Dogster and BibSonomy) and show how an attacker can exploit  public and private user profiles to learn private attributes such as user location and gender.
They show that declared social relationships and inferred group memberships are enough to predict undisclosed private information.
Using the classification model \emph{LINK-GROUP}, a combination of link and group-based classification models, they were able to accurately discover the information of private-profile users.
%However, they do not provide any sanitization technique to combat the privacy issue.

Heatherly et al.~\cite{Heatherly2013Prevent} describe three sanitization techniques to prevent undisclosed private information inference from a released social network dataset.  
First, they build classification models to accurately predict private data from the available details (attributes) of a user.
Then they apply the sanitization techniques to reduce the accuracy of the models.
In brief, the techniques are as follows: (i) remove some details (e.g., attributes) to decrease the classification accuracy of sensitive attributes; (ii) alter the link structure of the social graph by adding and removing links and (iii) provide a generalization of details. 
For example, if a user inputs a favorite activity as ``Boston Celtics'', the name will be replaced by a more generalized term ``Basketball''.
Experimenting on a Facebook dataset, the authors conclude that removal of  attributes and friendship links together in the published data is the best way to reduce classifier accuracy.

Dey et al.~\cite{Dey2012AgePrivacy} attempt to infer the age of over one million Facebook users in New York city.
Exploiting the Facebook social graph, they design an iterative algorithm which estimates a user's age based on her friends' ages (e.g., from inferred high school graduation year), friends of friends' edges and so on.
They find that for most users, including users who take maximal measures to prevent privacy leakage by hiding their friend lists, it is possible to estimate ages with an error of only a few years.
The authors recommend to hide high school graduation year and friend lists to other users who are not friends from users' profiles as a solution.
%(also if a user hides her friend lists, she should not appear her friends friend lists to prevent reverse friend loop up) 

\section{Mitigating Sybil Attacks}
\label{sybil}

The Sybil attack is a fundamental problem in distributed systems.
The term \emph{Sybil} was first introduced by Douceur~\cite{Douceur2002Sybil}, inspired from a 1973 book after the same name about the treatment of a person Sybil Dorsett, who manifests sixteen personalities.
In Sybil attacks, an attacker creates multiple identities and influence the working of the system. 

OSNs including Digg, YouTube, Facebook and BitTorrent have become vulnerable to Sybil attacks.
For example, Facebook anticipates that up to 83 million of its users may be illegitimate~\cite{BBC2012Facebook}, which is far more than what it anticipates (54 million) earlier~\cite{BBC2012Facebook1}.
The high number of  Sybils in the OSN is due to the fact that users can create accounts on OSNs quickly and freely; typically only an email address is enough for an account opening.

Sybil users affect the correct functioning of the system by contributing malicious contents.
By controlling a lot of identities, Sybil users increase their influence and power in the OSNs~\cite{Nazir2010Ghostbusting,ratkiewicz2011detecting}.
For example,  Sybil users  can outvote real users on YouTube and can promote clients' content to the top position by giving more up votes~\cite{RILEYYoutube}. 
In Facebook, users control Sybil identities to gain higher status in social games~\cite{Nazir2010Ghostbusting}. 
In Twitter, paid organizations conduct political campaigns disguising themselves as mass population~\cite{ratkiewicz2011detecting}. 
Researchers find that Sybils forward malware  and spam on social media~\cite{Gao2010Spamcampaign,Grier2010SUCSpam,irani2010study,Stringhini2010Spam,Thomas2011SuspendedAccounts}.
Sybil identities are used to acquire users' private contact lists~\cite{Bilge2009Contacts,Fong2011Privacy}.
Moreover, Sybils  manipulate Google social search results~\cite{Matt2011GoogleSearch} and location crowdsourcing results~\cite{MARINANDO2010Tuenti}.

Malicious activities from Sybil users are posing serious threats to OSN users, who trust the service and depend on it for online interactions.
In future the threat will be aggravated as nowadays more people are relying on OSNs for primary online communications~\cite{Lenhart2010Report,Murphy2010Email} and ready-to-get news~\cite{Kwak2010NewsMedia}.
Sybils cost OSN providers, too, in terms of monetary losses and time.
OSN providers  spend significant resources and times to detect, verify, and shut down Sybil identities.
For example, Tuenti, the largest OSN in Spain, dedicates 14 full-time employees to manually verify user reported Sybil identities~\cite{Cao2012SybilRank}.

However, mitigation of the Sybil attack is not easy.
Some open systems employ CAPTCHA~\cite{vonAhn2004Captcha}, IP address filtering, and computational puzzles to mitigate Sybil attacks~\cite{walsh2006experience,peterson2009antfarm,piatek2008one}.
Unfortunately, none of the solutions worked well in a real-world system.
Crowdsource CAPTHA cracking businesses~\cite{Acohido2009CapthaBusiness} employ cheap laborers from the undeveloped countries to break codes.
The IP address filtering allows only one account/identity per IP address, which causes serious problems for users behind NATs, and yet fails against the attackers controlling a subnet~\cite{YuSybilGuard2006}.
And finally, computational puzzles require a potential OSN registration to solve a computational problem with some controlled difficulty. 
Although it requires human efforts, an attacker might be equipped with more resources and could solve the problem.
Moreover, all of these solutions cannot reduce the number of Sybil identities in the system, at best they can limit the rate of Sybils' intrusion in the system.

Two categories of solutions are available to defend Sybils: Sybil detection and Sybil resistance.
%Sybil defense techniques can be categorized into Sybil detection and Sybil resistance schemes.
Sybil detection schemes~\cite{YuSybilGuard2006,Cao2012SybilRank,Wang2013Turing,Yang2011WildSybil} leverage the social graph structure to identify whether a given user is Sybil or non-Sybil (Section~\ref{det}).
On the other hand, Sybil resistance schemes do not explicitly label users' as Sybils or non-Sybils, rather they use application-specific knowledge to mitigate the influence of the Sybils in the network~\cite{Post2011Bazar,Post2011Bazar,Viswanath2012Canal} (Section~\ref{res}).
In a tutorial and survey Haifeng Yu~\cite{Yu2011Sybil} compiles social graph-based Sybil detection techniques.
In this paper, we report latest works on that category, as well as Sybil resistance schemes.

\subsection{Sybil detection}
\label{det}
Sybil detection techniques model an online social network (OSN) as an undirected graph $G=(V, E)$, where  a node $v \in V$ is a user in the network and an edge $e \in E$ between two nodes corresponds to a social connection between the users.
This connection could be a friendship relationship on Facebook or a colleague relationship on LinkedIn, and is assumed to be trusted.

\begin{figure}[htbp]
\centering
\includegraphics[height=5.5cm]{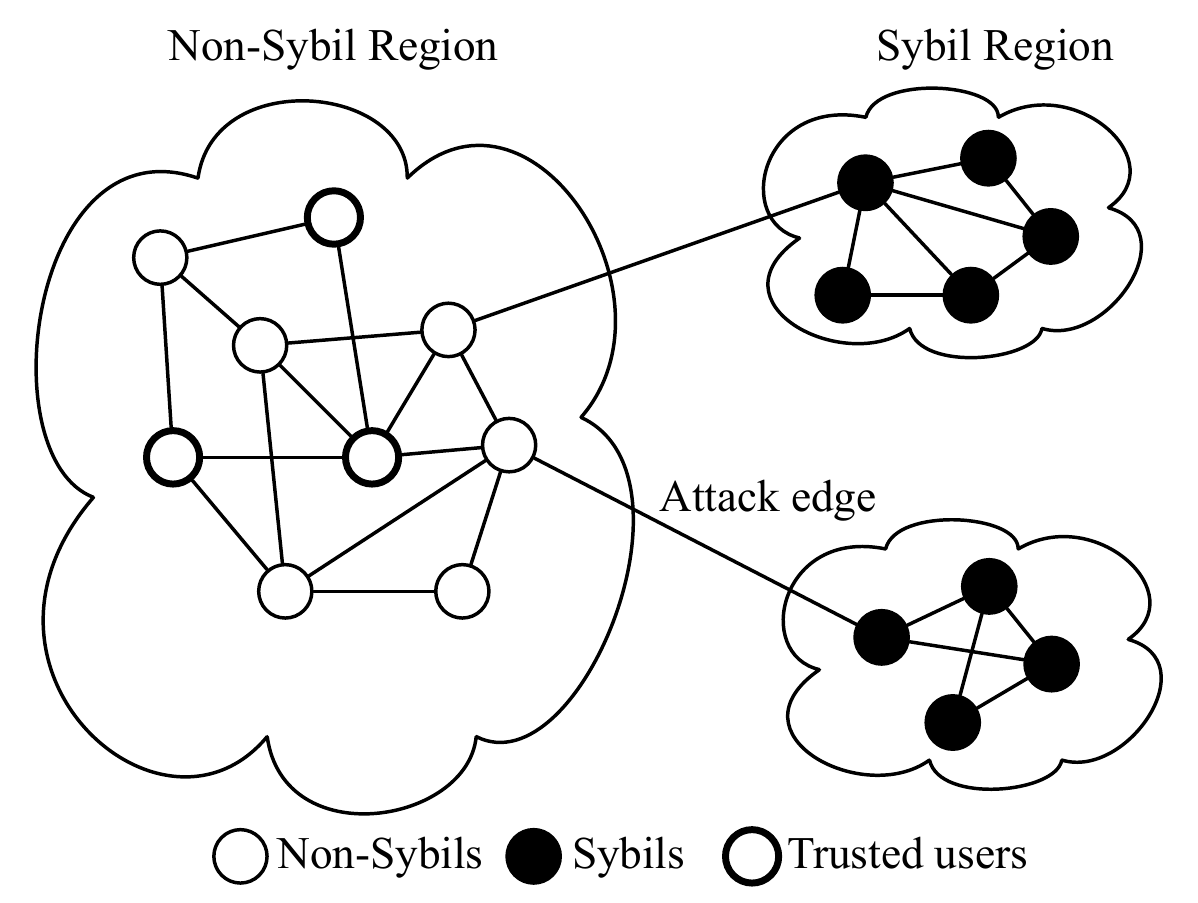}
\caption{The system model for Sybil detection.}
\label{fig:SystemModel}
\end{figure}

The social graph has $n=|V|$ nodes and $m=|E|$ edges.
By definition, if all nodes correspond to different persons, then the system should have $n$ users.
But, some persons have multiple identities.
These users are Sybil users and all the identities created by a Sybil user are called  \textit{Sybil identities}.
An edge between a Sybil user and a non-Sybil user may exist if a Sybil user is able to create a relationship (e.g., friend, colleague) with a non-Sybil user.
These types of edges are called \textit{attack edges} (see Figure~\ref{fig:SystemModel}).

Attackers can launch Sybil attacks by creating many Sybil identities and creating attack edges with non-Sybil users.
Detection systems against Sybil attacks provide mechanisms to detect whether  a user (node) $v \in V$ is Sybil or non-Sybil. 
Those mechanisms are based on the authority (e.g., the OSN provider) knows the topology of the network (a centralized solution), or a node only knows its social connections (a decentralized solution).
Some common assumptions of Sybil detection schemes are below.

\textit{Assumption 1}: %Although Sybil users create many Sybil identities in OSNs, they lack trust relationships. 
Attackers can create a large number of Sybil identities in OSNs and can create connections among those Sybil identities, but they  lack trust relationships because of their inability to  create an arbitrary number of social relationships to non-Sybil users.
Intuitively, a social relationship reflects trust and an out-of-band social interaction. 
So, it requires significant human efforts to establish such a relationship. 
The limited number of attack edges differentiates Sybil and non-Sybil regions in a social graph as shown in Figure~\ref{fig:SystemModel}. 

\textit{Assumption 2}: The non-Sybil region of a social graph is fast-mixing.
Mixing time determines how fast a random walk's probability of landing at each node reaches the stationary distribution~\cite{Boyd2005Mixing,Aiello2008Mixing}.
A limited number of the attack edges causes sparse cut between Sybil and non-Sybil regions.
Non-Sybil regions do not show sparse cut as non-Sybils are well connected.
As such, there should be a difference in terms of mixing time of the non-Sybil regions compare to the entire social graph.

\textit{Assumption 3}: The defense mechanism knows at least one non-Sybil.
This assumption is essential in a sense that without this knowledge the Sybil and non-Sybil regions become identical to the system.

Most of the Sybil detection techniques are based on social graphs.
Social graph-based approaches leverage random walks~\cite{YuSybilGuard2006,Danezis2009SybilInfer,Cao2012SybilRank,Wei2012SybilDefender}, social community~\cite{Viswanath2010Community}, and network centrality~\cite{Xu2010EdgeBetween} to detect Sybils in the network.
SybilGuard~\cite{YuSybilGuard2006} is a decentralized Sybil detection scheme, which uses Assumption $1$, Assumption $2$ and Assumption $3$.
A social graph with a small quotient cut has a large mixing time, which implies that a random walk should be long in order to converge to the stationary distribution.
So, the presence of too many Sybil nodes in the network disrupts the fast mixing property, in a sense that they increase social network mixing time by contributing small quotient cuts. 
Thus, a verifier, which is itself a non-Sybil node, can break this symmetry by examining the anomaly of the mixing time in the network.
In order to detect Sybils, a non-Sybil node (say a verifier) can perform a random route starting from itself and of a certain length \emph{w} (a theoretically identifiable quantity, but the paper experimentally shows that this is 2000 for a topology of one-million nodes).
A suspect (a node that is in question) is identified as non-Sybil if it's random route intersects with the verifier's random route.
As the underlying assumption is that the number of attack edges should be limited, the verifier's route should remain within the non-Sybil region with high probability, given the appropriate choice of \emph{w}.

SybilInfer's~\cite{Danezis2009SybilInfer} assumptions are also Assumption1, Assumption 2 and Assumption 3.
Moreover, it assumes that a modified random walk over a social network, that yields a uniform distribution over all nodes, is also fast mixing. 
The core of SybilInfer is a Bayesian inference that detects approximate cuts between non-Sybil and Sybil regions in social networks.
These identified cuts are used to infer the labels (Sybil or non-Sybil) of the nodes, with an associated probability.

SybilRank~\cite{Cao2012SybilRank} is also a random walk-based Sybil detection scheme, which uses all three assumptions and  ranks user according to their perceived likelihood of being Sybils.
Using early terminated power iteration, SybilRank computes landing probability of random short walks and from that it ranks users, so that substantial portion of the Sybil users have low rank.
The design of SybilRank is influenced by an observation on early terminated random walks in social graphs---if a walk of this kind starts from a non-Sybil node, then it has a high degree-normalized landing probability to land at non-Sybil node than a Sybil node.
SybilRank terms the probability of a random walk to land on a node as the node's \emph{trust}, ranks nodes based on that and filters lower ranked nodes as potential Sybil users.
Rather than keeping computationally intensive a large number of random walk traces used in other graph-based Sybil defense schemes~\cite{YuSybilGuard2006,Yu2010Limit}, it uses power iteration~\cite{Langville04deeperinside} in calculating the landing probability of random walks.

\begin{figure}[htbp]
\centering
\includegraphics[height=6cm]{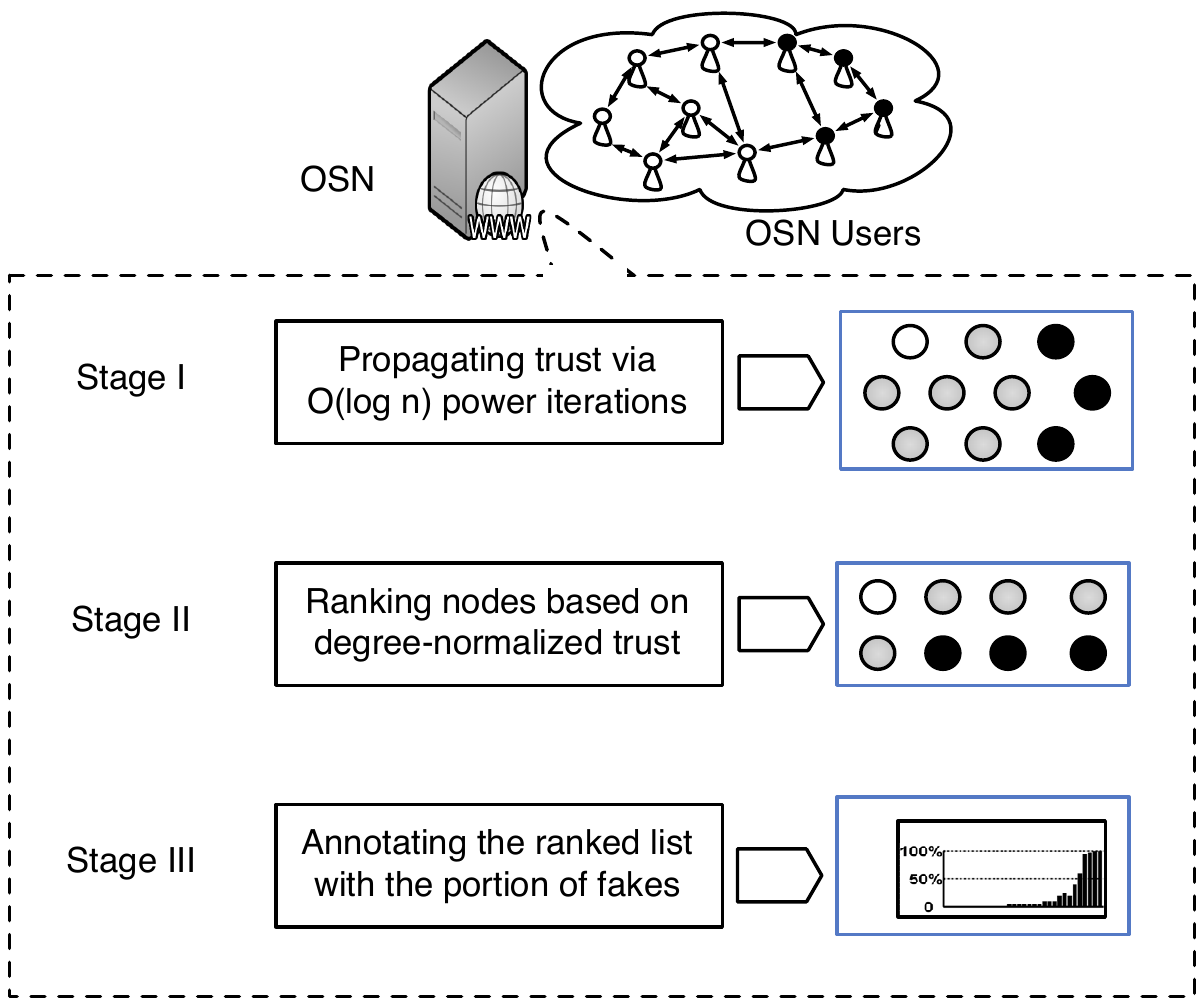}
\caption{Three steps performed by SybilRank in detecting Sybils. Black users are Sybils (~\cite{Cao2012SybilRank}).}
\label{fig:sybilRank}
\end{figure}

Operations performed by SybilRank are shown in  Figure~\ref{fig:sybilRank}.

Viswanath et al.~\cite{Viswanath2010Community} suggest to use community detection algorithms for Sybils' detection.
They  show that although other graph property based Sybil defense schemes have different working principles, the core of those works revolves around detecting local communities around a trusted node.
So, existing community detection algorithms could be used to defend the Sybils also.
Although, not explicitly mentioned, their approach is centralized, because community detection requires a central authority to have the knowledge of the entire topology.

Xu et al.~\cite{Xu2010EdgeBetween} propose Sybil detection based on the betweenness rank of the edges.
The betweenness of an edge is defined as the number of shortest paths in the social graph passing the edge~\cite{Brandes2001BetweennessAlgo}.
The scheme assumes that the number of attack edges is limited and Sybil and non-Sybil regions are separate clusters of nodes.
So, intuitively betweenness scores of the attack edges should be high as they connect the clusters.
Their scheme exploits this social network property and uses a Sybil Resisting Network Clustering (SRNC) algorithm to detect Sybils.
The algorithm computes the betweenness of each edge and identifies the edges with high betweenness as attack edges.

Social graph based approaches still have some challenges to overcome.
First, as graph-based Sybil detection schemes exploit trust relations, the success of the  identification highly depends on the trust related assumptions. 
If an assumption is not right in a network, social graph-based Sybil detection techniques might work poorly in that network. 
For example, the assumption that Sybils' have problems in creating social connections with legitimate users (non-Sybils) is not  well established.
Although study~\cite{Motoyama2011Dirtyjob} shows that most of a Sybil identity's  connections are also Sybil identities and Sybils' have less relationships with non-Sybil users, several other studies~\cite{Boshmaf2011Bots,irani2011reverse,Bilge2009Contacts}  show that users are not careful while accepting friendship requests and Sybil identities can easily  befriend with them.
Moreover, Sybil users are using advanced techniques to create more realistic Sybil identities, either by copying profile data from existing accounts, or by assigning real users to customize them.
Also, another assumption that a social network is fast-mixing may not be right for all social networks. 
Study~\cite{Mohaisen2010Mixing} shows that many of the social networks are not fast-mixing, especially where edges represent strong real-world trust (e.g., DBLP, Epinions, etc.).

Second, the performance of random walk-based Sybil detection techniques depends on the various relevant parameters of the random walks (e.g., the length of a random walk).
These factors will work for a fixed network size (as all the schemes have shown), but they have to be updated with the evolution of the social networks.

\subsection{Sybil Resistance}
\label{res}

Sybil resistance schemes do not explicitly label users' as Sybils and non-Sybils, rather they attempt to mitigate the impact that a Sybil user can have on others.
Sybil resistance schemes have been effectively used in applications from diverse domains including content rating systems~\cite{Tran09sumUp,Chiluka2012LTD}, spam protection~\cite{Mislove2008Ostra}, online auctions~\cite{Post2011Bazar}, reputation systems~\cite{DeFigueiredo2005TrustDavis}, and collaborative mobile applications~\cite{Quercia2010MobileSybil}.

Note two assumptions of  Sybil detection schemes: 1) non-Sybil region is fast mixing, 2) Sybils can not create an arbitrary number of social relationships with non-Sybils.
Sybil resistance schemes also assume that non-Sybils' have a limited number of social connections, but they do not rely on the fast mixing nature of the non-Sybil regions.
However, Sybil resistance schemes take an additional application related information such as users' interactions/transactions/votes etc.
Using the underlying social network of the users and system information, Sybil resistance schemes determine whether an action performed by a user should be allowed or denied.

\begin{figure}[htbp]
\centering
\includegraphics[height=4.5cm]{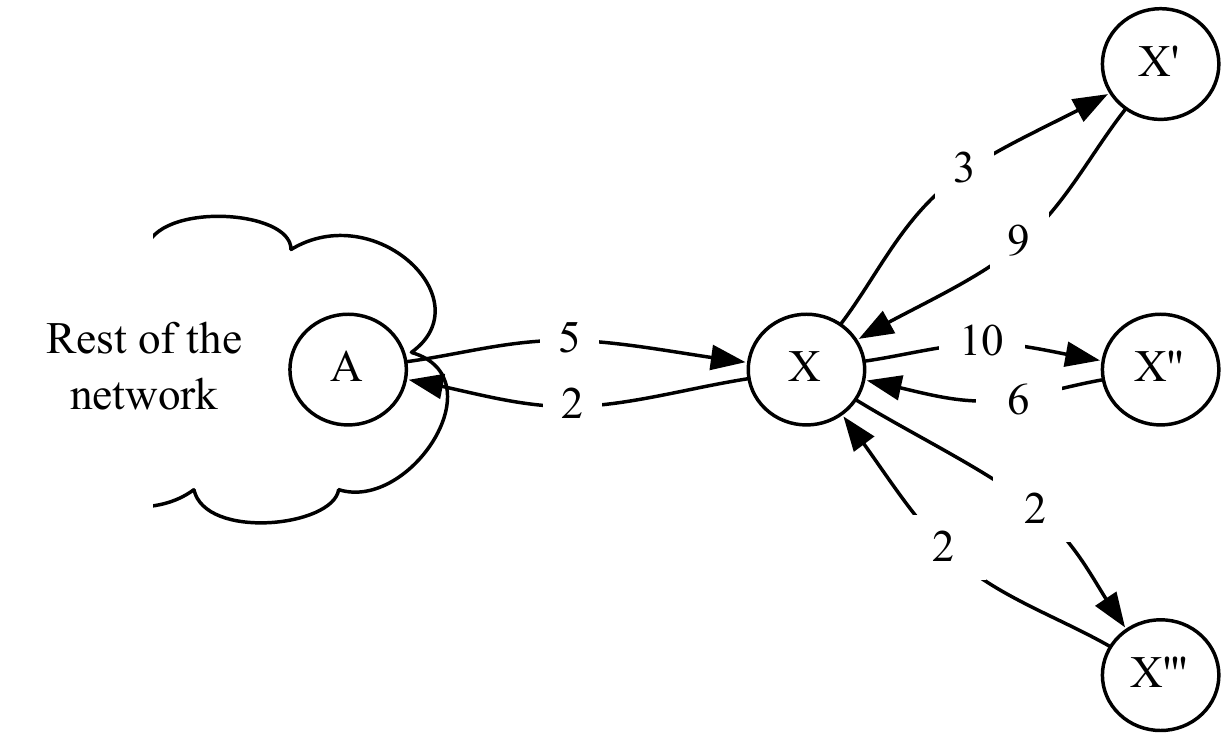}
\caption{Credit network based Sybil resistance~\cite{Viswanath2012Design}.
The network contains four Sybil identities as nodes {X, $X'$, $X''$, $X'''$} of a Sybil user.
A directed edge (X,Y) represents how much credit is available to X from Y.
If X wants to pay credits from other three nodes, the credits must be deducted from X's single legitimate link to A.
So, a Sybil's other identities do not provide any additional credits in the rest of the network.}
\label{fig:creditNetwrok}
\end{figure}

Most of the Sybil resistance schemes~\cite{Mislove2008Ostra,Post2011Bazar,Viswanath2012Canal} share a common approach in resisting Sybils---they use a \emph{credit network} built on the top of the social network of users~\cite{Viswanath2012Design}.
Originally proposed in the electronic commerce community, \textit{Credit Networks}~\cite{Dandekar2012SFC,Ghosh2007Credit} create mutual trust protocols in a situation where there is pairwise trust between two users, and a centralized trusted party is unavailable.
Nodes in a credit network trust each other by providing credits up to a certain limit.
Nodes use these credits to pay for services (e.g., sending a message, purchase items, vote casting) that they receive from one another.
These schemes assign credits to the network links, and allow an action between two nodes if there is a path between them that has enough credit to satisfy the operation.
As such, these schemes find a credit assignment strategy in the graph and apply the credit payment scheme to allow a limited number of illegitimate operations in the system.
A Sybil user has limited number of edges with non-Sybils (hence, limited credits available), which restricts her to gain additional advantages by creating multiple Sybil identities. 
This scenario is shown in figure~\ref{fig:creditNetwrok}, which is a core defense philosophy of some resistance schemes.
In the following, we provide a brief overview of  the Sybil resistance schemes.

\emph{Ostra}~\cite{Mislove2008Ostra} leverages existing trust relationships among users to thwart unwanted communication (e.g., spam).
It bounds the total number of unwanted communications a Sybil user can produce by assigning credit values to the trust links.
If a user sends a message to another user, Ostra finds a path with enough credit from the sender to the receiver.
If a path is available, credit is assigned along all the links in the path, which is refunded if the receiver considers the messages as not unwanted.
However, if no such path exists, Ostra blocks the communication, but the credit is paid.
In this way, Ostra ensures that a user with multiple identities cannot send a large number of unwanted communications, unless she also has additional trust relationships.

\emph{Bazaar}~\cite{Post2011Bazar} is targeted to strengthen the users' reputation in online marketplaces like eBay.
The opportunity to create accounts freely leads Sybil users to create multiple accounts and causes the waste of time and significant monetary losses for defrauded users. 
To mitigate Sybil users, Bazaar creates transaction network by linking users who have made a successful transaction.
The weight of  a link is the amount that has been successfully transferred due to the transaction.
Prior to a transaction, using a max flow based technique, Bazaar computes the reputation of the users doing the transaction and compares with the new transaction value.
If it finds available flow, it removes the value of the transaction between the users  as credits, and eventually adds back if the transaction is a fraud.
However, a new transaction is denied if essential flow is not found.

\emph{Canal}~\cite{Viswanath2012Canal} complements Ostra and Bazaar credit networks-based Sybil resistance schemes by applying landmark routing-based techniques in calculating credit payments over a large network.
One of the major problems of Ostra and Bazaar is that they require computing max-flow over a graph.
However, the huge size of present day network (Facebook has over billion of nodes in social graph) leads to significant computation complexity to compute the max-flow between two nodes in the network.
As such, this poses a bottleneck to those techniques to practically deploy in a real-world social network.
Canal efficiently computes an approximate max-flow (compromising accuracy with speed-up) path using existing landmark routing-based algorithm~\cite{Tsuchiya1988Landmark,Gubichev2010Landmark}.
The main components of Canal are \emph{universe creator processes} and \emph{path stitcher processes}.
Universe creator processes continuously select new landmarks and path stitcher processes continuously process incoming credit payment requests.
%To adjust the changes of credit payments in the network, it continuously calculates new landmarks in parallel with making flow calculations.
Using real-world network datasets the authors show that Canal can perform payment calculations efficiently (within a few milliseconds), even if the network contains hundreds of millions of links.

\emph{MobID}~\cite{Quercia2010MobileSybil} makes co-located mobile devices resilient to Sybil attackers.
Portable devices in close proximity of each other could collaborate various services (e.g., run localization algorithms to get a precise street map), which is severely disrupted by Sybil users (Sybils could inject false information).
MobID uses mobile networks to Sybil resilience.
More specifically, a device manages two small networks as it meets with other devices.
A network of friends contains non-Sybil devices and a network of foes contains suspicious devices.
Using two networks, MobID determines whether an unknown device is attempting a Sybil attack.
MobID ensures that a non-Sybil device accepts, and accepted by most other non-Sybil devices with high probability.
So, a non-Sybil device could successfully trade services with other non-Sybil devices.

\section{Mitigating Attacks From Large-scale Crawlers}
\label{crawl}

OSNs enhance social browsing experience by allowing users to view public profiles of others.
This way a user meets others, gets a chance to know strangers and eventually befriends some of them.
Unfortunately, attackers are there in the vast landscape of OSNs, who exploit this functionality.
Users' social data are always invaluable to marketers.
Professional data aggregators build databases using public views of profiles and social links and sale the databases to insurance companies, background-check agencies and credit-ratings agencies~\cite{Bonneau2009Prey}.
For example, crawling 100 million public profiles from Facebook created news recently~\cite{Bradley2012Crawl}.
Sometimes crawling is a violation of terms of service.
Facebook states that someone should not collect  ``...users' content or information, or otherwise access Facebook, using automated means (such as harvesting bots, robots, spiders, or scrapers) without our prior permission''~\cite{Facebook2015Terms}.

One solution of the problem could be the removal of the public profile view functionality.
But removal of the public profile view functionality is against the business model of OSNs.
Services like search and targeted advertisements bring new users and ultimately revenues to OSNs, but openly accessible contents are necessary for their operation~\cite{Wilson2010spikestrip}.
Moreover, removal of the public view functionality will undermine user experience, as it makes a connection, communication and sharing easy with unknown people in the network.

OSN operators such as Facebook and Twitter attempt to defend large-scale crawling by limiting the number of user profiles a user can see from an IP address in a time window~\cite{Stein2011FIS}.
However, tracking users with low level network identifiers (e.g., IP address, TCP port numbers or SSL session IDs) is fundamentally flawed as  a solution of this problem~\cite{Wilson2010spikestrip}.
Aggressive attackers may gather a large vector of those  identifiers by creating a large number of fake user accounts, gaining access to compromised accounts, virtualizing in a cloud, employing botnets, and forwarding requests to proxies.
Until now, researchers have leveraged encryption based technique~\cite{Wilson2010spikestrip} and crawler's observational behavior~\cite{mondal2012genie} to combat the problem.

ONS's anti-crawling techniques suffer from the fact that web clients can access a particular page using a common URL accessible to all clients~\cite{Wilson2010spikestrip}.
This can be exploited by a distributed crawler e.g., a crawling thread can download and parse a page for links using a session key and can deliver those links to another crawling thread to download and parse using different session keys.
So, if some crawlers get banned from the OSNs for malicious activities, the links they have parsed are still valid and a fresh start is possible from those links.
SpikeStrip~\cite{Wilson2010spikestrip} overcomes the problem by creating unique, per-session ``views'' of the protected website that forcibly tie each client to their session keys.
SpikeStrip is a web server add-on that leverages link encryption technique.
It allows OSN administrators to moderate data access, and it defends against large-scale crawling by securely identifying and rate limiting individual sessions.

When a crawler visits a page, it receives a new session key and a copy of the page whose links are all encrypted.
SpikeStrip appends each user's session key to those links and then encrypts the result using a server-side, secret symmetric key. 
It also appends a \textit{salt} to the link after encryption to make each link unique. 
As time passes, the crawler progressively covers more pages and collects links.
However, at a point, the crawler requires to change the session key due to the expiration of the session or due to a ban from the OSN. 
As SpikeStrip couples all URLs to the browser's session key, this switching of sessions invalidates all the links collected for future traversals.
Thus, a fresh start to reconstruct the collection should be started from the beginning.
The authors implemented mod\_spikestrip, a SpikeStrip implementation for Apache 2.x and showed that it imposes only 7\% performance penalty on Apache.

Genie~\cite{mondal2012genie} exploits browsing patterns of honest/real users and crawlers and thwarts large-scale crawls in OSNs using Credit Networks~\cite{Dandekar2012SFC,Ghosh2007Credit}.
The system design is based on three observations from real-world datasets: 
(i) there is a balance between the number of profiles a honest user views and views requested by other users to her profile, but crawlers view many more profiles than the number of times their profiles are viewed;
(ii) a honest user views profiles of socially close users.%other users who are nearby in the social network; 
(iii) a honest user repeatedly views a small set of profiles in the network, but unless re-crawling, the crawlers avoid repeating viewing of other users' profiles.
Genie leverages these observations and enforces a viewer to make a  ``credit payment'' in the credit network if a user wants to view a profile.
It allows a user (also might be a crawler) to view a profile if a max-flow between them has at least a threshold value.
The required credit payment to view a profile depends on the shortest path length from viewer to viewee; a user has to pay more to view the profile of a distant user in the social graph.
As a legitimate user usually views one or two hop distant profiles, and also other users also view her profile, her liquidity of credits remains almost the same. 
On the other hand, a crawler views a lot of distant profiles and gets fewer views.
Eventually it lacks credit liquidity to view the profiles of others.
As such, the credit network poses a strict rate limit on profile views of the crawlers.

Genie might see a large number of honest users' activities (profile viewing) flagged due to the existence of outliers in a social network.
This might limit the usability of social networks, because without viewing a profile an outlier will not be able to befriend others. 
Genie also might require a fast computation of shortest paths, as for each profile viewing request, it computes all the shortest paths from viewer to viewee.
Intuitively, this operation is too costly in a modern social network (more than one billion users), even considering the state of the art shortest path algorithms.

Both SpikeStrip and Genie limit crawlers' ability to quickly aggregate a significant portion of OSNs user data.
Unfortunately, equipped with a large number of user profiles (fake or compromised) and employing dedicated crawlers for a long time, attackers could still collect a huge amount of users' social data.

\section{Mitigating Social Spam}
\label{spams}

%Social spam is content that is designed to mislead users or content that the site's ``legitimate'' users don't wish to receive~\cite{Heymann2007Spam}.
Spam is  a news in web-based systems (e.g.,~\cite{Xie2006EDA,Ntoulas2006SpamWeb,Mehta2008Spam}).
However, OSNs have added a new flavor to it by acting as effective tools for spamming activities and propagation.
%Although OSNs serve as popular collaboration and communication tools for millions of users, at the same time they are also effective tools for spamming activities.
%Social spammers threaten to undermine resource sharing, interactivity, and openness of the OSNs by contributing phishing attacks~
Social spam (e.g., ~\cite{Zinman2007Britney},~\cite{Lin2007Splog}) is unwanted content  that is directed specifically at users of the OSN.
The worst consequences of social spam include  phishing attacks~\cite{Jagatic2007Phishing} and malware propagation~\cite{Boyd2006Malware}.
%commercial spam messages~\cite{Zinman2007Britney}, and promoting affiliate websites~\cite{Lin2007Splog}.

Spamming activity is pervasive in OSNs and spammers are successful.
For example, about 0.13\% of spam tweets in Twitter generate a page visit~\cite{Grier2010SUCSpam}, which is only 0.003\%-0.006\% for spam email~\cite{Kanich2008AnlysisSpam}.
%which is orders of magnitude higher than click-through rate of 0.003\%-0.006\% reported for spam email~\cite{Kanich2008AnlysisSpam}.
This high click-through is due to the fact that OSNs expose intrinsic trust relationship among online friends.
%, even though they may not know each other in real life. 
As such, users read  and click messages or links that are shared from their friends.
Study~\cite{Bilge2009Contacts} shows that 45\% of users on OSNs click on links posted by their friends' accounts.
%, even they do not know those people in real life.

Defending spam in OSNs can improve user experience.
OSN service providers will also be benefited as this will lessen the system workload in terms of dealing with unwanted communications and contents. 
Defense mechanisms against spam in OSNs can be classified into two categories: 1) spam content and profile detection, and 2) spam campaign detection.
Spam content and profile-level detection involve checking individual accounts or contents for an evidence of spam contents (Section~\ref{profile}). 
On the other hand, a spam ``campaign'' is a collection of malicious content having  a common goal, for example, selling backdoor products~\cite{Gao2010Spamcampaign} (Section~\ref{campaign}).

\subsection{Spam Content and Profile Detection}
\label{profile}
Some early spam profile detections~\cite{Spitzner2002Honeypots,Webb2008SocialHoneypots,Lee2010Spam} use social honeypots.
A honeypot is a trap deployed to capture examples of nefarious activities in networked systems~\cite{Spitzner2002Honeypots}.
For years, researchers have used honeypots to characterize malicious hacker activities~\cite{Spitzner2003Honeypots}, to obtain footprints of email address crawlers~\cite{Prince2005Honeypots}, and to create intrusion detection signatures ~\cite{Kreibich2004Honeypots}.
Social honeypots are used to monitor spammers' behaviors and store their information from the OSNs~\cite{Lee2010Spam}.

Webb et al.~\cite{Webb2008SocialHoneypots}  take the first step to characterize spam in OSNs using social honeypots.
They created 51 honeypot MySpace profiles  in different geographic locations for harvesting deceptive spam profiles on MySpace.
An automated program (commonly known as bots) works on behalf of a honeypot profile and collects all of the traffic it receives (via friend requests).
After four months of the deployment and operation, the bots collected 1,570 friend requests (and corresponding spam profiles).
Through statistical analysis the authors show the followings: (i) spam profiles follow distinct temporal patterns in spamming activity; (ii) 57.2\% of the ``About me'' contents of  the spam profiles  are duplicated; (iii) spam profiles redirect users to predefined web pages.
%use redirection techniques to funnel users to destination web pages.

In~\cite{Lee2010Spam}, the authors also collected spam profiles using social honeypots.
But this work is different from the previous one in that it not only collects and characterizes spam profiles, it extracts features from the gathered  spam profiles and builds classifiers to detect potential spam profiles. 
The authors consider four  categories of features such as demographics, content, activity and connections  from the spam profiles collected from MySpace and Twitter.
%They are: (i) user demographics:  age, gender, location, and other descriptive information about the user; 
%(ii) user-contributed content: ``About Me'' text, blog posts, comments posted on other users' profiles, tweets, etc.; 
%(iii) user activity: posting rate, and tweet frequency; 
%(iv) user connections: number of friends in the social network, followers and following.
The paper shows the performance results of ten classifiers  using the features. 
For MySpace dataset, each classifier's accuracy is  greater than 98.4\%, for Twitter dataset, each of the top 10 classifiers achieves an accuracy greater than 82.7\%. 
% For MySpace dataset, each classifier generates an accuracy greater than 98.4\%, an F1 measure over 0.98 and a false positive rate below 1.6\%.
 %For the Twitter dataset, each of the top 10 classifiers achieves an accuracy greater than 82.7\%, an F1 measure over 0.82, and a false positive rate less than 10.3\%.
 
One of the limitations of  these honeypot-based solutions~\cite{Webb2008SocialHoneypots,Lee2010Spam} is that they consider all profiles that sent friend requests to honeypots are spam profiles.
But in social networks, it is common to receive friend requests from unknown person, who might be legitimate users in the network. 
%Considering all of the requests as spam limits the true understanding of the collected data.
The solutions would be more rigorous if legitimate users were not considered.
Also, the methods are effective when spammers become friends with the honeypots.
%Unless social honeypots are popular (a difficult task to do), they will be able to target only a small subset of the spammers.
Otherwise the honeypots will be able to target only a small subset of the spammers.
Moreover, in social networks, friendship is not always required for spamming.
For example, in twitter, a spammer can use mention (e.g., @user) tag to send spam tweets to a user.
 
Stringhini et al.'s solution~\cite{Stringhini2010Spam} overcome some limitations of the previous two honeypot-based papers.
The authors deployed honeypots accounts on Facebook, Twitter and MySpace; 300 on each platform for about one year and logged the traffic (e.g., friend requests, messages, and invitations).
Combining all, the honeypots received 4,250 friend requests and 85,569 messages from the three platforms.
They build classifiers from the following six features: 
(i) FF ratio: the ratio of the number of friend requests sent by a user and the number of friends she has;
(ii) URL ratio:  the ratio of the number of messages containing URLs and total messages;
(iii) Message Similarity: similarity among the messages sent by a user;
(iv) Friend Choice: the ratio of the total number of names among the profiles' friends, and the number of distinct first names;
(v) Messages Sent: the number of messages sent by a profile as a feature; and 
(vi) Friend Number: the number of friends a profile has.

The authors manually inspected and labeled profiles as spam and used Random Forest algorithm for classification. 
A 10- fold cross validation on the training data set of Facebook yielded an estimated false positive ratio of 2\% and a false negative ratio of 1\%.
Twitter dataset yielded an estimated false positive ratio of 2.5\% and a false negative ratio of 3\%.
They detected 15,857 spam profiles on Twitter using the classifier and the Twitter spam team eventually suspended those accounts.

Benevenuto et al.~\cite{Benevenuto2009YoutubeSpam1,Benevenuto2009YoutubeSpam2} detect video polluters such as spammers and promoters in YouTube online video social networks using machine learning techniques.
The authors considered three attribute sets: user attributes, video attributes, and social network (SN) attributes in classification.
Four volunteers manually analyzed the videos and built a test set of the dataset labeling users as spammers, promoters and legitimate users.
They proposed a flat classification approach, which was able to detect correctly 96\% of the promoters, 57\% of spammers, and wrongly classifying only 5\% of the legitimate users. 
Interestingly, social network attributes performed the worst in classification---only one feature (UserRank) was within the top 30 features.
%YouTube spammers post irrelevant response videos targeting  a popular video which attracts a larger number of users. 
%Promoters also post a large number of irrelevant responses to boost the rank of the responded video, making it appear in the top lists maintained by the system.
%The authors crawled a total of 264,460 users, 381,616 responded videos and 701,950 video responses from YouTube and applied machine learning technique to detect spammers, promoters and legitimate users.
%First, four volunteers manually analyzed the videos and built a test set of the dataset labeling users as spammers, promoters and legitimate users.
%They got a test collection containing 829 users, of them 641 classified as legitimate users, 157 as spammers, and 31 as promoters. 
%The authors considered three attribute sets: video attributes, user attributes, and social network (SN) attributes in classification.
%They proposed a flat classification approach, which was able to detect correctly 96\% of the promoters, 57\% of spammers, and wrongly classifying only 5\% of the legitimate users. 
%Interestingly, social network attributes performed the worst in classification---only one feature (UserRank) was within the top 30 features.
%However, the accuracy of the classification  was low (it was able to detect 57\% of spammers).
%, it was one of the earliest steps toward detecting spam in video-based social networks.

\subsection{Spam Campaigns Detection}
\label{campaign}
Chu et al.~\cite{Chu2012SpamCampaign} detect social spam campaigns on Twitter using tweet URLs.
They collected a dataset of 50 million tweets from 22 million users.
They considered tweets having the same URL as a campaign and clustered the dataset into a number of campaigns.
%clustered tweets with the same final URL into a campaign and partitioned the dataset into a number of campaigns based on URLs.
%They got 5,183,656 campaigns, the largest one contained 7,350 accounts with 9,761 tweets posted.
The ground truth was produced through manual inspection using Twitter's spam rules  and automated URL checking in five services.
%They followed Twitter's spam rules during the manual inspection and a batch script performed the automated blacklisted URL checking  in five services. such as Google Safe Browsing, PhishingTank, URIBL, SURBL and Spamhaus.
%The ground truth set included 744 spam campaigns (containing around 70,000 accounts and 131,000 tweets) and  580 legitimate campaigns (containing around 150,000 accounts and 180,000 tweets).
They obtained a variety of features ranging from individual tweet/account levels to a collective campaign level and built a classification model to detect spam campaigns.
Using several classification algorithms they were able to detect spam campaigns with more than 80\% success rate.
The focus of this solution is spam tweets with URLs.
However, Twitter spammers can post tweets without any URL.
Even obfuscated URLs (e.g., somethingDOTcom) will make the detection inefficient.

Gao et al.~\cite{Gao2010Spamcampaign} conduct a rigorous and extensive study on detecting spam campaigns in Facebook wall posts.
They crawled 187 million Facebook wall posts from about  3.5 million users.
Inspired by a study~\cite{Kreibich2009Spamcraft} which shows that spamming bot-nets create email spam messages using  templates, they consider wall posts having similar texts as a spam campaign.
%propose to group wall posts (campaigns) with similar textual description together.
In particular, they model the wall posts as a graph: a post is a node and two nodes are connected by an edge if they have the same destination URL or their texts are very similar.
%all wall posts as nodes in a large graph, and add edges when two posts share either the same (possibly obfuscated) destination URL, or strong textual similarity.
%The idea is to find connected subgraphs, which could represent messages within the same spam campaign. 
As such, posts from the same spam campaign will make  connected subgraphs or clusters.
To detect which clusters are from spammers, they use ``distribute'' coverage and ``bursty'' natures of spam campaigns.
The ``distributed'' property is characterized based on the number of user accounts posting in the cluster under the intuition that spammers will use a significant number of registered accounts for a  campaign.
The intuition behind the ``bursty'' property is that most spam campaigns are the results of  coordinated actions of many accounts within short periods of time~\cite{Xie2008SpamSignature}.
Using threshold filters on these two properties they found 297 clusters of wall posts and classified them as potentially malicious spam campaigns. 
%(5 as the minimum number of users involved in each cluster and 5400 seconds as the maximum median interval between the timestamp of two consecutive wall posts) 
%Analyzing the spam campaigns the paper shows that  70.3\% of malicious wall posts direct the victim to a phishing site.
%Moreover, using time-duration of messages and the lifetimes of accounts who send them, the paper concludes that more than 97\% of spam messages are sent from compromised accounts, rather than fake accounts specifically created for spam delivery.

%The authors validated their approach by detecting malicious URLs that are embedded in one or more wall posts in a potentially malicious cluster. 
%They used several techniques such as URL de-obfuscation, redirection analysis, third-party tools, wall post keyword search (e.g., viagra),  groups of the URLs that exhibit highly uniform features and manual analysis in finding malicious clusters.
%Their results show that the threshold filter based approach could successfully detect spam campaigns with a high success rate---the true positive rate is 96.1\% and false positive rate is 3.9\%.

\section{Mitigating Distributed Denial-of-service attacks (DDoS) Attacks}
\label{ddos}

A denial-of-service (DOS) attack is characterized by an explicit attempt to monopolize a computer resource, so that an intended user cannot use the resource~\cite{Mirkovic2004DOS}.
A Distributed Denial-of-Service attack (DDoS)  deploys multiple attacking entities to simultaneously launch the attack (we refer readers~\cite{mirkovic2004taxonomy} for a taxonomy of web-based DDoS attacks and defenses). 
DDoS attacks in social networks are also common.
%might exploit the social network itself and motives behind those attacks might be very trivial.
For example, on August 6, 2009, Twitter, Facebook, LiveJournal, Google's Blogger, and YouTube were attacked by a DDoS attack~\cite{McCarthy2009DDoS}. 
Twitter experienced interrupted service for several hours, users were complaining of not being able to send their Tweets. 
Facebook users were experiencing longer periods of time (delays) in loading Facebook pages.
%This DDoS attack was launched to silence the account Cyxymu; an account owned by the Georgian economics professor, in response to the professor's criticism of Russia's conduct in the year-long war with Georgia~\cite{Timm2010Seven}!
%Although a DDoS attack is not a news and  traditional defense techniques are available~\cite{kandula2005botz,jung2002flash,gu2008botminer,walfish2006ddos}, social networks have amplified the threat by contributing the social fabric to the attacking model, and established themselves as new targets of  the DDoS attacks.

Several papers evaluated how a social network could be leveraged to launch a bot-net based DDoS on any target of the internet, including the social network itself.
%In OSNs, a DDoS attack might exploit the network itself to launch the attack.
Athanasopoulos et al.~\cite{Athanasopoulos2008Antisocial} introduce a  bot-net ``FaceBot'' that uses a social network to carry out a DDoS attack against any host on the internet (including the social network itself).
They created a real-world Facebook application, ``Photo of the Day'', that presents a different photo from National Geographic to Facebook users every day.
Every time a user clicks on the application, an image from the National Geographic appears. 
However, they placed special codes in the application's source code.
Every time a user views the photo, this code sends a HTTP request towards a victim host, which causes the victim to serve a request of 600 KBytes.
They used a web server as a victim and observed that the server recorded 6 Mbit per second of traffic.
%The authors conclude that using a highly popular Facebook application, it is possible to operate a DDoS attack, which could target the social network itself.
They introduce defense mechanisms which include providing application developers with a strict API that is capable of giving access to resources only related to the system.
%However, this requirement poses a serious constraint on application developers, who rely on their servers for operations of the applications.

Ur and Ganapathy~\cite{Ur2009Evaluating} showed how malicious social network users can leverage their connections with hubs to launch DDoS attacks.
They created MySpace profiles which befriended hubs in the network.
Those profiles posted ``hotlinks'' to  large media files hosted by a victim web server to Hubs' pages.
As hubs receive a large number of hits, a significant number of the visitors would click those hotlinks.
As a consequence, it staged a scenario where a flash crowd was sending requests to the victim web server---a denial of service was the result.
They proposed several mitigating techniques.
One approach is to restrict some privileges of a user when he becomes a hub (e.g., friends of a hub might no longer be able to post comments containing HTML tags to the hub's page). 
But this approach unfortunately restricts the user's freedom on the OSN.
So, they propose a focused automated monitoring on a hub or creating a hierarchy of a hub's friend, so that only close friends will be able to post on a Hub's profile (the intuition is that close friends will not exploit the hub).
Furthermore, they recommend a  reputation based system for social networks that scores user behavior. 
Only users with a higher reputation scores are allowed to post on the Hub's profile.

However, bot-net based DDoS attacks are difficult to mitigate, because of the difficulty to distinguish legitimate communications from those that are part of the attack. 
As social networks are flourishing, bot-net based DDoS attacks are becoming stronger, because more legitimate users are unwillingly becoming  part of an attack.

\section{Mitigating Malware Attacks}
\label{malware}

Malicious software (malware) is a program that is specifically designed to gain access, disrupt computer operation, gather sensitive information or damage a computer without the knowledge of the owner.
Participatory internet  technologies (e.g., AJAX) and applications (e.g., RSS)  have expedited malware attacks, because they enable the participation of the users.
OSNs (all of them use participatory technologies and applications) are providing themselves as infrastructures for propagating malware.
The ``Koobface'' is probably the best example of malware propagation using social networks~\cite{Facebook2012Koobface}.
It spread rapidly  through Facebook social networks.
The malware used Facebook credentials on a compromised computer and sent  messages to the owner's Facebook friends.
The messages redirected the owner's friends to a third-party website and they were asked to download an update of the Adobe Flash player. 
If they would download and install the file, Koobface would install and infect their system using the same process.

%A variety of other methods in which malware propagates through social networks is discussed in a survey of Gao et al.~\cite{Gao2011Security}.

In a survey, Gao et al.~\cite{Gao2011Security} discuss a number of methods  in which malware propagates through social networks.
For example, using cross-site request forgery (CSRF or XSRF) malware invites legitimate users to click on a link. 
If a user clicks, it opens an exploited page containing malicious scripts. 
Eventually, the malware submits a message with a URL for a wall post  on the user's profile and clicks on the ``Share'' button so that all of her friends can see this message as well as the link.
URL obfuscations are also widely used for malware attacks.
An attacker  uses commonly known URL shorteners to obfuscate the true location of a link and lures other users to click it. 

Unfortunately, malware propagation on social networks exhibits unique propagation vectors.
As such, existing Internet worm detection techniques (e.g.,~\cite{Ellis2004Worm}) cannot be applied to them.
In the context of OSNs, Xu et al.~\cite{Xu2010Malware} proposed an OSN malware detection system by leveraging both the propagation characteristics of the malware and the topological properties of OSNs.
They  introduced a ``maximum coverage algorithm'' that picks a subset of legitimate OSN users to whom the defense system attaches ``decoy friends'' to monitor the entire social graph.
When the decoy friends receive suspicious malware propagation evidence, the detection system performs local and network correlations to distinguish actual malware evidence from normal user communication. 
However, the challenge for this honeypot-based approach is to determine how many social honeypots (in this context decoy friends) large-scale OSNs (e.g., billions of Facebook users) should deploy.

%to what extent this honeypot based approach will be able to successfully detect malware in a large social network (e.g., billions of Facebook users) is a big concern.

\section{Summary and Discussion}
\label{summary}
Millions of Internet users are using OSNs for communication and collaboration.
Many companies rely on OSNs for promoting their products and influencing the market. 
It becomes harder and harder to imagine life without the use of OSN tools, whether for creating an image of oneself or organization, for selectively following news as filtered by the group of friends, or for keeping in touch.
However, the growing reliance on OSNs is impaired by an increasingly more sophisticated range of attacks that undermine the very usefulness of the OSNs.   

\ignore{
However, a myriad of privacy and security attacks are undermining the objectives of using OSNs.
Users are living in a social network ecosystem, where their privacy is being compromised by social applications, third-party data miners, crawlers and unfortunately sometimes by the OSNs themselves.
This ecosystem has become a perfect vehicle for spam, malware and rumor propagation and to launch distributed denial-of-service attacks.   
At the same time, malicious individuals and organizations are employing social robots and cheap crowdsourcing laborers to create fake and clone profiles.
These profiles are used to disrupt the proper functioning of an OSN, to gain illegitimate power, influence and consequently to  diminish the trust users put in the OSNs.
}

This paper reviews online social networks' privacy and security issues.
We have categorized various attacks on OSNs based on social network stakeholders and the forms of attack targeted at them.
%We have identified the attacks considering the stakeholders' interaction with other entities.
Specifically, we have categorized those attacks as attacks on users and attacks on the OSN.
We have discussed how the attacks are launched, what are the available defense techniques and what are the challenges involved in such defenses.

In online social networks, privacy and security issues are not separable.
In some contexts privacy and security goals may be the same, but there are other contexts where they may be orthogonal, and there are also contexts where they are in conflict.
For example, in an OSN, a user wants privacy when she is communicating with other users though the messaging service.
She will expect that non-recipients of the message will not be able to read it.
OSN services will ensure this by providing a secure communication channel.
In this context, the goals of security and privacy are the same.
Consider another context where there is a security goal of authenticating a user's account.
OSNs usually do this by sending an activation link as a message to the user's e-mail address.
This is not a privacy issue---OSNs are just securely authenticating that malicious users are not using the legitimate user's e-mail to register.
In this context, security and privacy goals are orthogonal.
However, anonymous views in OSNs (e.g., LinkedIn) present a context where security and privacy goals are in conflict.
Users may want to have privacy (e.g., anonymization) while viewing other users' profiles.
However, the viewee might want to secure her profile from anonymous viewing.
%This is a scenario where security and privacy goals are in conflict.
%Categorization of various attacks and their implications on privacy and security is a feasible way to understand privacy and security issues in OSNs.

The attacks discussed in this paper  are often closely intertwined.
User data collected though crawling attacks or via social applications may help an attacker to create background knowledge for launching de-anonymization attacks.
An attacker might possess an unprecedented number of user accounts using malware and Sybil attacks and could use those accounts for social spam propagation and distributed denial-of-service attacks. 
Social spam can also be used to propagate malware.
Some attacks might be a pre-requisite for another attacks.
For example, a de-anonymization attack can reveal the identity of an individual.
That identity could be used to launch an inference attack and to learn unspecified attributes of an individual.

There are also several functionality-oriented attacks that we did not discuss in this paper.
Functionality-oriented attacks attempt to exploit specific functionalities of a social network.
For example, Location-based Services (LSP) such as Foursquare, Loopt and Facebook Places utilize geo-location information to publish users' checked-in places.
In some LSP, users can accumulate ``points'' for ``checking in'' at certain venues or locations and  can get real-world discounts or freebies in exchange for these points.
There is a body of research that analyzes the technical feasibility of anonymous usage of location-based services so that users are not impacted by location sharing~\cite{Gruteser2003Location,Xiao2006Personalized}.
%Location cheating attacks are also common on LBS.
Moreover, real-world rewards and discounts give incentives for users in LSP to cheat on their locations, and hence research~\cite{Zhu2011Applaus,He2011Location} has focused on how to prevent users from location cheating.
%Reward-driven cheating has become a big concern for other social systems (e.g., user-contributed content systems, social blogging).
%For example, social blogging communities such as BlogCatalog reward their members with points when they post content, which leads to members posting irrelevant, useless content in order to gain rewards. \ainote{stated this way, this is not a problem of the social component of such systems, but of the incentive mechanism rather. I would lose the example with BlogCatalog -- or explain better if indeed is relevant for the discussion here}

OSNs and social applications are here to stay, and while they mature, new security and privacy attacks will take shape. 
Technical advances in this area can only be of limited effect if not supported by legislative measures for protecting the user from other users and from the service providers~\cite{Nissenbaum2004ContextOnline}. 
%\ainote{not sure if this is the note we want to end on. Or maybe there is a cite to support this? Certainly Nissembaum's initial paper on contextual integrity}

% Start of "Sample References" section

% Bibliography
\bibliographystyle{apalike}
\bibliography{Bibtex}

\begin{thebibliography}{}

\bibitem[Acohido, 2009]{Acohido2009CapthaBusiness}
Acohido, B. (2009).
\newblock Cybergangs use cheap labor to break codes on social sites.
\newblock
  \url{http://usatoday30.usatoday.com/tech/news/computersecurity/2009-04-22-captcha-code-breakers_N.htm/}.

\bibitem[Acquisti and Gross, 2006]{acquisti2006imagined}
Acquisti, A. and Gross, R. (2006).
\newblock Imagined communities: Awareness, information sharing, and privacy on
  the facebook.
\newblock In {\em Privacy enhancing technologies}, pages 36--58. Springer.

\bibitem[Adar et~al., 2007]{Adar2007AOLresearch}
Adar, E., Weld, D.~S., Bershad, B.~N., and Gribble, S.~S. (2007).
\newblock Why we search: visualizing and predicting user behavior.
\newblock In {\em Proceedings of the 16th international conference on World
  Wide Web}, WWW '07, pages 161--170, New York, NY, USA. ACM.

\bibitem[Adu-Oppong et~al., 2008]{Adu2008Social}
Adu-Oppong, F., Gardiner, C.~K., Kapadia, A., and Tsang, P.~P. (2008).
\newblock Social circles: Tackling privacy in social networks.
\newblock In {\em Symposium on Usable Privacy and Security (SOUPS)}.

\bibitem[Aiello et~al., 2008]{Aiello2008Likir}
Aiello, L., Milanesio, M., Ruffo, G., and Schifanella, R. (2008).
\newblock Tempering kademlia with a robust identity based system.
\newblock In {\em Peer-to-Peer Computing , 2008. P2P '08. Eighth International
  Conference on}, pages 30--39.

\bibitem[Aiello and Ruffo, 2012]{Aiello2012Lotusnet}
Aiello, L.~M. and Ruffo, G. (2012).
\newblock Lotusnet: tunable privacy for distributed online social network
  services.
\newblock {\em Computer Communications}, 35(1):75--88.

\bibitem[Alexa, 2014]{Alexa2014Users}
Alexa (2014).
\newblock The top 500 sites on the web.
\newblock \url{http://www.alexa.com/topsites}.

\bibitem[Arrington, 2006]{Arrington2006AOL}
Arrington, M. (2006).
\newblock Aol proudly releases massive amounts of private data.
\newblock
  \url{http://techcrunch.com/2006/08/06/aol-proudly-releases-massive-amounts-of-user-search-data/}.

\bibitem[Athanasopoulos et~al., 2008]{Athanasopoulos2008Antisocial}
Athanasopoulos, E., Makridakis, A., Antonatos, S., Antoniades, D., Ioannidis,
  S., Anagnostakis, K.~G., and Markatos, E.~P. (2008).
\newblock Antisocial networks: Turning a social network into a botnet.
\newblock In {\em 11th International Conference on Information Security}, pages
  146--160. Springer.

\bibitem[Backstrom et~al., 2007]{Backstrom2007WAT}
Backstrom, L., Dwork, C., and Kleinberg, J. (2007).
\newblock Wherefore art thou r3579x?: anonymized social networks, hidden
  patterns, and structural steganography.
\newblock In {\em Proceedings of the 16th international conference on World
  Wide Web}, WWW '07, pages 181--190, New York, NY, USA. ACM.

\bibitem[Baden et~al., 2009]{Baden2009Persona}
Baden, R., Bender, A., Spring, N., Bhattacharjee, B., and Starin, D. (2009).
\newblock Persona: an online social network with user-defined privacy.
\newblock In {\em Proceedings of the ACM SIGCOMM 2009 conference on Data
  communication}, SIGCOMM '09, pages 135--146, New York, NY, USA. ACM.

\bibitem[Banks and Wu, 2009]{Banks2009AllFriends}
Banks, L. and Wu, S. (2009).
\newblock All friends are not created equal: An interaction intensity based
  approach to privacy in online social networks.
\newblock In {\em Computational Science and Engineering, 2009. CSE '09.
  International Conference on}, volume~4, pages 970--974.

\bibitem[BBC, 2012]{BBC2012Facebook}
BBC (2012).
\newblock Facebook has more than 83 million illegitimate accounts.
\newblock \url{http://www.bbc.co.uk/news/technology-19093078}.

\bibitem[Benevenuto et~al., 2009a]{Benevenuto2009YoutubeSpam2}
Benevenuto, F., Rodrigues, T., Almeida, J., Goncalves, M., and Almeida, V.
  (2009a).
\newblock Detecting spammers and content promoters in online video social
  networks.
\newblock In {\em INFOCOM Workshops 2009, IEEE}, pages 1--2.

\bibitem[Benevenuto et~al., 2009b]{Benevenuto2009YoutubeSpam1}
Benevenuto, F., Rodrigues, T., Almeida, V., Almeida, J., and Gon\c{c}alves, M.
  (2009b).
\newblock Detecting spammers and content promoters in online video social
  networks.
\newblock In {\em Proceedings of the 32nd international ACM SIGIR conference on
  Research and development in information retrieval}, SIGIR '09, pages
  620--627, New York, NY, USA. ACM.

\bibitem[Besmer et~al., 2009]{Besmer2009Framework}
Besmer, A., Lipford, H.~R., Shehab, M., and Cheek, G. (2009).
\newblock Social applications: exploring a more secure framework.
\newblock In {\em Proceedings of the 5th Symposium on Usable Privacy and
  Security}, SOUPS '09, pages 2:1--2:10, New York, NY, USA. ACM.

\bibitem[Bilge et~al., 2009]{Bilge2009Contacts}
Bilge, L., Strufe, T., Balzarotti, D., and Kirda, E. (2009).
\newblock All your contacts are belong to us: automated identity theft attacks
  on social networks.
\newblock In {\em Proceedings of the 18th international conference on World
  wide web}, WWW '09, pages 551--560, New York, NY, USA. ACM.

\bibitem[Bonneau et~al., 2009]{Bonneau2009Prey}
Bonneau, J., Anderson, J., and Danezis, G. (2009).
\newblock Prying data out of a social network.
\newblock In {\em Social Network Analysis and Mining, 2009. ASONAM '09.
  International Conference on Advances in}, pages 249--254.

\bibitem[Bonneau and Preibusch, 2010]{Bonneau2010PrivacyJungle}
Bonneau, J. and Preibusch, S. (2010).
\newblock The privacy jungle: On the market for data protection in social
  networks.
\newblock In {\em Economics of information security and privacy}, pages
  121--167. Springer.

\bibitem[Boshmaf et~al., 2011]{Boshmaf2011Bots}
Boshmaf, Y., Muslukhov, I., Beznosov, K., and Ripeanu, M. (2011).
\newblock The socialbot network: when bots socialize for fame and money.
\newblock In {\em Proceedings of the 27th Annual Computer Security Applications
  Conference}, ACSAC '11, pages 93--102, New York, NY, USA. ACM.

\bibitem[Boyd, 2004]{Boyd2004Friendster}
Boyd, D. (2004).
\newblock Friendster and publiclly articulated social networking.
\newblock In {\em Extended Abstracts of the Conference on Human Factors and
  Computing Systems (CHI 2004)}, pages 1279--1282.

\bibitem[Boyd and Heer, 2006]{Boyd2006Malware}
Boyd, D. and Heer, J. (2006).
\newblock Profiles as conversation: Networked identity performance on
  friendster.
\newblock In {\em System Sciences, 2006. HICSS '06. Proceedings of the 39th
  Annual Hawaii International Conference on}, volume~3, pages 59c--59c.

\bibitem[Boyd et~al., 2005]{Boyd2005Mixing}
Boyd, S., Ghosh, A., Prabhakar, B., and Shah, D. (2005).
\newblock Gossip algorithms: design, analysis and applications.
\newblock In {\em INFOCOM 2005. 24th Annual Joint Conference of the IEEE
  Computer and Communications Societies. Proceedings IEEE}, volume~3, pages
  1653 -- 1664 vol. 3.

\bibitem[Bradley, 2012]{Bradley2012Crawl}
Bradley, T. (2012).
\newblock 45,000 facebook accounts compromised: What to know.
\newblock \url{http://bit.ly/TUY3i8}.

\bibitem[Brandes, 2001]{Brandes2001BetweennessAlgo}
Brandes, U. (2001).
\newblock A faster algorithm for betweenness centrality.
\newblock {\em Journal of Mathematical Sociology}, 40(2):163--177.

\bibitem[Buchegger et~al., 2009]{Buchegger2009Peerson}
Buchegger, S., Schi\"{o}berg, D., Vu, L.-H., and Datta, A. (2009).
\newblock Peerson: P2p social networking: early experiences and insights.
\newblock In {\em Proceedings of the Second ACM EuroSys Workshop on Social
  Network Systems}, SNS '09, pages 46--52, New York, NY, USA. ACM.

\bibitem[Campan and Truta, 2009]{Campan2009Data}
Campan, A. and Truta, T.~M. (2009).
\newblock Data and structural k-anonymity in social networks.
\newblock In {\em Privacy, Security, and Trust in KDD}, pages 33--54. Springer.

\bibitem[Cao et~al., 2012]{Cao2012SybilRank}
Cao, Q., Sirivianos, M., Yang, X., and Pregueiro, T. (2012).
\newblock Aiding the detection of fake accounts in large scale social online
  services.
\newblock In {\em Proceedings of the 9th USENIX conference on Networked Systems
  Design and Implementation}, NSDI'12, pages 15--15, Berkeley, CA, USA. USENIX
  Association.

\bibitem[Carminati et~al., 2009]{Carminati2009Rule}
Carminati, B., Ferrari, E., Heatherly, R., Kantarcioglu, M., and Thuraisingham,
  B. (2009).
\newblock A semantic web based framework for social network access control.
\newblock In {\em Proceedings of the 14th ACM symposium on Access control
  models and technologies}, SACMAT '09.

\bibitem[Carminati et~al., 2006]{Carminati2006Trust}
Carminati, B., Ferrari, E., and Perego, A. (2006).
\newblock Rule-based access control for social networks.
\newblock In {\em Proceedings of the 2006 international conference on On the
  Move to Meaningful Internet Systems}, pages 1734--1744.

\bibitem[Cellan-Jones, 2012]{BBC2012Facebook1}
Cellan-Jones, R. (2012).
\newblock Facebook 'likes' and adverts' value doubted.
\newblock \url{http://www.bbc.co.uk/news/technology-18813237}.

\bibitem[Chiluka et~al., 2012]{Chiluka2012LTD}
Chiluka, N., Andrade, N., Pouwelse, J., and Sips, H. (2012).
\newblock Leveraging trust and distrust for sybil-tolerant voting in online
  social media.
\newblock In {\em Proceedings of the 1st Workshop on Privacy and Security in
  Online Social Media}, PSOSM '12, pages 1:1--1:8, New York, NY, USA. ACM.

\bibitem[Choi et~al., 2006]{Choi2006Trust}
Choi, H.~C., Kruk, S.~R., Grzonkowski, S., Stankiewicz, K., Davis, B., and
  Breslin, J. (2006).
\newblock Trust models for community aware identity management.
\newblock In {\em Proceedings of the Identity, Reference and Web Workshop, in
  conjunction with WWW 2006}, page 140Ð154.

\bibitem[Chu et~al., 2012]{Chu2012SpamCampaign}
Chu, Z., Widjaja, I., and Wang, H. (2012).
\newblock Detecting social spam campaigns on twitter.
\newblock In {\em In Proceedings of Conference on Applied Cryptography and
  Network Security}, Lecture Notes in Computer Science, pages 455--472.
  Springer Berlin Heidelberg.

\bibitem[Conti et~al., 2011]{Conti2011VPS}
Conti, M., Hasani, A., and Crispo, B. (2011).
\newblock Virtual private social networks.
\newblock In {\em Proceedings of the first ACM conference on Data and
  application security and privacy}, CODASPY '11, pages 39--50, New York, NY,
  USA. ACM.

\bibitem[Cummings et~al., 2002]{Cummings2002QOS}
Cummings, J.~N., Butler, B., and Kraut, R. (2002).
\newblock The quality of online social relationships.
\newblock {\em Commun. ACM}, 45(7):103--108.

\bibitem[Cutillo et~al., 2009a]{Cutillo2009Decentralization}
Cutillo, L., Molva, R., and Strufe, T. (2009a).
\newblock Privacy preserving social networking through decentralization.
\newblock In {\em Wireless On-Demand Network Systems and Services, 2009. WONS
  2009. Sixth International Conference on}, pages 145--152.

\bibitem[Cutillo et~al., 2009b]{Cutillo2009Safebook}
Cutillo, L., Molva, R., and Strufe, T. (2009b).
\newblock Safebook: A privacy-preserving online social network leveraging on
  real-life trust.
\newblock {\em Communications Magazine, IEEE}, 47(12):94--101.

\bibitem[Dam, 2009]{Dam2009Facebook}
Dam, W.~B. (2009).
\newblock School teacher suspended for facebook gun photo.
\newblock
  \url{http://www.foxnews.com/story/2009/02/05/schoolteacher-suspended-for-facebook-gun-photo/}.

\bibitem[Dandekar et~al., 2012]{Dandekar2012SFC}
Dandekar, P., Goel, A., Wellman, M.~P., and Wiedenbeck, B. (2012).
\newblock Strategic formation of credit networks.
\newblock In {\em Proceedings of the 21st international conference on World
  Wide Web}, WWW '12, pages 559--568, New York, NY, USA. ACM.

\bibitem[Danezis, 2009]{Danezis2009Privacypolicy}
Danezis, G. (2009).
\newblock Inferring privacy policies for social networking services.
\newblock In {\em Proceedings of the 2nd ACM workshop on Security and
  artificial intelligence}, AISec '09, pages 5--10, New York, NY, USA. ACM.

\bibitem[Danezis and Mittal., 2009]{Danezis2009SybilInfer}
Danezis, G. and Mittal., P. (2009).
\newblock Sybilinfer: Detecting sybil nodes using social networks.
\newblock In {\em Network and Distributed System Security Symposium (NDSS)}.

\bibitem[DeFigueiredo and Barr, July]{DeFigueiredo2005TrustDavis}
DeFigueiredo, D. and Barr, E. (July).
\newblock Trustdavis: a non-exploitable online reputation system.
\newblock In {\em E-Commerce Technology, 2005. CEC 2005. Seventh IEEE
  International Conference on}, pages 274--283.

\bibitem[Devriese and Piessens, 2010]{Devriese2010Multiexecution}
Devriese, D. and Piessens, F. (2010).
\newblock Noninterference through secure multi-execution.
\newblock In {\em Proceedings of the 2010 IEEE Symposium on Security and
  Privacy}, SP '10, pages 109--124, Washington, DC, USA. IEEE Computer Society.

\bibitem[Dey et~al., 2012]{Dey2012AgePrivacy}
Dey, R., Tang, C., Ross, K., and Saxena, N. (2012).
\newblock Estimating age privacy leakage in online social networks.
\newblock In {\em INFOCOM, 2012 Proceedings IEEE}, pages 2836--2840.

\bibitem[Douceur, 2002]{Douceur2002Sybil}
Douceur, J. (2002).
\newblock The sybil attack.
\newblock In Druschel, P., Kaashoek, F., and Rowstron, A., editors, {\em
  Peer-to-Peer Systems}, volume 2429 of {\em Lecture Notes in Computer
  Science}, pages 251--260. Springer Berlin Heidelberg.

\bibitem[Dwyer, 2011]{Dwyer2011Privacy}
Dwyer, C. (2011).
\newblock Privacy in the age of google and facebook.
\newblock {\em Technology and Society Magazine, IEEE}, 30(3):58--63.

\bibitem[Egele et~al., 2012]{Egele2012Pox}
Egele, M., Moser, A., Kruegel, C., and Kirda, E. (2012).
\newblock Pox: Protecting users from malicious facebook applications.
\newblock {\em Computer Communications}, 35(12).

\bibitem[Elahi et~al., 2008]{Elahi2008Rule}
Elahi, N., Chowdhury, M., and Noll, J. (2008).
\newblock Semantic access control in web based communities.
\newblock In {\em Proceedings of the 2008 The Third International
  Multi-Conference on Computing in the GlobalInformation Technology}, pages 131
  --136.

\bibitem[Ellis et~al., 2004]{Ellis2004Worm}
Ellis, D.~R., Aiken, J.~G., Attwood, K.~S., and Tenaglia, S.~D. (2004).
\newblock A behavioral approach to worm detection.
\newblock In {\em Proceedings of the 2004 ACM workshop on Rapid malcode}, WORM
  '04, pages 43--53, New York, NY, USA. ACM.

\bibitem[Engelmore, 1988]{EngelmoreAI}
Engelmore, R., editor (1988).
\newblock {\em Readings from the AI magazine}.
\newblock American Association for Artificial Intelligence, Menlo Park, CA,
  USA.

\bibitem[Facebook, 2012]{Facebook2012Koobface}
Facebook (2012).
\newblock Facebook's continued fight against koobface.
\newblock \url{http://on.fb.me/y5ibe1}.

\bibitem[Facebook, 2015]{Facebook2015Terms}
Facebook (2015).
\newblock Statement of rights and responsibilities.
\newblock \url{https://www.facebook.com/legal/terms}.

\bibitem[Fang and LeFevre, 2010]{Fang2010Wizard}
Fang, L. and LeFevre, K. (2010).
\newblock Privacy wizards for social networking sites.
\newblock In {\em Proceedings of the 19th international conference on World
  wide web}, WWW '10, pages 351--360, New York, NY, USA. ACM.

\bibitem[Felt and Evans, 2008]{Felt2008Privacy}
Felt, A. and Evans, D. (2008).
\newblock Privacy protection for social networking apis.
\newblock In {\em 2008 Web 2.0 Security and Privacy (W2SPÕ08)}.

\bibitem[Fiesler and Bruckman, 2014]{fiesler2014copyright}
Fiesler, C. and Bruckman, A. (2014).
\newblock Copyright terms in online creative communities.
\newblock In {\em CHI'14 Extended Abstracts on Human Factors in Computing
  Systems}, pages 2551--2556. ACM.

\bibitem[Flaxman, 2008]{Aiello2008Mixing}
Flaxman, A. (2008).
\newblock Expansion and lack thereof in randomly perturbed graphs.
\newblock {\em Algorithms and Models for the Web-Graph}, 4936:24--35.

\bibitem[Fong, 2011a]{Fong2011Privacy}
Fong, P. (2011a).
\newblock Preventing sybil attacks by privilege attenuation: A design principle
  for social network systems.
\newblock In {\em Security and Privacy (SP), 2011 IEEE Symposium on}, pages
  263--278.

\bibitem[Fong, 2011b]{Fong2011Relation}
Fong, P.~W. (2011b).
\newblock Relationship-based access control: protection model and policy
  language.
\newblock In {\em Proceedings of the first ACM conference on Data and
  application security and privacy}, pages 191--202.

\bibitem[Gao et~al., 2011]{Gao2011Security}
Gao, H., Hu, J., Huang, T., Wang, J., and Chen, Y. (2011).
\newblock Security issues in online social networks.
\newblock {\em Internet Computing, IEEE}, 15(4):56--63.

\bibitem[Gao et~al., 2010]{Gao2010Spamcampaign}
Gao, H., Hu, J., Wilson, C., Li, Z., Chen, Y., and Zhao, B.~Y. (2010).
\newblock Detecting and characterizing social spam campaigns.
\newblock In {\em Proceedings of the 10th ACM SIGCOMM conference on Internet
  measurement}, IMC '10, pages 35--47, New York, NY, USA. ACM.

\bibitem[Ghosh et~al., 2007]{Ghosh2007Credit}
Ghosh, A., Mahdian, M., Reeves, D., Pennock, D., and Fugger, R. (2007).
\newblock Mechanism design on trust networks.
\newblock In Deng, X. and Graham, F., editors, {\em Internet and Network
  Economics}, volume 4858 of {\em Lecture Notes in Computer Science}, pages
  257--268. Springer Berlin Heidelberg.

\bibitem[Giunchiglia et~al., 2008]{Giunchiglia2008RelBAC}
Giunchiglia, F., Zhang, R., and Crispo, B. (2008).
\newblock Relbac: Relation based access control.
\newblock In {\em Fourth International Conference on Semantics, Knowledge and
  Grid}, pages 3 --11.

\bibitem[Graffi et~al., 2011]{Graffi2011LifeSocial}
Graffi, K., Gross, C., Stingl, D., Hartung, D., Kovacevic, A., and Steinmetz,
  R. (2011).
\newblock Lifesocial.kom: A secure and p2p-based solution for online social
  networks.
\newblock In {\em Consumer Communications and Networking Conference (CCNC),
  2011 IEEE}, pages 554--558.

\bibitem[Grier et~al., 2010]{Grier2010SUCSpam}
Grier, C., Thomas, K., Paxson, V., and Zhang, M. (2010).
\newblock @spam: the underground on 140 characters or less.
\newblock In {\em Proceedings of the 17th ACM conference on Computer and
  communications security}, CCS '10, pages 27--37, New York, NY, USA. ACM.

\bibitem[Gross and Acquisti, 2005a]{Gross2005InformationRevelation}
Gross, R. and Acquisti, A. (2005a).
\newblock Information revelation and privacy in online social networks.
\newblock In {\em Proceedings of the 2005 ACM workshop on Privacy in the
  electronic society}, WPES '05, pages 71--80, New York, NY, USA. ACM.

\bibitem[Gross and Acquisti, 2005b]{Gross2005Revelation}
Gross, R. and Acquisti, A. (2005b).
\newblock Information revelation and privacy in online social networks.
\newblock In {\em Proceedings of the 2005 ACM workshop on Privacy in the
  electronic society}, pages 71--80.

\bibitem[Gruteser and Grunwald, 2003]{Gruteser2003Location}
Gruteser, M. and Grunwald, D. (2003).
\newblock Anonymous usage of location-based services through spatial and
  temporal cloaking.
\newblock In {\em Proceedings of the 1st international conference on Mobile
  systems, applications and services}, MobiSys '03, pages 31--42, New York, NY,
  USA. ACM.

\bibitem[Gubichev et~al., 2010]{Gubichev2010Landmark}
Gubichev, A., Bedathur, S., Seufert, S., and Weikum, G. (2010).
\newblock Fast and accurate estimation of shortest paths in large graphs.
\newblock In {\em Proceedings of the 19th ACM international conference on
  Information and knowledge management}, CIKM '10, pages 499--508, New York,
  NY, USA. ACM.

\bibitem[Guha et~al., 2008]{Guha2008Noyb}
Guha, S., Tang, K., and Francis, P. (2008).
\newblock Noyb: privacy in online social networks.
\newblock In {\em Proceedings of the first workshop on Online social networks},
  WOSN '08, pages 49--54, New York, NY, USA. ACM.

\bibitem[Hay et~al., 2008]{Hay2008RSR}
Hay, M., Miklau, G., Jensen, D., Towsley, D., and Weis, P. (2008).
\newblock Resisting structural re-identification in anonymized social networks.
\newblock {\em Proc. VLDB Endow.}, 1(1):102--114.

\bibitem[He et~al., 2011]{He2011Location}
He, W., Liu, X., and Ren, M. (2011).
\newblock Location cheating: A security challenge to location-based social
  network services.
\newblock In {\em Distributed Computing Systems (ICDCS), 2011 31st
  International Conference on}, pages 740--749. IEEE.

\bibitem[Heatherly et~al., 2013]{Heatherly2013Prevent}
Heatherly, R., Kantarcioglu, M., and Thuraisingham, B. (2013).
\newblock Preventing private information inference attacks on social networks.
\newblock {\em Knowledge and Data Engineering, IEEE Transactions on},
  25(8):1849--1862.

\bibitem[Heymann et~al., 2007]{Heymann2007Spam}
Heymann, P., Koutrika, G., and Garcia-Molina, H. (2007).
\newblock Fighting spam on social web sites: A survey of approaches and future
  challenges.
\newblock {\em IEEE Internet Computing}, 11(6):36--45.

\bibitem[Holt, 2013]{Holt2013Twitter}
Holt, R. (2013).
\newblock Twitter in numbers.
\newblock
  \url{http://www.telegraph.co.uk/technology/twitter/9945505/Twitter-in-numbers.html}.

\bibitem[Hwang et~al., 2012]{Hwang2012Bot}
Hwang, T., Pearce, I., and Nanis, M. (2012).
\newblock Socialbots: Voices from the fronts.
\newblock {\em interactions}, 19(2):38--45.

\bibitem[Irani et~al., 2011]{irani2011reverse}
Irani, D., Balduzzi, M., Balzarotti, D., Kirda, E., and Pu, C. (2011).
\newblock Reverse social engineering attacks in online social networks.
\newblock In {\em Detection of Intrusions and Malware, and Vulnerability
  Assessment}, pages 55--74. Springer.

\bibitem[Irani et~al., 2010]{irani2010study}
Irani, D., Webb, S., and Pu, C. (2010).
\newblock Study of static classification of social spam profiles in myspace.
\newblock In {\em Proceedings of the 4th International Conference on Weblogs
  and Social Media}.

\bibitem[Jagatic et~al., 2007]{Jagatic2007Phishing}
Jagatic, T.~N., Johnson, N.~A., Jakobsson, M., and Menczer, F. (2007).
\newblock Social phishing.
\newblock {\em Commun. ACM}, 50(10):94--100.

\bibitem[Jansen and Booth, 2010]{Jansen2010AOLresearch}
Jansen, B.~J. and Booth, D. (2010).
\newblock Classifying web queries by topic and user intent.
\newblock In {\em CHI '10 Extended Abstracts on Human Factors in Computing
  Systems}, CHI EA '10, pages 4285--4290, New York, NY, USA. ACM.

\bibitem[Jurek, 2011]{Matt2011GoogleSearch}
Jurek, M. (2011).
\newblock Google explores +1 button to influence search results.
\newblock
  \url{http://www.tekgoblin.com/2011/08/29/google-explores-1-button-to-influence-search-results/
  }.

\bibitem[Kanich et~al., 2008]{Kanich2008AnlysisSpam}
Kanich, C., Kreibich, C., Levchenko, K., Enright, B., Voelker, G.~M., Paxson,
  V., and Savage, S. (2008).
\newblock Spamalytics: an empirical analysis of spam marketing conversion.
\newblock In {\em Proceedings of the 15th ACM conference on Computer and
  communications security}, CCS '08, pages 3--14, New York, NY, USA. ACM.

\bibitem[Kayes and Iamnitchi, 2013a]{Kayes13Aegis}
Kayes, I. and Iamnitchi, A. (2013a).
\newblock Aegis: A semantic implementation of privacy as contextual integrity
  in social ecosystems.
\newblock In {\em 11th International Conference on Privacy, Security and Trust
  (PST)}.

\bibitem[Kayes and Iamnitchi, 2013b]{Kayes13Out_of_the_wild}
Kayes, I. and Iamnitchi, A. (2013b).
\newblock Out of the wild: On generating default policies in social ecosystems.
\newblock In {\em IEEE ICC'13 - Workshop on Beyond Social Networks: Collective
  Awareness}.

\bibitem[Kelly, 2008]{Kelly2008Risk}
Kelly, S. (2008).
\newblock Identity `at riskÕ on facebook.
\newblock \url{http://news.bbc.co.uk/2/hi/programmes/click_online/7375772.stm}.

\bibitem[Klimt and Yang, 2004]{klimt2004enron}
Klimt, B. and Yang, Y. (2004).
\newblock The enron corpus: A new dataset for email classification research.
\newblock In {\em 15th European Conference on Machine Learning}, pages
  217--226. Springer.

\bibitem[Kourtellis et~al., 2010]{Kourtellis2010Middleware}
Kourtellis, N., Finnis, J., Anderson, P., Blackburn, J., Borcea, C., and
  Iamnitchi, A. (2010).
\newblock Prometheus: User-controlled p2p social data management for
  socially-aware applications.
\newblock In {\em 11th International Middleware Conference}.

\bibitem[Kreibich and Crowcroft, 2004]{Kreibich2004Honeypots}
Kreibich, C. and Crowcroft, J. (2004).
\newblock Honeycomb: creating intrusion detection signatures using honeypots.
\newblock {\em SIGCOMM Comput. Commun. Rev.}, 34(1):51--56.

\bibitem[Kreibich et~al., 2009]{Kreibich2009Spamcraft}
Kreibich, C., Kanich, C., Levchenko, K., Enright, B., Voelker, G.~M., Paxson,
  V., and Savage, S. (2009).
\newblock Spamcraft: An inside look at spam campaign orchestration.
\newblock {\em Proc. of 2nd USENIX LEET}.

\bibitem[Krishnamurthy et~al., 2008]{Krishnamurthy2008Twitter}
Krishnamurthy, B., Gill, P., and Arlitt, M. (2008).
\newblock A few chirps about twitter.
\newblock In {\em Proceedings of the first workshop on Online social networks},
  pages 19--24.

\bibitem[Krishnamurthy and Wills, 2008]{Krishnamurthy2008Characterizing}
Krishnamurthy, B. and Wills, C.~E. (2008).
\newblock Characterizing privacy in online social networks.
\newblock In {\em Proceedings of the first workshop on Online social networks},
  pages 37--42.

\bibitem[Kruk, 2004]{Kruk2004FOAF}
Kruk, S. (2004).
\newblock Foaf-realm: control your friends access to the resource.
\newblock In {\em In Proceedings of the 1st Workshop on Friend of a Friend}.

\bibitem[Kwak et~al., 2010]{Kwak2010NewsMedia}
Kwak, H., Lee, C., Park, H., and Moon, S. (2010).
\newblock What is twitter, a social network or a news media?
\newblock In {\em Proceedings of the 19th international conference on World
  wide web}, WWW '10, pages 591--600, New York, NY, USA. ACM.

\bibitem[Langville and Meyer, 2004]{Langville04deeperinside}
Langville, A.~N. and Meyer, C.~D. (2004).
\newblock Deeper inside pagerank.
\newblock {\em Internet Mathematics}, 1:2004.

\bibitem[Lee et~al., 2010]{Lee2010Spam}
Lee, K., Caverlee, J., and Webb, S. (2010).
\newblock Uncovering social spammers: social honeypots + machine learning.
\newblock In {\em Proceedings of the 33rd international ACM SIGIR conference on
  Research and development in information retrieval}, SIGIR '10, pages
  435--442, New York, NY, USA. ACM.

\bibitem[Lenhart et~al., 2010]{Lenhart2010Report}
Lenhart, A., Purcell, K., Smith, A., and Zickuhr, K. (2010).
\newblock Social media and young adults.
\newblock \url{http://bit.ly/cQdgi3}.

\bibitem[Lewis and Gale, 1994]{Lewis1994SAT}
Lewis, D.~D. and Gale, W.~A. (1994).
\newblock A sequential algorithm for training text classifiers.
\newblock In {\em Proceedings of the 17th annual international ACM SIGIR
  conference on Research and development in information retrieval}, SIGIR '94,
  pages 3--12, New York, NY, USA. Springer-Verlag New York, Inc.

\bibitem[Lin et~al., 2007]{Lin2007Splog}
Lin, Y.-R., Sundaram, H., Chi, Y., Tatemura, J., and Tseng, B.~L. (2007).
\newblock Splog detection using self-similarity analysis on blog temporal
  dynamics.
\newblock In {\em Proceedings of the 3rd international workshop on Adversarial
  information retrieval on the web}, AIRWeb '07, pages 1--8, New York, NY, USA.
  ACM.

\bibitem[Lindamood et~al., 2009]{Lindamood2009IPI}
Lindamood, J., Heatherly, R., Kantarcioglu, M., and Thuraisingham, B. (2009).
\newblock Inferring private information using social network data.
\newblock In {\em Proceedings of the 18th international conference on World
  wide web}, WWW '09, pages 1145--1146, New York, NY, USA. ACM.

\bibitem[Lipford et~al., 2008]{Lipford2008Understanding}
Lipford, H.~R., Besmer, A., and Watson, J. (2008).
\newblock Understanding privacy settings in facebook with an audience view.
\newblock {\em UPSEC}, 8:1--8.

\bibitem[Liu and Terzi, 2008]{Liu2008anonymization}
Liu, K. and Terzi, E. (2008).
\newblock Towards identity anonymization on graphs.
\newblock In {\em Proceedings of the 2008 ACM SIGMOD international conference
  on Management of data}, SIGMOD '08, pages 93--106, New York, NY, USA. ACM.

\bibitem[Liu et~al., 2011]{Liu2011Facebook}
Liu, Y., Gummadi, K.~P., Krishnamurthy, B., and Mislove, A. (2011).
\newblock Analyzing facebook privacy settings: user expectations vs. reality.
\newblock In {\em Proceedings of the 2011 ACM SIGCOMM conference on Internet
  measurement conference}, pages 61--70.

\bibitem[Lotan et~al., 2011]{lotan2011arab}
Lotan, G., Graeff, E., Ananny, M., Gaffney, D., Pearce, I., et~al. (2011).
\newblock The arab spring| the revolutions were tweeted: Information flows
  during the 2011 tunisian and egyptian revolutions.
\newblock {\em International Journal of Communication}, 5:31.

\bibitem[Lucas and Borisov, 2008]{Lucas2008FlybyNight}
Lucas, M.~M. and Borisov, N. (2008).
\newblock Flybynight: mitigating the privacy risks of social networking.
\newblock In {\em Proceedings of the 7th ACM workshop on Privacy in the
  electronic society}, WPES '08, pages 1--8, New York, NY, USA. ACM.

\bibitem[Luo et~al., 2009]{Luo2009FaceCloak}
Luo, W., Xie, Q., and Hengartner, U. (2009).
\newblock Facecloak: An architecture for user privacy on social networking
  sites.
\newblock In {\em Computational Science and Engineering, 2009. CSE '09.
  International Conference on}, volume~3, pages 26--33.

\bibitem[Madejski et~al., 2011]{Madejski2011Failure}
Madejski, M., Johnson, M.~L., and Bellovin, S.~M. (2011).
\newblock The failure of online social network privacy settings.
\newblock {\em Department of Computer Science, Columbia University}.

\bibitem[Mail, 2011]{Daily2001Banker}
Mail, D. (2011).
\newblock Bank worker fired for facebook post comparing her £$7$-an-hour wage
  to lloyds boss's £$4,000$-an-hour salary.
\newblock \url{http://dailym.ai/fjRTlC}.

\bibitem[Marinando, 2010]{MARINANDO2010Tuenti}
Marinando (2010).
\newblock Tuenti, spain's leading social network, switches on local for a
  location-based future.
\newblock
  \url{http://techcrunch.com/2010/03/25/tuenti-spains_leading_social_network_mixes_local_into_social_in_a_very_big_way/}.

\bibitem[Masoumzadeh and Joshi, 2011]{Masoumzadeh2011Rule}
Masoumzadeh, A. and Joshi, J. (2011).
\newblock Ontology-based access control for social network systems.
\newblock {\em IJIPSI}, 1(1):59--78.

\bibitem[McCarthy, 2009]{McCarthy2009DDoS}
McCarthy, C. (2009).
\newblock Twitter crippled by denial-of-service attack.
\newblock \url{http://news.cnet.com/8301-13577_3-10304633-36.html}.

\bibitem[Mehta et~al., 2008]{Mehta2008Spam}
Mehta, B., Nangia, S., Gupta, M., and Nejdl, W. (2008).
\newblock Detecting image spam using visual features and near duplicate
  detection.
\newblock In {\em Proceedings of the 17th international conference on World
  Wide Web}, WWW '08, pages 497--506, New York, NY, USA. ACM.

\bibitem[Mills, 2008]{Mills2008Risk}
Mills, E. (2008).
\newblock Facebook suspends app that permitted peephole.
\newblock \url{http://news.cnet.com/8301-10784\_3-9977762-7.html}.

\bibitem[Mirkovic et~al., 2004]{Mirkovic2004DOS}
Mirkovic, J., Dietrich, S., Dittrich, D., and Reiher, P. (2004).
\newblock {\em Internet Denial of Service: Attack and Defense Mechanisms (Radia
  Perlman Computer Networking and Security)}.
\newblock Prentice Hall PTR, Upper Saddle River, NJ, USA.

\bibitem[Mirkovic and Reiher, 2004]{mirkovic2004taxonomy}
Mirkovic, J. and Reiher, P. (2004).
\newblock A taxonomy of ddos attack and ddos defense mechanisms.
\newblock {\em ACM SIGCOMM Computer Communication Review}, 34(2):39--53.

\bibitem[Mishra et~al., 2007]{Mishra2007Clustering}
Mishra, N., Schreiber, R., Stanton, I., and Tarjan, R.~E. (2007).
\newblock Clustering social networks.
\newblock In {\em Proceedings of the 5th international conference on Algorithms
  and models for the web-graph}, WAW'07, pages 56--67, Berlin, Heidelberg.
  Springer-Verlag.

\bibitem[Mislove et~al., 2008]{Mislove2008Ostra}
Mislove, A., Post, A., Druschel, P., and Gummadi, K.~P. (2008).
\newblock Ostra: leveraging trust to thwart unwanted communication.
\newblock In {\em Proceedings of the 5th USENIX Symposium on Networked Systems
  Design and Implementation}, NSDI'08, pages 15--30, Berkeley, CA, USA. USENIX
  Association.

\bibitem[Mohaisen et~al., 2010]{Mohaisen2010Mixing}
Mohaisen, A., Yun, A., and Kim, Y. (2010).
\newblock Measuring the mixing time of social graphs.
\newblock In {\em Proceedings of the 10th ACM SIGCOMM conference on Internet
  measurement}, IMC '10, pages 383--389, New York, NY, USA. ACM.

\bibitem[Mondal et~al., 2012]{mondal2012genie}
Mondal, M., Viswanath, B., Clement, A., Druschel, P., Gummadi, K.~P., Mislove,
  A., and Post, A. (2012).
\newblock Defending against large-scale crawls in online social networks.
\newblock In {\em Proceedings of the 8th ACM International Conference on
  emerging Networking EXperiments and Technologies (CoNEXT'12)}, Nice, France.

\bibitem[Motoyama et~al., 2011]{Motoyama2011Dirtyjob}
Motoyama, M., McCoy, D., Levchenko, K., Savage, S., and Voelker, G.~M. (2011).
\newblock Dirty jobs: the role of freelance labor in web service abuse.
\newblock In {\em Proceedings of the 20th USENIX conference on Security},
  SEC'11, pages 14--14, Berkeley, CA, USA. USENIX Association.

\bibitem[Murphy, 2010]{Murphy2010Email}
Murphy, S. (2010).
\newblock Teens ditch e-mail for texting and facebook.
\newblock \url{http://www.nbcnews.com/id/38585236/}.

\bibitem[Narayanan et~al., 2011]{Narayanan2011Link}
Narayanan, A., Shi, E., and Rubinstein, B.~I. (2011).
\newblock Link prediction by de-anonymization: How we won the kaggle social
  network challenge.
\newblock In {\em Neural Networks (IJCNN), The 2011 International Joint
  Conference on}, pages 1825--1834. IEEE.

\bibitem[Narayanan and Shmatikov, 2009]{Narayanan2009De-anonymizing}
Narayanan, A. and Shmatikov, V. (2009).
\newblock De-anonymizing social networks.
\newblock In {\em Security and Privacy, 2009 30th IEEE Symposium on}, pages
  173--187.

\bibitem[Nazir et~al., 2010]{Nazir2010Ghostbusting}
Nazir, A., Raza, S., Chuah, C.-N., and Schipper, B. (2010).
\newblock Ghostbusting facebook: detecting and characterizing phantom profiles
  in online social gaming applications.
\newblock In {\em Proceedings of the 3rd conference on Online social networks},
  WOSN'10, pages 1--1, Berkeley, CA, USA. USENIX Association.

\bibitem[Newman, 2010]{Newman2010Book}
Newman, M. E.~J. (2010).
\newblock {\em Networks: An Introduction}.
\newblock Oxford University Press.

\bibitem[Nielsen, 2012]{Nielsen2009OSN}
Nielsen (2012).
\newblock Social networks \& blogs now $4^{th}$ most popular online activity,
  ahead of personal e-mail.
\newblock
  \url{http://www.nielsen.com/us/en/press-room/2009/social_networks__.html}.

\bibitem[Nissenbaum, 2004]{Nissenbaum2004Context}
Nissenbaum, H. (2004).
\newblock Privacy as contextual integrity.
\newblock {\em Washington Law Review}, 79(1):119--158.

\bibitem[Nissenbaum, 2011]{Nissenbaum2004ContextOnline}
Nissenbaum, H. (2011).
\newblock A contextual approach to privacy online.
\newblock {\em Daedalus}, 140(4):32--48.

\bibitem[Ntoulas et~al., 2006]{Ntoulas2006SpamWeb}
Ntoulas, A., Najork, M., Manasse, M., and Fetterly, D. (2006).
\newblock Detecting spam web pages through content analysis.
\newblock In {\em Proceedings of the 15th international conference on World
  Wide Web}, WWW '06, pages 83--92, New York, NY, USA. ACM.

\bibitem[Nunes et~al., 2008]{Nunes2008AOLresearch}
Nunes, S., Ribeiro, C., and David, G. (2008).
\newblock Use of temporal expressions in web search.
\newblock In {\em Proceedings of the IR research, 30th European conference on
  Advances in information retrieval}, ECIR'08, pages 580--584, Berlin,
  Heidelberg. Springer-Verlag.

\bibitem[Paul et~al., 2012]{Paul2012C4PS}
Paul, T., Stopczynski, M., Puscher, D., Volkamer, M., and Strufe, T. (2012).
\newblock C4ps - helping facebookers manage their privacy settings.
\newblock In {\em Social Informatics}, pages 188--201.

\bibitem[Peterson and Sirer, 2009]{peterson2009antfarm}
Peterson, R.~S. and Sirer, E.~G. (2009).
\newblock Antfarm: efficient content distribution with managed swarms.
\newblock In {\em Proceedings of the 6th USENIX symposium on Networked systems
  design and implementation}, pages 107--122.

\bibitem[Piatek et~al., 2008]{piatek2008one}
Piatek, M., Isdal, T., Krishnamurthy, A., and Anderson, T. (2008).
\newblock One hop reputations for peer to peer file sharing workloads.
\newblock In {\em NSDI'08}.

\bibitem[Post et~al., 2011]{Post2011Bazar}
Post, A., Shah, V., and Mislove, A. (2011).
\newblock Bazaar: strengthening user reputations in online marketplaces.
\newblock In {\em Proceedings of the 8th USENIX conference on Networked systems
  design and implementation}, NSDI'11, pages 14--14, Berkeley, CA, USA. USENIX
  Association.

\bibitem[Prince et~al., 2005]{Prince2005Honeypots}
Prince, M., Dahl, B., Holloway, L., Keller, A., and Langheinrich, E. (2005).
\newblock Understanding how spammers steal your e-mail address: An analysis of
  the first six months of data from project honey pot.
\newblock In {\em Second Conference on Email and Anti-Spam}.

\bibitem[Quercia and Hailes, 2010]{Quercia2010MobileSybil}
Quercia, D. and Hailes, S. (2010).
\newblock Sybil attacks against mobile users: Friends and foes to the rescue.
\newblock In {\em INFOCOM, 2010 Proceedings IEEE}, pages 1--5.

\bibitem[Ratkiewicz et~al., 2011]{ratkiewicz2011detecting}
Ratkiewicz, J., Conover, M., Meiss, M., Gon{\c{c}}alves, B., Flammini, A., and
  Menczer, F. (2011).
\newblock Detecting and tracking political abuse in social media.
\newblock In {\em Proc. of ICWSM}.

\bibitem[Reynaert et~al., 2012]{Reynaert2012PESAP}
Reynaert, T., De~Groef, W., Devriese, D., Desmet, L., and Piessens, F. (2012).
\newblock Pesap: A privacy enhanced social application platform.
\newblock In {\em Privacy, Security, Risk and Trust (PASSAT), 2012
  International Conference on and 2012 International Confernece on Social
  Computing (SocialCom)}, pages 827--833.

\bibitem[Riley, 2007]{RILEYYoutube}
Riley, D. (2007).
\newblock Stat gaming services come to youtube.
\newblock \url{http://www.bbc.co.uk/news/technology-18813237}.

\bibitem[Russell et~al., 1995]{Russell1995Artificial}
Russell, S.~J., Norvig, P., Canny, J.~F., Malik, J.~M., and Edwards, D.~D.
  (1995).
\newblock {\em Artificial intelligence: a modern approach}, volume~74.
\newblock Prentice hall Englewood Cliffs.

\bibitem[Saltzer and Schroeder, 1975]{Saltzer1975Least}
Saltzer, J. and Schroeder, M. (1975).
\newblock The protection of information in computer systems.
\newblock {\em Proceedings of the IEEE}, 63(9):1278--1308.

\bibitem[Shakimov et~al., 2011]{Shakimov2011visavis}
Shakimov, A., Lim, H., Caceres, R., Cox, L., Li, K., Liu, D., and Varshavsky,
  A. (2011).
\newblock Vis- a-vis: Privacy-preserving online social networking via virtual
  individual servers.
\newblock In {\em Communication Systems and Networks (COMSNETS), 2011 Third
  International Conference on}, pages 1--10.

\bibitem[Shehab et~al., 2010]{Shehab2010PolicyMgr}
Shehab, M., Cheek, G., Touati, H., Squicciarini, A., and Cheng, P.-C. (2010).
\newblock User centric policy management in online social networks.
\newblock In {\em IEEE International Symposium on Policies for Distributed
  Systems and Networks (POLICY)}, pages 9 --13.

\bibitem[Shetty and Adibi, 2005]{Shetty2005ENron}
Shetty, J. and Adibi, J. (2005).
\newblock Discovering important nodes through graph entropy the case of enron
  email database.
\newblock In {\em Proceedings of the 3rd international workshop on Link
  discovery}, LinkKDD '05, pages 74--81, New York, NY, USA. ACM.

\bibitem[Simpson, 2008]{Simpson2008FineGrained}
Simpson, A. (2008).
\newblock On the need for user-defined fine-grained access control policies for
  social networking applications.
\newblock In {\em Proceedings of the workshop on Security in Opportunistic and
  SOCial networks}, SOSOC '08, pages 1:1--1:8, New York, NY, USA. ACM.

\bibitem[Singh et~al., 2009]{Singh2009xBook}
Singh, K., Bhola, S., and Lee, W. (2009).
\newblock xbook: redesigning privacy control in social networking platforms.
\newblock In {\em Proceedings of the 18th conference on USENIX security
  symposium}, SSYM'09, pages 249--266, Berkeley, CA, USA. USENIX Association.

\bibitem[Spitzner, 2002]{Spitzner2002Honeypots}
Spitzner, L. (2002).
\newblock {\em Honeypots tracking hackers}.
\newblock Addison-Wesley, 1 edition.

\bibitem[Spitzner, 2003]{Spitzner2003Honeypots}
Spitzner, L. (2003).
\newblock The honeynet project: trapping the hackers.
\newblock {\em Security Privacy, IEEE}, 1(2):15--23.

\bibitem[Squicciarini et~al., 2010]{Squicciarini2010Prima}
Squicciarini, A., Paci, F., and Sundareswaran, S. (2010).
\newblock Prima: an effective privacy protection mechanism for social networks.
\newblock In {\em Proceedings of the 5th ACM Symposium on Information, Computer
  and Communications Security}, pages 320--323.

\bibitem[Staff, 2010]{Staff2010Sale}
Staff, E. (2010).
\newblock Verisign: 1.5m facebook accounts for sale in web forum.
\newblock \url{http://www.pcmag.com/article2/0,2817,2363004,00.asp}.

\bibitem[Steel and Fowler, 2010]{Steel2010RIsk}
Steel, E. and Fowler, G.~A. (2010).
\newblock Facebook in privacy breach.
\newblock
  \url{http://online.wsj.com/article/SB10001424052702304772804575558484075236968.html}.

\bibitem[Stein et~al., 2011]{Stein2011FIS}
Stein, T., Chen, E., and Mangla, K. (2011).
\newblock Facebook immune system.
\newblock In {\em Proceedings of the 4th Workshop on Social Network Systems},
  SNS '11, pages 8:1--8:8, New York, NY, USA. ACM.

\bibitem[Strater and Lipford, 2008]{Strater2008StrategisandStruggles}
Strater, K. and Lipford, H.~R. (2008).
\newblock Strategies and struggles with privacy in an online social networking
  community.
\newblock In {\em Proceedings of the 22nd British HCI Group Annual Conference
  on People and Computers: Culture, Creativity, Interaction - Volume 1},
  BCS-HCI '08, pages 111--119, Swinton, UK, UK. British Computer Society.

\bibitem[Stringhini et~al., 2010]{Stringhini2010Spam}
Stringhini, G., Kruegel, C., and Vigna, G. (2010).
\newblock Detecting spammers on social networks.
\newblock In {\em Proceedings of the 26th Annual Computer Security Applications
  Conference}, ACSAC '10, pages 1--9, New York, NY, USA. ACM.

\bibitem[Stringhini et~al., 2013]{Stringhini2013Follower}
Stringhini, G., Wang, G., Egele, M., Kruegel, C., Vigna, G., Zheng, H., and
  Zhao, B.~Y. (2013).
\newblock Follow the green: Growth and dynamics in twitter follower markets.
\newblock In {\em Proceedings of the 2013 Conference on Internet Measurement
  Conference}, IMC '13, pages 163--176, New York, NY, USA. ACM.

\bibitem[Strufe, 2010]{Strufe2010PPB}
Strufe, T. (2010).
\newblock Profile popularity in a business-oriented online social network.
\newblock In {\em Proceedings of the 3rd Workshop on Social Network Systems},
  SNS '10, pages 2:1--2:6, New York, NY, USA. ACM.

\bibitem[Sweeney, 2000]{Sweeney2000Uniqueness}
Sweeney, L. (2000).
\newblock Uniqueness of simple demographics in the us population.
\newblock {\em Carnegie Mellon University Laboratory for International Data
  Privacy}.

\bibitem[Sweeney, 2002]{Sweeney2002k}
Sweeney, L. (2002).
\newblock k-anonymity: A model for protecting privacy.
\newblock {\em International Journal of Uncertainty, Fuzziness and
  Knowledge-Based Systems}, 10(05):557--570.

\bibitem[Thomas et~al., 2011]{Thomas2011SuspendedAccounts}
Thomas, K., Grier, C., Song, D., and Paxson, V. (2011).
\newblock Suspended accounts in retrospect: an analysis of twitter spam.
\newblock In {\em Proceedings of the 2011 ACM SIGCOMM conference on Internet
  measurement conference}, IMC '11, pages 243--258, New York, NY, USA. ACM.

\bibitem[Tran et~al., 2009]{Tran09sumUp}
Tran, N., Min, B., Li, J., and Subramanian, L. (2009).
\newblock Sybil-resilient online content voting.
\newblock In {\em In Proceedings of the 6th Symposium on Networked System
  Design and Implementation (NSDI}.

\bibitem[Tsuchiya, 1988]{Tsuchiya1988Landmark}
Tsuchiya, P.~F. (1988).
\newblock The landmark hierarchy: a new hierarchy for routing in very large
  networks.
\newblock {\em SIGCOMM Comput. Commun. Rev.}, 18(4):35--42.

\bibitem[Ur and Ganapathy, 2009]{Ur2009Evaluating}
Ur, B.~E. and Ganapathy, V. (2009).
\newblock Evaluating attack amplification in online social networks.
\newblock In {\em Proceedings of the 2009 Web 2.0 Security and Privacy
  Workshop}. Citeseer.

\bibitem[Viswanath et~al., 2012a]{Viswanath2012Design}
Viswanath, B., Mondal, M., Clement, A., Druschel, P., Gummadi, K., Mislove, A.,
  and Post, A. (2012a).
\newblock Exploring the design space of social network-based sybil defenses.
\newblock In {\em Communication Systems and Networks (COMSNETS), 2012 Fourth
  International Conference on}, pages 1--8.

\bibitem[Viswanath et~al., 2012b]{Viswanath2012Canal}
Viswanath, B., Mondal, M., Gummadi, K.~P., Mislove, A., and Post, A. (2012b).
\newblock Canal: scaling social network-based sybil tolerance schemes.
\newblock In {\em Proceedings of the 7th ACM european conference on Computer
  Systems}, EuroSys '12, pages 309--322, New York, NY, USA. ACM.

\bibitem[Viswanath et~al., 2010]{Viswanath2010Community}
Viswanath, B., Post, A., Gummadi, K.~P., and Mislove, A. (2010).
\newblock An analysis of social network-based sybil defenses.
\newblock In {\em Proceedings of the ACM SIGCOMM 2010 conference}, SIGCOMM '10,
  pages 363--374, New York, NY, USA. ACM.

\bibitem[von Ahn et~al., 2004]{vonAhn2004Captcha}
von Ahn, L., Blum, M., and Langford, J. (2004).
\newblock Telling humans and computers apart automatically.
\newblock {\em Commun. ACM}, 47(2):56--60.

\bibitem[Wagner et~al., 2012]{Wagner2012socialbot}
Wagner, C., Mitter, S., K{\"o}rner, C., and Strohmaier, M. (2012).
\newblock When social bots attack: Modeling susceptibility of users in online
  social networks.
\newblock In {\em Proceedings of the WWW}, volume~12.

\bibitem[Walsh and Sirer, 2006]{walsh2006experience}
Walsh, K. and Sirer, E.~G. (2006).
\newblock Experience with an object reputation system for peer-to-peer
  filesharing.
\newblock In {\em NSDI'06}.

\bibitem[Wang et~al., 2013]{Wang2013Turing}
Wang, G., Mohanlal, M., Wilson, C., Xiao~Wang, M.~M., Zheng, H., and Zhao,
  B.~Y. (2013).
\newblock Social turing tests: Crowdsourcing sybil detection.
\newblock In {\em In Proceedings of The 20th Annual Network \& Distributed
  System Security Symposium (NDSS)}.

\bibitem[Webb et~al., 2008]{Webb2008SocialHoneypots}
Webb, S., Caverlee, J., and Pu, C. (2008).
\newblock Social honeypots: Making friends with a spammer near you.
\newblock In {\em Proceedings of the Fifth Conference on Email and Anti-Spam
  (CEAS 2008), Mountain View, CA}.

\bibitem[Wei et~al., arch]{Wei2012SybilDefender}
Wei, W., Xu, F., Tan, C., and Li, Q. (March).
\newblock Sybildefender: Defend against sybil attacks in large social networks.
\newblock In {\em INFOCOM, 2012 Proceedings IEEE}, pages 1951--1959.

\bibitem[Wilson et~al., 2010]{Wilson2010spikestrip}
Wilson, C., Sala, A., Bonneau, J., Zablit, R., and Zhao, B.~Y. (2010).
\newblock Don't tread on me: moderating access to osn data with spikestrip.
\newblock In {\em Proceedings of the 3rd conference on Online social networks},
  WOSN'10, pages 5--5, Berkeley, CA, USA. USENIX Association.

\bibitem[Wu et~al., 2010]{Wu2010Anonymization}
Wu, X., Ying, X., Liu, K., and Chen, L. (2010).
\newblock {\em A survey of privacy-preservation of graphs and social networks},
  pages 421--453.
\newblock Springer.

\bibitem[Xiao and Tao, 2006]{Xiao2006Personalized}
Xiao, X. and Tao, Y. (2006).
\newblock Personalized privacy preservation.
\newblock In {\em Proceedings of the 2006 ACM SIGMOD international conference
  on Management of data}, pages 229--240. ACM.

\bibitem[Xie et~al., 2006]{Xie2006EDA}
Xie, M., Yin, H., and Wang, H. (2006).
\newblock An effective defense against email spam laundering.
\newblock In {\em Proceedings of the 13th ACM conference on Computer and
  communications security}, CCS '06, pages 179--190, New York, NY, USA. ACM.

\bibitem[Xie et~al., 2008]{Xie2008SpamSignature}
Xie, Y., Yu, F., Achan, K., Panigrahy, R., Hulten, G., and Osipkov, I. (2008).
\newblock Spamming botnets: signatures and characteristics.
\newblock {\em SIGCOMM Comput. Commun. Rev.}, 38(4):171--182.

\bibitem[Xu et~al., 2010a]{Xu2010EdgeBetween}
Xu, L., Chainan, S., Takizawa, H., and Kobayashi, H. (2010a).
\newblock Resisting sybil attack by social network and network clustering.
\newblock In {\em Proceedings of the 2010 10th IEEE/IPSJ International
  Symposium on Applications and the Internet}, SAINT '10, pages 15--21,
  Washington, DC, USA. IEEE Computer Society.

\bibitem[Xu et~al., 2010b]{Xu2010Malware}
Xu, W., Zhang, F., and Zhu, S. (2010b).
\newblock Toward worm detection in online social networks.
\newblock In {\em Proceedings of the 26th Annual Computer Security Applications
  Conference}, ACSAC '10, pages 11--20, New York, NY, USA. ACM.

\bibitem[Yang et~al., 2011]{Yang2011WildSybil}
Yang, Z., Wilson, C., Wang, X., Gao, T., Zhao, B.~Y., and Dai, Y. (2011).
\newblock Uncovering social network sybils in the wild.
\newblock In {\em Proceedings of the 2011 ACM SIGCOMM conference on Internet
  measurement conference}, IMC '11, pages 259--268, New York, NY, USA. ACM.

\bibitem[Yu, 2011]{Yu2011Sybil}
Yu, H. (2011).
\newblock Sybil defenses via social networks: A tutorial and survey.
\newblock {\em SIGACT News}, 42(3):80--101.

\bibitem[Yu et~al., 2010]{Yu2010Limit}
Yu, H., Gibbons, P.~B., Kaminsky, M., and Xiao, F. (2010).
\newblock Sybillimit: a near-optimal social network defense against sybil
  attacks.
\newblock {\em IEEE/ACM Trans. Netw.}, 18(3):885--898.

\bibitem[Yu et~al., 2006]{YuSybilGuard2006}
Yu, H., Kaminsky, M., Gibbons, P.~B., and Flaxman, A. (2006).
\newblock Sybilguard: defending against sybil attacks via social networks.
\newblock In {\em Proceedings of the 2006 conference on Applications,
  technologies, architectures, and protocols for computer communications},
  SIGCOMM '06, pages 267--278, New York, NY, USA. ACM.

\bibitem[Zhang et~al., 2010]{zhang2010privacy}
Zhang, C., Sun, J., Zhu, X., and Fang, Y. (2010).
\newblock Privacy and security for online social networks: challenges and
  opportunities.
\newblock {\em Network, IEEE}, 24(4):13--18.

\bibitem[Zheleva and Getoor, 2008]{Zheleva2008Preserving}
Zheleva, E. and Getoor, L. (2008).
\newblock Preserving the privacy of sensitive relationships in graph data.
\newblock In {\em Privacy, security, and trust in KDD}, pages 153--171.
  Springer.

\bibitem[Zheleva and Getoor, 2009]{Zheleva2009ToJoin}
Zheleva, E. and Getoor, L. (2009).
\newblock To join or not to join: the illusion of privacy in social networks
  with mixed public and private user profiles.
\newblock In {\em Proceedings of the 18th international conference on World
  wide web}, WWW '09, pages 531--540, New York, NY, USA. ACM.

\bibitem[Zheleva and Getoor, 2011]{zheleva2011privacy}
Zheleva, E. and Getoor, L. (2011).
\newblock {\em Privacy in social networks: A survey}, pages 277--306.
\newblock Springer.

\bibitem[Zhou and Pei, 2011]{Zhou2011k}
Zhou, B. and Pei, J. (2011).
\newblock The k-anonymity and l-diversity approaches for privacy preservation
  in social networks against neighborhood attacks.
\newblock {\em Knowledge and Information Systems}, 28(1):47--77.

\bibitem[Zhou et~al., 2008]{Zhou2008Anonymization}
Zhou, B., Pei, J., and Luk, W. (2008).
\newblock A brief survey on anonymization techniques for privacy preserving
  publishing of social network data.
\newblock {\em SIGKDD Explor. Newsl.}, 10(2):12--22.

\bibitem[Zhu and Cao, 2011]{Zhu2011Applaus}
Zhu, Z. and Cao, G. (2011).
\newblock Applaus: A privacy-preserving location proof updating system for
  location-based services.
\newblock In {\em INFOCOM, 2011 Proceedings IEEE}, pages 1889--1897. IEEE.

\bibitem[Zinman and Donath, 2007]{Zinman2007Britney}
Zinman, A. and Donath, J. (2007).
\newblock Is britney spears spam.
\newblock In {\em Fourth Conference on Email and Anti-Spam, Mountain View, CA}.

\end{thebibliography}
                                % Sample .bib file with references that match those in
                                % the 'Specifications Document (V1.5)' as well containing
                                % 'legacy' bibs and bibs with 'alternate codings'.
                                % Gerry Murray - March 2012

% History dates

\end{document}